\documentclass[12pt]{article}
\pdfoutput=1
\usepackage{jheppub}
\usepackage{amsmath,bbm,array,amsfonts,graphicx,wrapfig,lscape,float,mathtools,multirow,longtable,rotating,subfigure,ulem,cancel}
\usepackage[section]{placeins}

\newcommand{\be}{\begin{equation}}
\newcommand{\ee}{\end{equation}}
\newcommand{\beq}{\begin{equation}}
\newcommand{\beql}[1]{\begin{equation}\label{#1}}
\newcommand{\eeq}{\end{equation}}
\newcommand{\ba}{\begin{array}}
\newcommand{\ea}{\end{array}}
\newcommand{\bea}{\begin{eqnarray}}
\newcommand{\beal}[1]{\begin{eqnarray}\label{#1}}
\newcommand{\eea}{\end{eqnarray}}
\newcommand{\ben}{\begin{enumerate}}
\newcommand{\een}{\end{enumerate}}
\newcommand{\bean}{\begin{eqnarray*}}
\newcommand{\eean}{\end{eqnarray*}}

\newcommand{\btab}[1]{\begin{tabular}{#1}}
\newcommand{\etab}{\end{tabular}}

\newcommand{\comment}[1]{}

\newcommand{\qed}{\nobreak \ifvmode \relax \else
      \ifdim\lastskip<1.5em \hskip-\lastskip
      \hskip1.5em plus0em minus0.5em \fi \nobreak
      \vrule height0.75em width0.5em depth0.25em\fi}

\numberwithin{equation}{section}

\newcommand{\dummy}{${\begin{array}{*{20}{l}}{~}\\{~}\end{array}}$}
\newcommand{\pal}{\ldots \text{palindrome}\ldots}
\newcommand{\ns}{$\text{not shown} $}

\title{Quiver Theories and Formulae for Nilpotent Orbits of Exceptional Algebras}

\author{Amihay Hanany, }
\author{Rudolph Kalveks}

\affiliation{
Theoretical Physics Group, The Blackett Laboratory,
Imperial College London, \\
Prince Consort Road, London SW7 2AZ, United Kingdom
}

\emailAdd{a.hanany@imperial.ac.uk, rudolph.kalveks09@imperial.ac.uk}

\preprint{Imperial/TP/17/AH/05}

\abstract{ 
We treat the topic of the closures of the nilpotent orbits of the Lie algebras of Exceptional groups through their descriptions as moduli spaces, in terms of Hilbert series and the highest weight generating functions for their representation content. We extend the set of known Coulomb branch quiver theory constructions for Exceptional group minimal nilpotent orbits, or reduced single instanton moduli spaces, to include all orbits of Characteristic Height 2, drawing on extended Dynkin diagrams and the unitary monopole formula. We also present a representation theoretic formula, based on localisation methods, for the normal nilpotent orbits of the Lie algebras of any Classical or Exceptional group. We analyse lower dimensioned Exceptional group nilpotent orbits in terms of Hilbert series and the Highest Weight Generating functions for their decompositions into characters of irreducible representations and/or Hall Littlewood polynomials. We investigate the relationships between the moduli spaces describing different nilpotent orbits and propose candidates for the constructions of some non-normal nilpotent orbits of Exceptional algebras.
\\

~\today}

\begin{document}

\maketitle

\listoftables

\listoffigures

\section{Introduction}
\label{sec:intro}

The relationships between supersymmetric (``SUSY") quiver gauge theories and the nilpotent orbits of Classical Lie groups  were examined in the companion paper \cite{Hanany:2016gbz} (which elaborates on the motivation for these studies). It was shown how (i) any Classical group nilpotent orbit can be constructed as the moduli space of an ${\cal N}=2$ Higgs branch quiver theory in $4d$, and (ii) any $A$ series nilpotent orbit, or any $BCD$ series near to minimal nilpotent orbit, can be constructed as the moduli space of an ${\cal N}=4$ Coulomb branch quiver theory in $2+1$ dimensions, based on a Dynkin diagram.

In the case of Exceptional groups, the Higgs branch method of constructing nilpotent orbits is not available since Exceptional groups do not act as matrices on their fundamental vector spaces. Furthermore, while Coulomb branch quiver theory constructions for minimal nilpotent orbits have been known for some time \cite{Intriligator:1996ex, Cremonesi:2013lqa, Cremonesi:2014xha}, and while maximal nilpotent orbits correspond to modified Hall Littlewood polynomials transforming in the singlet representation of a group $G$ \cite{Cremonesi:2014kwa}, quiver theory constructions for other nilpotent orbits of Exceptional groups have not been given in the Literature. 

The purpose of this note is to examine Coulomb branch quiver theory constructions for the nilpotent orbits of Exceptional groups, beyond the minimal nilpotent orbit, and also to develop representation theoretic methods for calculating properties of these moduli spaces. This in turn facilitates the exploration of the branching relationships between Exceptional group and Classical nilpotent orbits (and their quiver theories).

As in \cite{Hanany:2016gbz}, we approach the topic of Exceptional group nilpotent orbits through their constructions as moduli spaces, with Hilbert series (``HS") that can be analysed using the tools of the Plethystics Program \cite{Feng:2007ur,Hanany:2014dia}. Each such HS counts holomorphic functions on the closure of a nilpotent orbit \cite{Namikawa:2016xe}. For brevity, unless the context dictates otherwise, this paper generally refers to \emph{closures of nilpotent orbits} simply as ``nilpotent orbits".

We summarise in appendix \ref{apxNO}, relevant aspects of the theory of nilpotent orbits from the mathematical literature \cite{Dynkin:1957um, Collingwood:1993fk} and give simple algorithms for identifying nilpotent orbits, by enumerating homomorphisms from $SU(2)$ to $G$ using character maps and selection rules, and for calculating their dimensions. The character map for a nilpotent orbit of $G$ follows directly from its Characteristic \cite{Dynkin:1957um}, and we use Characteristics to label nilpotent orbits. The nilpotent orbits of a group can be organised into a Hasse diagram \cite{Kraft:1982fk}, which displays their partial ordering, in terms of dimensions and moduli space inclusion relations. We summarise the standard terminology used for the classification of nilpotent orbits, according to their properties.

Each nilpotent orbit of $G$ is associated with a set of holomorphic functions transforming in irreps of $G$. Our approach is to describe these sets in terms of HS. Presented in refined form, such HS faithfully encode the class function content of nilpotent orbits, up to isomorphisms. We decompose these HS into their representation content, which can be described in terms of highest weight generating functions (``HWGs"), based either on the irreducible representations (``irreps") of $G$, or on the modified Hall Littlewood polynomials (``$mHL$") of $G$. The key transformations are summarised in appendix \ref{apx:HST}. Thus, while the (closures of the) nilpotent orbits of a Classical or Exceptional group are constructed from its Lie algebra $\mathfrak g$, they can also be referred to by the group $G$. The reader is referred to \cite{Hanany:2015hxa, Hanany:2016gbz} for a fuller exposition of our methods of working with plethystic generating functions and Weyl integration to decompose HS into their constituent characters or $mHL$ functions.

In section \ref{sec:Coulomb}, we give Coulomb branch constructions for near to minimal Exceptional group orbits using the unitary monopole formula. The quivers for these constructions can be found by a variety of means; either from affine Dynkin diagrams, or from the canonical data associated with nilpotent orbits via their Characteristics. All these nilpotent orbits have a Characteristic Height 2. This is similar to the situation for Coulomb branch constructions of Classical group nilpotent orbits\footnote{ For the $A$ series only, a wider range of constructions is available via $3d$ mirror symmetry \cite{Intriligator:1996ex}.}.

In section \ref{sec:GmodH}, we show how normal nilpotent orbits can be constructed using a localisation formula, working directly from the group theoretic parameters encoded by their Characteristics. Our method effectively generalises the expression for the $mHL$ of $G$, formulated in terms of the roots and Weyl group of $G$, to an expression for (the normalisation of) any nilpotent orbit, including non-Richardson orbits.\footnote{ Whilst the mathematical Literature contains schema for normalisations of nilpotent orbits \cite{mcgovern_1989, Collingwood:1993fk}, these lack the explicit Hilbert series grading incorporated in the NON formula.} We use this Nilpotent Orbit Normalisation (or ``$NON$") formula to calculate normal nilpotent orbits of many Exceptional groups, although the high dimensions of their Weyl groups restrict the set of calculations that is feasible at this time.

Several issues emerge from this analysis of Exceptional groups, relating in particular to their non-normal nilpotent orbits and some extra root maps from $SU(2)$ homomorphisms, which appear to give rise to dualities. While the computational challenges hinder a definitive resolution, these issues are discussed in the concluding section, where potential avenues for future work are identified.

\paragraph{Notation and Terminology}
We freely use the terminology and concepts of the Plethystics Program, including the Plethystic Exponential (``PE"), its inverse, the Plethystic Logarithm (``PL"), the Fermionic Plethystic Exponential (``PEF") and, its inverse, the Fermionic Plethystic Logarithm(``PFL"). For our purposes:
\begin{equation} 
\label{eq:intro1}
\begin{aligned}
PE\left[ {\sum\limits_{i = 1}^d {{A_i}} ,t} \right] & \equiv \prod\limits_{i = 1}^d {\frac{1}{{\left( {1 - {A_i}t} \right)}}},\\
PE\left[ { - \sum\limits_{i = 1}^d {{A_i}} ,t} \right] & \equiv \prod\limits_{i = 1}^d {\left( {1 - {A_i}t} \right)},\\
PE\left[ {\sum\limits_{i = 1}^d {{A_i}} , - t} \right] & \equiv \prod\limits_{i = 1}^d {\frac{1}{{\left( {1 + {A_i}t} \right)}}} ,\\
 PE\left[ { - \sum\limits_{i = 1}^d {{A_i}} , - t} \right] & \equiv PEF\left[ {\sum\limits_{i = 1}^d {{A_i}} ,t} \right] & \equiv \prod\limits_{i = 1}^d {\left( {1 + {A_i}t} \right)},\\
 \end{aligned}
\end{equation}
where $A_i$ are monomials in weight or root coordinates or fugacities. The reader is referred to \cite{Benvenuti:2006qr} or \cite{Hanany:2014dia} for further detail.

We present the characters of a group $G$ either in the generic form ${\cal X}_{G}(x_i)$, or as $[irrep]_{G}$, or using Dynkin labels as ${[ {{n_1}, \ldots , {n_r}}]_{G}}$, where $r$ is the rank of $G$. We may refer to \emph{series}, such as $1 + f + {f^2} + \ldots $, by their \emph{generating functions} $1/\left( {1 - f} \right)$. We use distinct coordinates/variables to help distinguish the different types of generating function, as indicated in table \ref{tab1}.
\begin{table}[htp]
\small
\begin{center}
\begin{tabular}{|c|c|c|}
\hline
$\text{Generating~Function}$&$ \text{Notation}$&$ \text{Definition} $\\
\hline
$\text{Refined~HS~(Weight~coordinates)}$&$ {{g^{G}_{HS}}( {{x},{t}} )}$&$\sum\limits_{n = 0}^\infty {{a_n}({x})}{t^n} $\\
$\text{Refined~HS~(Simple root~coordinates)}$&$ {{g^{G}_{HS}}( {{z},{t}} )}$&$\sum\limits_{n = 0}^\infty {{a_n}({z})}{t^n} $\\
$\text{Unrefined~HS}$&$ {{g^{G}_{HS}}\left( t \right)}$&$ \sum\limits_{n = 0}^\infty {{a_n}} {t^n} \equiv \sum\limits_{n = 0}^\infty {{a_n}({1})}{t^n} $\\
\hline
$\text{HWG (Character) for HS}$&$g_{HWG}^{G}( {{m},{t}} ) $&$ \sum\limits_{{n_1}, \ldots ,{n_r} = 0}^\infty {{a_{{n_1}, \ldots ,{n_r}}}( {{t}} )} ~m_1^{{n_1}} \ldots m_r^{{n_r}}$\\
$\text{HWG (mHL) for HS}$&$g_{HWG}^{G}( {{h},{t}} ) $&
$ \sum\limits_{{n_1}, \ldots ,{n_r} = 0}^\infty  {{a_{{n_1}, \ldots ,{n_r}}}(t)} h_1^{{n_1}} \ldots h_r^{{n_r}} $ \\
\hline
$\text{Character}$&$ {{g^{G}_{\cal X}}( {{x},{m}} )}$&$\sum\limits_{{n_1}, \ldots ,{n_r} = 0}^\infty {[ {{n_1}, \ldots ,{n_r}}]_G(x)} ~m_1^{{n_1}} \ldots m_r^{{n_r}} $\\
\hline
$\text{(modified)~Hall~Littlewood}$&$ {{g^{G}_{(m)HL}}( {{x},{h,t}} )}$&$\sum\limits_{{n_1}, \ldots ,{n_r} = 0}^\infty {(m)HL^G_{[ {{n_1}, \ldots ,{n_r}}]}} \left( {{x},t} \right)~h_1^{{n_1}} \ldots h_r^{{n_r}}$\\
\hline
\end{tabular}
\end{center}
\label{tab1}
\caption{Types of Generating Function}
\end{table}
These different types of generating function are related and can be considered as a hierarchy in which the refined HS, HWG, character and mHL generating functions fully encode the group theoretic information about a moduli space. We typically label unimodular Cartan subalgebra (``CSA") coordinates for weights within characters by $x \equiv (x_1 \ldots x_r)$ and simple root coordinates by $z \equiv (z_1 \ldots z_r)$, dropping subscripts if no ambiguities arise. The Cartan matrix $A_{ij}$ mediates the canonical relationship between simple root and CSA coordinates as ${z_i} = \prod\limits_j {x_j^{{A_{ij}}}}$ and ${x_i} = \prod\limits_j {z_j^{{A^{ - 1}}_{ij}}}$. We generally label field (or R-charge) counting variables with $t$, adding subscripts if necessary.

Finally, we deploy highest weight notation \cite{Hanany:2014dia}, which uses fugacities to track highest weight Dynkin labels, and describes the structure of a HS in terms of the highest weights of its constituent irreps. We typically denote such Dynkin label counting variables by $m \equiv (m_1 \ldots m_r)$ for representations based on characters, and by $h \equiv (h_1 \ldots h_r)$ for representations based on (modified) Hall-Littlewood polynomials ${(m)HL^G_{[n]}}$, although we may also use other letters, where this is helpful. We define these counting variables to have a complex modulus of less than unity and follow established practice in referring to them as ``fugacities", along with the monomials formed from the products of CSA or root coordinates.
\section{Coulomb Branch Constructions}
\label{sec:Coulomb}
\subsection{Monopole Formula}
\label{subsec:monopole}
By way of recapitulation, the monopole construction introduced in \cite{Borokhov:2002cg}, with an explicit formula for conformal dimension given in \cite{Gaiotto:2008ak} and subsequently refined in \cite{Cremonesi:2013lqa}, provides a systematic method for the construction of the moduli spaces of the Coulomb branches of particular SUSY quiver theories, being ${\cal N}=4$ superconformal gauge theories in $2+1$ dimensions with $8$ supercharges. The Coulomb branches of these theories are HyperK\"ahler manifolds. The monopole formula draws upon a lattice of charges, often referred to as a GNO lattice \cite{Goddard:1976qe}, that is applied to a linked system of gauge and flavour nodes defined by a quiver diagram.

We focus herein on Coulomb branch constructions that are based on quivers with unitary gauge groups, so it is useful to specialise to a \emph{unitary monopole formula}, as distinct from versions that use other gauge groups \cite{Cremonesi:2014kwa}. In the absence of external flavour charges, the unitary monopole formula is given by the schema, refined from \cite{Cremonesi:2013lqa}:

\begin{equation}
\label{eq:mon1}
g_{HS:{\mathrm{Coulomb}}}^{G}\left( {z,t} \right) \equiv \sum\limits_q^{} P_q^{U\left( N \right)} (t){~}{z^{J(q)}}{~}{t^{\Delta (q)}}.
\end{equation}
In \ref{eq:mon1}, $q$ is a collective coordinate for the set of $U(N)$ monopole charges of the quiver gauge nodes (``overall monopole flux"),  $P_q^{U\left( N \right)}$ is a combined symmetry factor from the Casimirs of the $U(N)$ quiver gauge nodes under each monopole flux, $z$ is a collective coordinate for the simple root fugacities of a group $G$, and $t$ is an R-charge counting fugacity.

Each gauge node is associated with adjoint valued fields from the vector multiplet and the links between nodes correspond to complex scalars from the hypermultiplets of the SUSY theory. The monopole formula assembles the Coulomb branch of the quiver theory by projecting monopole charge configurations from the GNO lattice to the root space lattice of $G$, under a grading determined by the conformal dimension $\Delta \left( q \right)$ of the monopole flux $q$.

This \emph{conformal dimension} (equivalent to R-charge or the spin of an $SU(2)_R$ global symmetry) is found by applying the following general schema \cite{Gaiotto:2008ak} to the quiver diagram:
\begin{equation}
\label{eq:mon0}
\Delta \left( q \right) = \underbrace {\frac{1}{2}\sum\limits_{i } {\sum\limits_{{\rho _i} \in R_i}^{} {\left| {{\rho _i}(q)} \right|} } }_{\scriptstyle contribution~of~{\cal N} = 4\atop
\scriptstyle hyper~multiplets} - \underbrace {\sum\limits_{\alpha  \in \Phi_+ }^{}{\left| {\alpha (q)} \right|}}_{\scriptstyle contribution~of~{\cal N} = 4\atop
\scriptstyle{{vector~multiplets}}}.
\end{equation}
The positive R-charge contribution in the first term comes from the matter fields that link adjacent nodes in the quiver diagram. These are bifundamental chiral operators within the ${\cal N}=4$ hypermultiplets. The second term describes a negative R-charge contribution from the ${\cal N}=4$ vector multiplets; this arises due to symmetry breaking, whenever the monopole flux $q_i$ over a gauge node $i$ combines a number of different charges. To explicate the unitary monopole formula, assuming $G$ has rank $r$ :

\begin{enumerate}

\item The gauge nodes are indexed by $i$, where $i$ runs from 1 to $r$, with each $U(N_i)$ gauge node carrying a \emph{monopole flux} $q_i \equiv (q_{i,1}, \ldots ,q_{i,N_i})$ comprising one or more \emph{monopole charges} $q_{i,j}$. The fluxes are assigned the collective coordinate $q \equiv (q_1, \ldots, q_r)$. The limits of summation for the monopole charges are $\infty  \ge {q_{i,1}} \ge \ldots {q_{i,k}}  \ge \ldots {q_{i,{N_i}}} \ge  - \infty $ for $i=1,\ldots r$.

\item The {\small${P_q^{U(N)} \equiv \prod\limits_{i=1}^r {P_{{q_i}}^{U({N_i})}}}$} symmetry factor equals the product of the symmetry factors from each gauge node. Each $P_{q_i}^{U(N_i)}$ encodes the degrees $\{d_{i,j}\}$ of the Casimirs of the residual $U(N_i)$ symmetries at each gauge node under its monopole flux $q_i$:

\begin{equation}
\label{eq:mon2}
P_{q}^{U\left( N \right)} \equiv \prod\limits_{i,j} {\frac{1}{{\left( {1 - {t^{{d_{i,j}(q)}}}} \right)}}}  = \prod\limits_{i = 1}^r {\prod\limits_{j = 1}^{{N_i}} {\prod\limits_{k = 1}^{{\lambda _{ij}}\left( {{q_i}} \right)} {\frac{1}{{1 - {t^k}}}} } }.
\end{equation}

See Appendix \ref{apxPUN} for some low rank examples.

\item The monopole flux over the gauge nodes is counted by the fugacity $z \equiv (z_1, \ldots, z_r)$. The monomial $z^{J(q)}$, which combines the monopole fluxes $q_i$ into total charges for each $z_i$, expands as ${z^{J(q)}} \equiv \prod\limits_{i = 1}^r {z_i^{\sum\limits_{j = 1}^{{N_i}} {{q_{i,j}}} }}$.

\item A gauge node may also be attached to one or more flavour nodes. In the absence of external charges, any flavour nodes carry a zero monopole charge.

\item The conformal dimension $\Delta(q)$ associated with the monopole flux $q$ (taking external flavour charges as zero) is given by the explicit formula  \cite{Cremonesi:2013lqa}:
\begin{equation}
\label{eq:mon3}
\begin{aligned}
\Delta (q) = \frac{1}{2}\underbrace {\sum\limits_{j > i}^r {\sum\limits_{m,n} {\left| {{q_{i,m}}{A_{ij}} - {q_{j,n}}{A_{ji}}} \right|} } }_{\text{gauge - gauge hypers}} 
& + \frac{1}{2}\underbrace {\sum\limits_{j > i}^{} {\sum\limits_{m,n} {\left| {{q_{i,m}}{A_{ij}}} \right|} } }_{{\text{gauge - flavour hypers}}}\\
& - \underbrace {\sum\limits_{i = 1}^r {\sum\limits_{m > n}^{} {\left| {{q_{i,m}} - {q_{i,n}}} \right|} } }_{\text{gauge vplets}},
\end{aligned}
\end{equation}
where (i) the summations are taken over all the monopole charges in the flux $q$ and (ii) the linking pattern between nodes is defined by the $A_{ij}$ off-diagonal terms of a linking matrix, which are only non-zero for linked nodes.

\end{enumerate}

It is remarkable that with a little further specialisation, the unitary monopole formula \ref{eq:mon1}, together with \ref{eq:mon2} and \ref{eq:mon3}, exactly generates the moduli spaces of certain class functions over the root lattice of a Classical or Exceptional group. This specialisation involves basing the gauge nodes of the quiver on the Dynkin diagram of some chosen group $G$, taking the $z$ as fugacities for the simple roots of $G$ and extracting the linking factors $A_{ij}$ from the Cartan matrix for $G$ (extended to incorporate any flavour nodes). Thus, for theories with simply laced quivers of ADE type, $A_{ij} = 0$ or $-1$, for $i \neq j$.

Various choices are possible for the $U(N)$ charges on the gauge nodes and also for the number and linking of flavour nodes to gauge nodes, providing that the quiver diagram remains \emph{balanced}. As elaborated in \cite{Hanany:2016gbz}, the \emph{balance} of a gauge node $(i)$, introduced in \cite{Gaiotto:2008ak}, can be defined as:

\begin{equation}
\label{eq:coulomb1}
\begin{aligned}
\text{Balance}_{G}^{(i)}=  - \sum\limits_{j}^{} {A}^{ij}{N_j} ,
\end{aligned}
\end{equation}
where the sum is effectively taken over the gauge/flavour nodes to which each gauge node is linked. When all the gauge nodes in a quiver are balanced, $\forall i: \text{Balance}_{G}^{(i)}=0$, the conformal dimension $\Delta (q)$ of each overall monopole flux takes a non-negative integer value and this meets the criteria for a \emph{good} theory \cite{Gaiotto:2008ak}. The Coulomb branch moduli spaces constructed from these balanced quivers yield class functions over the root space of $G$.  Amongst these balanced quivers there exists a subset of quivers whose Coulomb branches have moduli spaces that are closures of certain nilpotent orbits. The issue is one of identifying these balanced quivers.

\subsection{Quivers for Exceptional Group Nilpotent Orbits}
\label{subsec:CoulombExcept}

Based on early work in \cite{Intriligator:1996ex}, it was shown in \cite{Cremonesi:2013lqa} how the unitary monopole formula can be combined with a quiver based on the affine Dynkin diagram of a simply laced group $G$ to construct the reduced single instanton moduli space ("RSIMS") or minimal nilpotent orbit of $G$, by choosing the $U(N)$ gauge groups to have ranks defined by the Coexter labels of $G$. In \cite{Cremonesi:2014xha} this program was extended to the RSIMS of non-simply laced $BCFG$ groups, by working with dual Coexter labels\footnote{This distinction is critical for non simply laced groups for which the dual Coxeter labels differ from the Coxter labels by factors depending on root lengths.}, and by dressing the hypermultiplet linking factors to reflect different root lengths, using off-diagonal elements of the Cartan matrix of $G$. In \cite{Hanany:2015hxa} it was shown that quivers based on \emph{twisted} affine Dynkin diagrams can be used to construct the moduli spaces of near to minimal nilpotent orbits of Classical groups. One of the findings herein, is that such a construction based on the twisted affine Dynkin diagram of $F_4$ yields the next to minimal nilpotent orbit of $F_4$.

The success of the constructions based on affine Dynkin diagrams results from the fact that the dual Coxeter numbers of an affine Dynkin diagram form a kernel of the affine Cartan matrix \cite{Fuchs:1997bb} and so satisfy the condition for a zero balance \ref{eq:coulomb1}. Setting one of the nodes of the affine Dynkin diagram as a flavour node selects the root system of $G$, but does not affect the balance of the quiver.

It was also observed in \cite{Hanany:2015hxa} that the Characteristics (or root maps) and weight maps of nilpotent orbits can be combined to form balanced quivers, by using weight map labels to define the $U(N)$ ranks of the gauge nodes and Characteristics to define the numbers of flavours attached to each gauge node. In the case of Classical groups, such quivers associated to the Characteristics of minimal and near to minimal nilpotent orbits coincide with those obtained from the affine/twisted affine constructions; additionally, quivers associated to the Characteristics of some higher dimensioned nilpotent orbits also yield Coulomb branch moduli spaces that match the nilpotent orbits (constructed, for example, on the Higgs branch).

In order for such a matching to occur, it is clearly necessary that the dimension of the nilpotent orbit should match the dimension of the Coulomb branch construction. Now, the (complex) dimension of a Coulomb branch construction using the unitary monopole formula is equal to twice the sum of the ranks of the $U(N)$ gauge nodes \cite{Cremonesi:2014vla}; this in turn limits this construction method to those nilpotent orbits whose dimensions are equal to twice the sums of their weight map labels. We refer to this as the \emph{weight map condition}. In these cases, calculations show that the monopole formula yields moduli spaces matching the nilpotent orbits.

In \cite{Hanany:2015hxa} it was shown that this approach yields not only the affine constructions, but also the 12 and 18 dimensional nilpotent orbits of $C_3$ and $C_4$, respectively. We find herein that the Coulomb branches of quivers based on Characteristics yield, in addition to the minimal nilpotent orbits of all Exceptional groups and the 22 dimensional (next to minimal) nilpotent orbit of $F_4$, the 32 dimensional (next to minimal) nilpotent orbit of $E_6$, the 52 (next to minimal) and 54 dimensional nilpotent orbits of $E_7$, and the 92 (next to minimal) dimensional nilpotent orbit of $E_8$.

An alternative formulation of the condition for the Characteristic of a nilpotent orbit of $G$ to yield a Coulomb branch quiver for its construction, can be given in terms of the Characteristic root height $[\theta ]$ of the highest root $\theta$. This follows from \ref{eq:homo5a} by setting the coefficients $a_i$ to the Coxeter labels of $G$, which express the highest root in terms of simple roots. The empirical condition\footnote{Identified by Giulia Ferlito, Imperial College, London.} for a quiver based on the Characteristic of a nilpotent orbit to yield its Coulomb branch construction is simply that $[\theta]=2$. We refer to $[\theta]$ as the \emph{Characteristic Height}.

\begin{figure}[htbp]
\begin{center}
\includegraphics[scale=0.42]{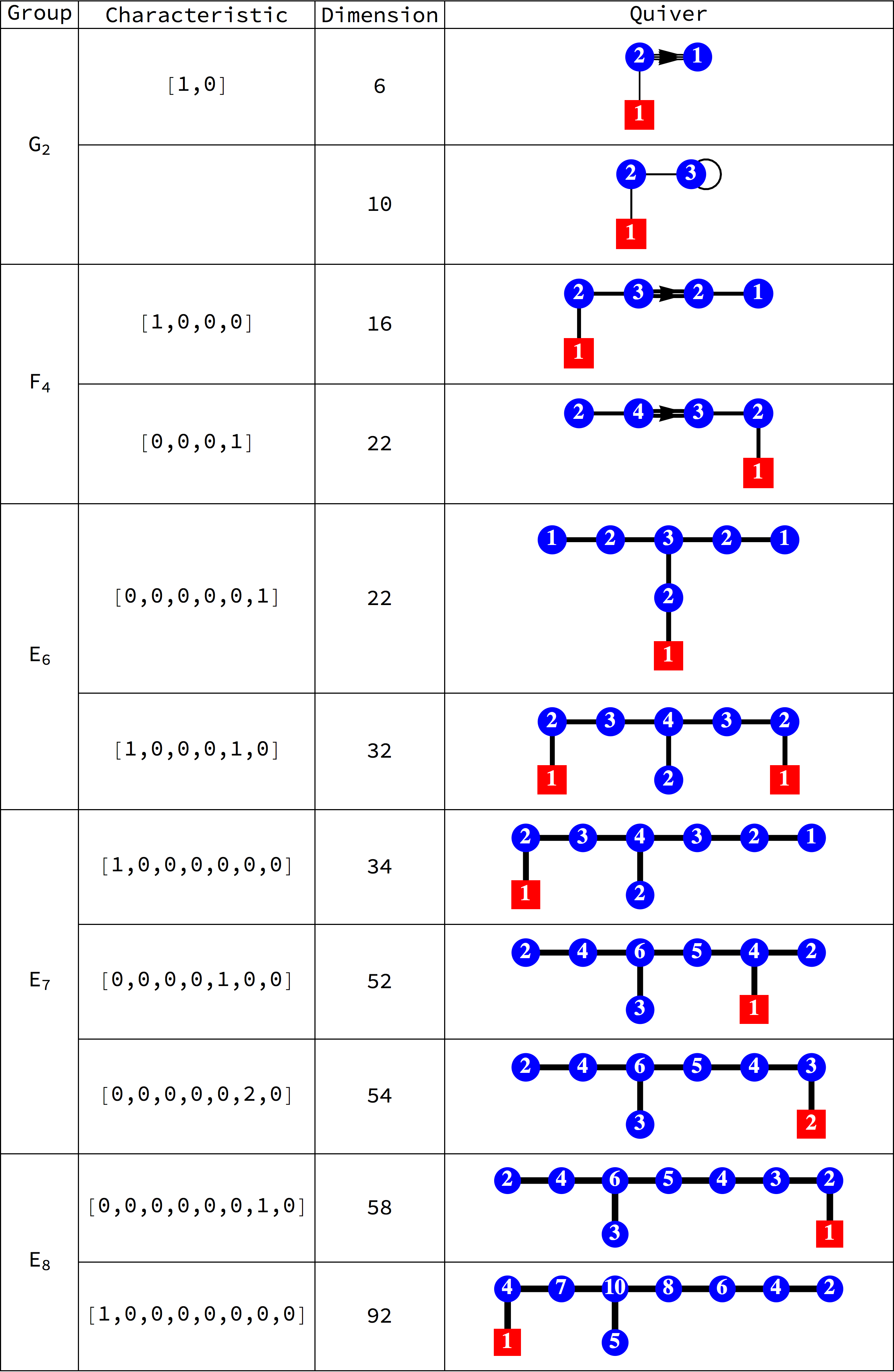}\\
\caption[Exceptional Group Quivers with Characteristic Height 2]{Exceptional Group Quivers with Characteristic Height 2. Round (blue) nodes denote unitary gauge nodes of the indicated rank. Square (red) nodes denote numbers of flavour nodes. The Characteristics coincide with the numbers of flavour nodes attached to each gauge node. The dimension of a Coulomb branch nilpotent orbit construction equals twice the sums of the ranks of its gauge nodes. Note that the quiver for the $G_2$ 10 dimensional nilpotent orbit is not based on its Characteristic (see discussion).}
\label{fig:CBgrid}
\end{center}
\end{figure}

The balanced quivers for Coulomb branch constructions of Exceptional group nilpotent orbits are shown in figure \ref{fig:CBgrid} and their Hilbert series and HWGs are included within the tables in the next section. All these constructions obey both the weight map and the Characteristic Height conditions. For example, the next to minimal nilpotent orbit of $F_4$ has a Characteristic of $[0001]$ and $F_4$ has Coxeter labels of $\{2,3,4,2\}$, so $[\theta]=2$. Also, the dimension of the Hilbert series of this nilpotent orbit is 22, which equals twice the sum of the weight map labels $\{2,4,3,2\}$.

For higher dimensioned Exceptional group nilpotent orbits, $[\theta]>2$ and so Coulomb branch constructions \emph {direct from Characteristics} are not available. For example, the twisted affine Dynkin diagram of $G_2$ has $[\theta]=3$ and so there is no Coulomb branch construction for the next to minimal 8 dimensional nilpotent orbit of $G_2$.\footnote{As a distinct case, a Coulomb branch construction is known for the 10 dimensional orbit of $G_2$ \cite{Cremonesi:2014vla}. This draws on a modification of the Dynkin diagram to include a self-linked gauge node, as in figure \ref{fig:CBgrid}; the calculation of $\Delta(q)$ uses $A_2$ hypermultiplets and there is no vector multiplet on the self-linked node.}

\FloatBarrier

\subsection{Monopole Formula Evaluation}
\label{subsec:CoulombMonopole}

Some digression on the mechanics of Hilbert series calculations using the monopole formula is worthwhile, since these can pose computational challenges. The difficulties result from the absolute differences between monopole charges within the conformal dimension function, which require a piecewise treatment of the summation over the gauge nodes. Thus, while it can be a relatively straightforward matter to find the first few terms of a series expansion in $t$ of $g_{HS:Coulomb}^{G}(z,t)$, obtaining the rational generating function for the refined Hilbert series is more delicate.

The approach taken herein to finding these refined Hilbert series involves splitting the summation \ref{eq:mon1} into a set of summations over hypersurfaces within the lattice, each defined by limits on the monopole charges $q$, such that conformal dimension $\Delta (q)$ reduces to a (generally different) linear function of $q$ on each hypersurface. The algebra of these hypersurfaces (or sets of lattice points known as \emph{monoids}) is discussed in \cite{Hanany:2016pfm}. From a computational viewpoint, there are various aspects to consider:

\begin{enumerate}

\item The number of such hypersurfaces is driven by the structure of the hypermultiplets. If we consider two adjacent gauge nodes with ranks $N_1$ and $N_2$, then the number of distinct ways of ordering their monopole charges is given by the binomial coefficient $\frac{{{N_1} + {N_2}!}}{{{N_1}!{N_2}!}}$. Each such ordering defines the limits of a hypersurface on the $q$ lattice over which the contribution to conformal dimension $\Delta (q)$ from the fields linking the two nodes is a linear function.

\item These hypersurfaces intersect along hypersurfaces of lower dimension (e.g. common edges between faces, common points between edges, etc.), and it is important to allow for such intersections, so as to avoid over or under counting of lattice points.

\item Also, the $P_q^{U(N)}$ symmetry factors require the above calculations to be carried out for each distinct degeneracy pattern of  the ordered monopole charges. For example, a $U(3)$ gauge node has the 4 possible degeneracy patterns $ \{a,b,c \},\{a,a,b \},\\ \{a,b,b\},\{a,a,a \}$. Generally, a $U(N)$ gauge node has $2^{(N-1)}$ such degeneracies, and, when linked to its adjacent nodes, each degeneracy defines a different hypersurface.

\item In order to minimise the number of hypersurfaces to be evaluated, it can be helpful to separate the components of the monopole formula according to their node dependence and to tackle the summations one gauge node at a time.

\item There is a choice to make in the order of evaluation of the nodes. In the case of non-simply laced quivers, it can be simpler to carry out the summations over the short roots first in order to keep the limits of summation as integer variables (rather than ceiling or floor functions). In the case of $DE$ quivers, it can be simpler to calculate the central node last.

\item The number of hypersurfaces to be evaluated for each gauge node is compounded by the different possible positions of the zero charge of the flavour node relative to the gauge node monopole charges. This requires the use of piecewise conditional functions, unless this is avoidable by leaving the summation over a single gauge node attached to the flavour node until last.

\item For $ADE$ groups it may be possible to shift the zero charge of the flavour node onto one of the gauge nodes (for example the lowest monopole charge of the central node of a $DE$ quiver diagram), as discussed in \cite{Cremonesi:2014vla, Hanany:2015hxa}. Such a shift may make it possible to use the symmetries of the quiver diagram to reduce the number of calculations and/or to minimise the compounding of the number of hypersurfaces resulting from the zero charge of the flavour node.

\end{enumerate}
Examples, including minimal nilpotent orbits of Exceptional groups, have been presented in \cite{Cremonesi:2013lqa, Cremonesi:2014xha, Hanany:2015hxa, Hanany:2016gbz}. We complement these by providing some details of the calculations of the next to minimal nilpotent orbits of $F_4$ and $E_6$.

\subsection{Monopole Formula for $F_4$ 22 Dimensional Nilpotent Orbit}
\label{subsec:CoulombF4NO22}

The Coulomb branch construction for the 22 dimensional (next to minimal) nilpotent orbit of $F_4$ is based on the twisted affine Cartan matrix and dual Coxeter labels:
\begin{equation}
\label{eq:mon3.33}
\begin{tabular}{c|c c c c c|c|}
${{z_1}}$&$ 2$&${ - 1}$&$0$&$0$&${ - 1}$&$ 2$ \\
${{z_2}}$&$ { - 1}$&$2$&${ - 2}$&$0$&$0$&$ 4$ \\
${{z_3}}$&$ 0$&${ - 1}$&$2$&${ - 1}$&$0$&$ 3$ \\
${{z_4}}$&$ 0$&$0$&${ - 1}$&$2$&$0$&$ 2 $ \\
${{z_0}}$&$ { - 1}$&$0$&$0$&$0$&$2$&$ 1$ \\
\end{tabular},
\end{equation}
where the extra root has the fugacity $z_0$ and the dual Coxeter labels have been expressed as a column vector.

In the case of $g_{HS:Coulomb}^{{F_4}[0001]}(z,t)$, the most effective approach found was to calculate the nodes in the order $(4) \to (3) \to (2) \to (1)$, using a piecewise logic to keep track of the position of the zero flavour charge amongst the gauge node monopole charges. 

Thus, by combining \ref{eq:mon1}, \ref{eq:mon2} and \ref{eq:mon3}, and by expanding components, the monopole formula can be rearranged into the sequence of sums over gauge nodes:
\begin{equation}
\label{eq:mon3.35}
\small
\begin{aligned}
g_{HS:Coulomb}^{{F_4}[0001]}(z,t) &= \mathop \sum \limits_{\infty \ge  {q_{11}} \ge {q_{1,2}} \ge  - \infty }^{} {~} P_{{q_1}}^{U\left( 2 \right)}(t) {~}{z_1}^{{q_{1,1}} + {q_{1,2}}} {~}{t^{\Delta _1 \left( q \right)}} {~}g_{}^{[2]}({q_1},z,t),\\
\text{where}\\
g_{}^{[2]}({q_1},z,t)  &= \mathop \sum \limits_{\infty \ge {q_{2,1}} \ge {q_{2,2}} \ge {q_{2,3}} \ge {q_{2,4}} \ge  - \infty }^{}  {~} P_{{q_2}}^{U\left( 4 \right)}(t) {~}{z_2}^{{q_{2,1}} + {q_{2,2}} + {q_{2,3}} + {q_{2,4}}} {~}{t^{{\Delta _2}\left( q \right)}} {~}g_{}^{[3]}({q_2},z,t),\\
\text{where}\\
g_{}^{[3]}({q_2},z,t) &= \mathop \sum \limits_{\infty \ge {q_{3,1}} \ge {q_{3,2}} \ge {q_{3,3}} \ge  - \infty }^{}  {~} P_{{q_3}}^{U\left( 3 \right)}(t) {~}{z_3}^{{q_{3,1}} + {q_{3,2}} + {q_{3,3}}} {~}{t^{{\Delta _3}\left( q \right)}} {~}g_{}^{[4]}({q_3},z,t),\\
\text{where}\\
g_{}^{[4]}({q_3},z,t)  &= \mathop \sum \limits_{ \infty \ge {q_{41}} \ge {q_{4,2}} \ge  - \infty }^{}  {~} P_{{q_4}}^{U\left( 2 \right)}(t) {~}{z_4}^{{q_{4,1}} + {q_{4,2}}} {~}{t^{{\Delta _4}\left( q \right)}},\\
\end{aligned}
\end{equation}
and the components of conformal dimension are given by:
\begin{equation}
\label{eq:mon3.36}
\small
\begin{aligned}
{\Delta _1}\left( q \right) &=  - \left| {{q_{1,1}} - {q_{1,2}}} \right|,\\
{\Delta _2}\left( q \right) &= \frac{1}{2}\left( {\sum\limits_{i = 1}^2 {\sum\limits_{j = 1}^4 {\left| {{q_{1,i}} - {q_{2,j}}} \right|} } } \right) - \sum\limits_{i < j}^{} {\left| {{q_{2,i}} - {q_{2,j}}} \right|},\\
{\Delta _3}\left( q \right) &=  \frac{1}{2}\left( {\sum\limits_{i = 1}^4 {\sum\limits_{j = 1}^3 {\left| {2{q_{2,i}} - {q_{3,j}}} \right|} } } \right) - \sum\limits_{i < j}^{} {\left| {{q_{3,i}} - {q_{3,j}}} \right|}, \\
{\Delta _4}\left( q \right) &= \frac{1}{2}\left( {\sum\limits_{i = 1}^3 {\sum\limits_{j = 1}^2 {\left| {{q_{3,i}} - {q_{4,j}}} \right|} }  + \sum\limits_{i = 1}^2 {\left| {{q_{4,i}}} \right|} } \right) - \left| {{q_{4,1}} - {q_{4,2}}} \right|. \\
\end{aligned}
\end{equation}
It should be noted that the $g_{}^{[j]}({q},z,t)$ are piecewise functions that take different values according to the position of the zero flavour charge relative to the monopole charges within the fluxes $q$. For example, $g_{}^{[2]}({q_1},z,t)$ is a different function in the three cases $\{q_{1,1} \ge q_{1,2} \ge 0,q_{1,1} \ge 0 \ge q_{1,2} , 0 \ge q_{1,1} \ge q_{1,2} \}$.

The Hilbert series for $g_{HS:Coulomb}^{{F_4}[0001]}(z,t)$ and its Highest Weight Generating functions are included in tables \ref{tab:F4NO1} and \ref{tab:F4NO4} in the next section, along with some further commentary.

\subsection{Monopole Formula for $E_6$ 32 Dimensional Nilpotent Orbit}
\label{subsec:CoulombE6NO32}

The Coulomb branch construction of the refined HS for the 32 dimensional (next to minimal) orbit of $E_6$ uses the quiver based on its Characteristic, as shown in figure \ref{fig:CBgrid}. The most effective approach found was to calculate the central node last, using a piecewise logic to keep track of the position of the zero flavour charge amongst the gauge node monopole charges and making use of the outer automorphism symmetry of the diagram.

By combining \ref{eq:mon1}, \ref{eq:mon2} and \ref{eq:mon3}, and by expanding, the monopole formula:
\begin{equation}
\label{eq:mon3.34}
\begin{aligned}
g_{HS:Coulomb}^{{E_6}[100010]}(z,t)  = \sum\limits_q^{} {P_q^{U\left( N \right)} (t)}  {~} {z^{J(q)}} {~}{t^{\Delta (q)}},
\end{aligned}
\end{equation}
can be rearranged into the sequence of sums over gauge nodes:

\begin{equation}
\label{eq:mon3.21}
\small
\begin{aligned}
g_{HS:Coulomb}^{{E_6}[100010]}(z,t)  &=  \sum \limits_{\infty \ge {q_{3,1}} \ge {q_{3,2}} \ge {q_{3,3}} \ge {q_{3,4}} \ge  - \infty }^{}  {~} P_{{q_3}}^{U\left( 4 \right)}(t)  {~} {z_3}^{{q_{3,1}} + {q_{3,2}} + {q_{3,3}} + {q_{3,4}}} {~} {t^{\Delta _3}}  {~} g_{}^{[2]}  ({q_3},z,t)\\
& {~}{~}{~}{~}{~}{~}{~}{~}{~}{~}{~}{~}{~}{~}{~}{~} {~}{~}{~}{~} \times  {~}{~}{~}{~}{~}{~}{~}{~}{~}{~}{~}{~}{~}{~}{~}{~} {~}{~}{~}{~} g_{}^{[4]}({q_3},z,t)  {~}g_{}^{[6]}({q_3},z,t),\\
\text{where}\\
g_{}^{[2]}({q_3},z,t) & =  \sum \limits_{\infty \ge {q_{2,1}} \ge {q_{2,2}} \ge {q_{2,3}} \ge  - \infty }^{}  {~} P_{{q_2}}^{U\left( 3 \right)}(t)  {~} {z_2}^{{q_{2,1}} + {q_{2,2}} + {q_{2,3}}}  {~} {t^{\Delta _2}}  {~} g_{}^{[1]}({q_2},z,t),\\
\text{where}\\
g_{}^{[1]}({q_2},z,t) & =  \sum \limits_{\infty \ge {q_{11}} \ge {q_{1,2}} \ge  - \infty }^{}  {~} P_{{q_1}}^{U\left( 2 \right)}(t)  {~} {z_1}^{{q_{1,1}} + {q_{1,2}}} {~}{t^{\Delta _1}},\\
\text{and}\\
g_{}^{[6]}({q_3},z,t) & =  \sum \limits_{\infty \ge {q_{6,1}} \ge {q_{6,2}} \ge  - \infty }^{}  {~} P_{{q_6}}^{U\left( 2 \right)}(t)  {~} {z_2}^{{q_{6,1}} + {q_{6,2}}} {~} {t^{\Delta _6}},\\
\text{and}\\
g_{}^{[4]}({q_3},z,t) & = {\left. {g_{}^{[2]}({q_3},z,t)} \right|_{\left\{ {{z_1} \to {z_5},{z_2} \to {z_4}} \right\}}},\\
\end{aligned}
\end{equation}

and the components of conformal dimension are given by:
\begin{equation}
\label{eq:mon3.22}
\small
\begin{aligned}
{\Delta _1}\left( q \right) &= {\frac{1}{2}\left( {\sum\limits_{i = 1}^2 {\sum\limits_{j = 1}^3 {\left| {{q_{1,i}} - {q_{2,j}}} \right|} }  + \sum\limits_{i = 1}^2 {\left| {{q_{1,i}}} \right|} } \right) - \left| {{q_{1,1}} - {q_{1,2}}} \right|},\\
{\Delta _2}\left( q \right) &= {\frac{1}{2}\left( {\sum\limits_{i = 1}^3 {\sum\limits_{j = 1}^4 {\left| {{q_{2,i}} - {q_{3,j}}} \right|} } } \right) - \sum\limits_{i < j}^{} {\left| {{q_{2,i}} - {q_{2,j}}} \right|} },\\
{\Delta _3}\left( q \right) &= { - \sum\limits_{i < j}^{} {\left| {{q_{3,i}} - {q_{3,j}}} \right|} }, \\
{\Delta _6}\left( q \right) &= {\frac{1}{2}\left( {\sum\limits_{i = 1}^4 {\sum\limits_{j = 1}^2 {\left| {{q_{3,i}} - {q_{6,j}}} \right|} } } \right) - \sum\limits_{i < j}^{} {\left| {{q_{6,i}} - {q_{6,j}}} \right|} }. \\
\end{aligned}
\end{equation}

Once again, $g_{}^{[2]}({q_3},z,t)$ and $g_{}^{[1]}({q_2},z,t)$ are piecewise functions that take different values according to the position of the zero flavour charge relative to the monopole charges within the fluxes. The Hilbert series for $g_{HS:Coulomb}^{{E_6}[100010]}(z,t)$ and its Highest Weight Generating function are included in tables \ref{tab:E6NO1} and \ref{tab:E6NO6} in the next section, along with some further commentary.

While a similar approach to obtaining the refined Hilbert series of the Coulomb branches of the quivers for the next to minimal nilpotent orbits of $E_7$ and $E_8$ is in principle feasible, the number of hypersurfaces involved leads to a considerable computational burden, and the calculation of these refined Hilbert series can be more practical using localisation methods, as discussed in the next section.


\section{Localisation Constructions}
\label{sec:GmodH}

\subsection{Nilpotent Orbit Normalisation formula}
\label{sec:NON}

We continue by presenting a formula for the normalisation of (the closure of) a nilpotent orbit $g_{NON}^{G[\rho]}$, which can in principle be restricted to yield a formula for the nilpotent orbit $g_{NO}^{G[\rho]}$ itself. We refer to this as the \emph{Nilpotent Orbit Normalisation} (``$NON$") formula; it is given in \ref{eq:gmodh6}. It is defined by the fixed points under the Weyl group, of plethystic functions, which are parameterised by subsets of roots and background charges, over the root space of $G$.

By way of motivation, a more general localisation formula, which is an extension of a localisation formula for generalised Hall Littlewood functions of $SU(N)$ \cite{Cremonesi:2014uva}, and from which many relevant generating functions emerge as special instances, is given by:

\begin{equation}
\label{eq:gmodh1}
g_{HS}^G\left( {x,t,[n]} \right) \equiv \sum\limits_{w \in {W_{G/H}}}^{} w  \cdot {\rm{ }}\left( { {x^{\left[ n \right]}}\prod\limits_{\alpha  \in \tilde \Phi _{G/H}^ +  \subseteq \Phi _{G/H}^ + } {\frac{1}{{1 - {z^\alpha }t}}} \prod\limits_{\beta  \in \Phi _{G/H}^ + }^{} {\frac{1}{{1 - {z^{ - \beta }}}}} } \right).
\end{equation}
As usual, $x$ represents the weight space fugacities and $z=x^A$ represents the root space fugacities of some Lie group $G$, with Dynkin labels $[n]$ and positive root space $\Phi^+_G$.  The group $H$, with positive root space $\Phi^+_H$, is a semi-simple regular subgroup of $G$, such that the quotient $G/H$ contains the positive roots $\Phi^+_{G/H}$=$\Phi^+_G \ominus \Phi^+_H$, and ${\tilde \Phi _{G/H}^ + }$ is some subset of $\Phi^+_{G/H}$ (specific to the instance).  The summation is over the action of representative elements $w$ of the cosets $W_{G/H}$, which act as $x \to w \cdot x$ and $z(x) \to z(w \cdot x)$. A key requirement of the construction is that the summand should be invariant under $W_{H}$. Since $\Phi^+_{G/H}$ is $W_{H}$ invariant by construction, this requires that $x^{[n]}$ and ${\tilde \Phi _{G/H}^ + }$ should each be $W_{H}$ invariant.

The family of plethystic functions to which the $NON$ formula belongs includes the Weyl character formula and the modified Hall Littlewood formula. We can note some special cases of \ref{eq:gmodh1}:

\begin{enumerate}

\item $H=\emptyset, t=0$ recovers the Weyl character formula \cite{Fuchs:1997bb} for the character of an irrep with Dynkin label $[n]$:
\begin{equation}
\label{eq:gmodh2}
{\chi}_{\left [ n \right ]}^G\left( {x} \right) = \sum\limits_{w \in W_{G}}^{} {w \cdot \left( {{ x^{\left[ n \right]}}} \prod\limits_{\beta  \in \Phi _G^ + } {\frac{1}{{1 - {z^{ - \beta }}}}}\right)}.
\end{equation}

\item $H=\emptyset, {\tilde \Phi _{G/H}^ + }= {\Phi _G^ +} $ recovers the formula for the modified Hall Littlewood polynomials of $G$ with Dynkin label $[n]$:
\begin{equation}
\label{eq:gmodh3}
mHL_{[n]}^G\left( {x,t} \right) = \sum\limits_{w \in W_{G}}^{}{w \cdot \left( { x^{\left[ n \right]}} \prod\limits_{\beta  \in \Phi _G^ + }^{} {\frac{1}{{\left( {1 - {z^\beta }t} \right)\left( {1 - {z^{ - \beta }}} \right)}}}\right)}.
\end{equation}

\item $H=G_0, [n]=[0],{\tilde \Phi _{G/H}^ + } =\{ \theta \} $, where $G_0$ is the stability group of the (highest weight Dynkin labels of the) highest root  $ \theta $, recovers a character generating function for a RSIMS \cite{Hanany:2015hxa} :\footnote{The $G_0$ stability group of $\theta$ is used in an equivalent manner in \cite{Keller:2011ek, Keller:2012da}, where the RSIMS generating function is implemented as a sum over long roots.}

\begin{equation}
\label{eq:gmodh4}
g_{HS:RSIMS}^G\left( {x,t} \right) = \sum\limits_{w \in W_{G/G_0}}^{} {}  {w \cdot \left(  \frac{1}{{1 - {z^\theta}t}}\prod\limits_{\beta  \in \Phi _{G/G_0}^ + }^{} {\frac{1}{{1 - {z^{ - \beta }}}}} \right)}.
\end{equation}

\end{enumerate}

It is a key finding of this study that, with appropriate choice of parameters, the localisation formula \ref{eq:gmodh1} can be adapted to yield a formula for the normalisation of a nilpotent orbit. Considerations motivated by the above special cases, along with explicit checking versus Higgs/Coulomb branch calculations, identify a \emph{Nilpotent Orbit Normalisation} formula that appears common to all \emph{normal} nilpotent orbits.

\paragraph{Basic $NON$ formula} The parameters of the basic $NON$ formula are fixed directly from the Characteristic of a nilpotent orbit via a simple algorithm. The formula follows from \ref{eq:gmodh1}, by precise choices of $H$ and $ {\tilde \Phi }$, and by setting $[n]$ to $[0]$ - this selects the singlet from a family of charged functions associated with a given nilpotent orbit:

\begin{equation}
\label{eq:gmodh6}
g_{HS:NON}^{G[\rho ]} \left( {x,t} \right) \equiv \sum\limits_{w \in W_{G/G_0}}^{} {} {w \cdot \left( \prod\limits_{\alpha  \in \tilde \Phi _{G/G_0}^ +} {\frac{1}{{1 - {z^\alpha }t}}} \prod\limits_{\beta  \in \Phi _{G/G_0}^ + }{\frac{1}{{1 - {z^{ - \beta }}}}}\right)},
\end{equation}
where $\tilde \Phi _{G/{G_0}}^ +  \equiv \Phi _{G/{G_0}}^ + { \ominus }{\Phi _G^{[1]}}$, as elaborated below. In cases where the nilpotent orbit is normal, $g_{NO}^{G[\rho]}= g_{NON}^{G[\rho]}$. In cases where the nilpotent orbit is non-normal, $g_{NO}^{G[\rho ]}$ can be found by \emph{restricting} $g_{NON}^{G[\rho]}$ to the nilpotent cone $\cal N $ \footnote{For Classical groups, this restriction is achievable with the Higgs branch formula; for Exceptional groups, its analytical implementation can be a non-trivial exercise, as will be discussed later.}:

\begin{equation}
\label{eq:gmodh5}
g_{HS:NO}^{G[\rho ]}\left( {x,t} \right) ={\left. {g_{HS:NON}^{G[\rho ]}\left( {x,t} \right)} \right|_{{\cal N}}}.
\end{equation}

The $SU(2)$ homomorphism $\rho$ (see Appendices \ref{apxNO} and \ref{apxHom}) induces a grading of the root system of $G$. Adapting notation introduced in \cite{Dynkin:1957um}, we define the following subsets of roots, under a grading by their Characteristic root height \ref{eq:homo5a}:

\begin{equation}
\label{eq:gmodh6a}
 {\Phi _G^{[k]}} \equiv \left\{ {\alpha \in \Phi_G^+ :  [\alpha ] = k} \right\},
 \end{equation}
and then we define the following unions of these subsets:

\begin{equation}
\label{eq:gmodh6b}
\begin{aligned}
\Phi _G^ +  &= \bigcup\limits_{k \ge 0} {\Phi _G^{\left[ k \right]}};{~}{~}{~}
\Phi _{G/{G_0}}^ +  &=  \bigcup\limits_{k \ge 1} {\Phi _G^{\left[ k \right]}};{~}{~}{~}
\tilde \Phi _{G/{G_0}}^ +  &= \bigcup\limits_{k \ge 2} {\Phi _G^{\left[ k \right]}}.
\end{aligned}
 \end{equation}

Each $SU(2)$ homomorphism selects a subset $\tilde \Phi _{G/{G_0}}^ +$ of positive roots for symmetrisation with $t$ within the $NON$ formula. This subset invariably includes the highest root $\theta$, plus some system of positive roots connected to the highest root in the Hasse diagram (for roots). The excluded roots include those of $G_0$, which is the subgroup of $G$ for which $k=0$, along with any in ${\Phi _G^{[1]}}$.

The Weyl denominator identity,  $ \sum\limits_{w \in W_{H}}^{} {\prod\limits_{\beta  \in \Phi _{H}^ + } {\frac{1}{{1 - {z^{ - \beta }}}}} } = 1$, which follows from \ref{eq:gmodh2}, permits rearrangement of \ref{eq:gmodh6} into the equivalent form:
\begin{equation}
\label{eq:gmodh7}
g_{HS:NON}^{G[\rho]}\left( {x,t} \right) = \sum\limits_{w \in W_{G}}^{} {}{w \cdot \left( \prod\limits_{\alpha  \in \tilde \Phi _{G/G_0}^ + } {\frac{1}{1 - {z^\alpha }t}} \prod\limits_{\beta  \in \Phi _G^ + } {\frac{1}{1 - {z^{ - \beta }}}} \right)}.
\end{equation}
For computational purposes, \ref{eq:gmodh6} is often simpler, involving smaller denominator terms and fewer Weyl group reflections.\footnote{The simplification of \ref{eq:gmodh7} to the quotient group construction in \ref{eq:gmodh6} requires that $\Phi _G^{[1]}$ be invariant under $W_{G_0}$. This appears to be the case for all Characteristics derived from $SU(2)$ homomorphisms of $G$.}

We can easily check that the $NON$ formula matches known results for canonical types of nilpotent orbit. In particular, choosing a Characteristic of $[22\ldots 2]$ entails that both $\Phi _G^{[0]}$ and $\Phi _G^{[1]}$ are empty and so \ref{eq:gmodh6} reduces to \ref{eq:gmodh3}, corresponding to $mHL_{[0]}^G$, the maximal nilpotent orbit of $G$. Also, it is straightforward to check that the Characteristic of a minimal nilpotent orbit leads to $\Phi _G^{[2]}$ containing just the highest root, so that ${\tilde \Phi _{G/G_0}^ + }  =\{ \theta \}$ and \ref{eq:gmodh7} reduces to \ref{eq:gmodh4}.

\paragraph{$NON$ formula: Even and Richardson Orbits}
In the case of an even orbit, whose Characteristic contains only the labels 0 and 2, $\Phi _G^{[1]} = \emptyset $ and the $NON$ formula simplifies:
\begin{equation}
\label{eq:gmodh8}
g_{NON(even)}^{G[\rho]}\left( {x,t} \right) = \sum\limits_{w \in W_{G/G_0}}^{} {}{w \cdot \left( \prod\limits_{\alpha  \in \Phi _{G/G_0}^ + } {\frac{1}{{\left( {1 - {z^\alpha }t} \right)\left( {1 - {z^{ - \alpha }}} \right)}}}\right)}.
\end{equation}
All Richardson nilpotent orbits can in principle be treated within this category. If the Richardson orbit is even, its $H \equiv G_0$ subgroup follows directly from the zeros of the Characteristic of $G$. If the Richardson orbit is not even, an $H$ subgroup embedding still exists, even if its identity cannot be read directly from the Characteristic.

\paragraph{$NON$ formula: Induced Orbits}
A different and important rearrangement can be made to the $NON$ formula to induce a given nilpotent orbit (or its normalisation) from the nilpotent orbit of a subgroup, whenever its Characteristic contains at least one 2. Essentially, we define a $G/H/G_0$ quotient group structure, by taking $H$ as the semi-simple subgroup defined by the Dynkin diagram of $G$ that remains after any nodes corresponding to 2 in the Characteristic have been eliminated. As a result, the Characteristic for the nilpotent orbit in $H$ contains only 0 and 1.

Starting from \ref{eq:gmodh6}, we set $G/G_0 \to G/H \otimes H/G_0$, so that $ \Phi _G^{[1]}$ falls within $ \Phi _{H}$. We obtain:
\begin{equation}
\label{eq:gmodh14}
\begin{aligned}
g_{HS:NON}^{G[\rho]} \left( {x,t} \right) &= \sum\limits_{w \in W_{G/G_0}}^{} {}{w \cdot \left( \prod\limits_{\alpha  \in \tilde \Phi _{G/G_0}^ +} {\frac{1}{{1 - {z^\alpha }t}}} \prod\limits_{\beta  \in \Phi _{G/G_0}^ + } {\frac{1}{{1 - {z^{ - \beta }}}}}\right)} \\
  &= \sum\limits_{w \in W_{G/H}}^{} {w \cdot \left( g_{HS:NON}^{H[\rho]}\left( {x,t} \right)~{\prod\limits_{\alpha  \in \Phi _{G/H}^ + }   {\frac{1}{\left( {1 - {z^\alpha }t} \right)\left(1 - {z^{ - \alpha }} \right)}} }\right)} 
\end{aligned}
\end{equation}
where
\begin{equation}
\label{eq:gmodh15}
\begin{aligned}
g_{HS:NON}^{H [\rho]}\left( {x,t} \right) & = \sum\limits_{W_{H/G_0}}^{}{w \cdot \left( {\prod\limits_{\gamma  \in \tilde \Phi _{H/G_0}^ +} {\frac{1}{{1 - {z^\gamma }t}}} \prod\limits_{\delta  \in \Phi _{H/G_0}^ + } {\frac{1}{{1 - {z^{ - \delta }}}}} }\right)} .
\end{aligned}
\end{equation}

Since \ref{eq:gmodh15} takes the same form as \ref{eq:gmodh6}, the nilpotent orbit (or its normalisation) $g_{NON}^{G[\rho]}$ is shown to be induced from the nilpotent orbit (or its normalisation) $g_{NON}^{H[\rho]}$. One feature of the induction method  \ref{eq:gmodh14} is that it opens the door to hybrid constructions in which an Exceptional orbit can be induced from a Classical orbit that has been calculated using the Higgs branch formula. For example, a candidate for $g_{Induced}^{F_4 [1012]}$ can be induced in this manner from the non-normal $g_{Higgs}^{B_3 [101]}$, which we write as:
\begin{equation}
\label{eq:gmodh15a}
\begin{aligned}
g_{Induced}^{{F_4 [1012]}}\left( {x,t} \right) = g_{HS:NON}^{{F_4 [0002]}}\left( {x,t} \right) \left[ {g_{Higgs}^{{B_3 [101]}}\left( {x,t} \right)} \right].
\end{aligned}
\end{equation}

The fugacity maps between the weight space $x$ coordinates of $G$ and $H$ can be obtained by equating the respective simple root fugacities $z$ that are involved in the branching.\footnote{When carrying out induction calculations it is important to equate root space (not weight space) fugacities of $G$ and $H$.}

\paragraph{Charged $NON$ formula}
Finally, it is helpful to generalise version \ref{eq:gmodh7} of the $NON$ formula to deal with root systems that are modulated by background charges, as in \ref{eq:gmodh1}. Define the \emph{charged} $NON$ formula:
\begin{equation}
\label{eq:gmodh16}
g_{HS:NON}^{G[\rho]}\left( {x,t} \right)\left[ {{x^{[n]}}} \right] \equiv  \sum\limits_{w \in W_{G}}^{} {w \cdot \left( x^{[n]} \prod\limits_{\alpha  \in \tilde \Phi _{G/G_0}^ + } {\frac{1}{{1 - {z^\alpha }t}}} \prod\limits_{\beta  \in \Phi _{G}^ + }{\frac{1}{{1 - {z^{ - \beta }}}}}\right)},
\end{equation}
where ${x^{[n]}}$ is a weight given by the CSA coordinates $x$ and Dynkin labels $[n]$. Note that, in order to permit general Dynkin labels, the quotient group $W_{G/G_0}$ structure is not generally available.\footnote{A quotient group structure can be introduced only to the extent that the Weyl group symmetries of $[n]$ permit.}

The charged functions \ref{eq:gmodh16} constitute an orthogonal basis (under an appropriate measure) only in special cases. Specifically, $t \to 0$ yields the Weyl Character formula and $\Phi_G^{[0]} =\emptyset = \Phi_G^{[1]}$ yields charged functions of the maximal nilpotent orbit, which equal the modified Hall Littlewood functions. Unfortunately, the charged functions associated to a nilpotent orbit do not generally constitute an orthogonal basis. This limits their utility, although they can be used to provide a description of relations between non-normal nilpotent orbits and their normalisations, as discussed below.

\subsection{HWGs from $NON$ formula for Hilbert series}
A refined Hilbert series $g_{HS}^{G}({x,t})$ from the $NON$ formula (or from a Coulomb branch or other construction) can be transformed in various ways, as discussed in \cite{Hanany:2014dia, Hanany:2015hxa, Hanany:2016gbz}. The HS can be unrefined as $g_{HS}^{G}({1,t} )$, or converted in a faithful transformation either to a character HWG, or to an orthogonal $mHL$ HWG, by using Weyl integration to project it onto a character generating function:

\begin{equation}
\label{eq:gmodh17}
\begin{aligned}
g_{HWG:NON}^{G\left[ \rho  \right]}\left( {m,t} \right) & =\oint\limits_G {d{\mu ^{G - }}} {~} \prod\limits_{i = 1}^r {\frac{1}{{1 - {m_i}/{x_i}}}}{~}g_{HS:NON}^{G\left[ \rho  \right]}\left( {x,t} \right)\\
{\tilde g_{HWG:NON}^{G\left[ \rho  \right]}}\left( {h,t} \right) & = \oint\limits_G {d{\mu ^{G - }}}{~} \prod\limits_{\alpha  \in {\Phi ^ + }} {\left( {1 - {z^{ \alpha }}t} \right)}{~} \prod\limits_{i = 1}^r {\frac{1}{{1 - {h_i}/{x_i}}}} {~} g_{HS:NON}^{G\left[ \rho  \right]}\left( {x,t} \right)\\
\end{aligned}
\end{equation}
Note that ${\tilde g_{HWG:NON}^{G\left[ \rho  \right]}}\left( {h,t} \right)$ needs to be glued to the $1/{v_{[n]}^G (t)}$ normalisation factors via a further transformation, as discussed in \cite{Hanany:2016gbz}, to obtain ${ g_{HWG:NON}^{G\left[ \rho  \right]}}\left( {h,t} \right)$.

\subsection{Relationship of $NON$ formula to $T^*(G/H)$ theory}
It is instructive to relate the $NON$ formula to a result that appears in \cite{Hanany:2016djz} for the Highest Weight Generating function of the representation content of a $T^*(G/H)$ theory. This moduli space selects a subset of the representations of $G$ from within a generating function for the characters of $G$, by gauging a reductive subgroup $H$:
\begin{equation}
\label{eq:gmodh9}
g^{T^*(G/H)}\left( {m} \right) \equiv \oint\limits_H {{d \mu}^H(y)} \text{ } g_{{\chi}}^G(y,m),
\end{equation}
where $g_{{\chi}}^G( {y,m})$ is the character generating function for $G$ and ${d \mu}^H(y)$ is the Haar measure for $H$.  It can be shown that $g^{T^*(G/H)}\left( {m} \right)$ emerges as a special case from the HWG $g_{HWG:NON}^{G\left[ {even} \right]}\left( {m,t} \right)$, as follows.

First, we define a Levi subgroup of $G$, $H \equiv G_0 \otimes U(1)^{\text{rank}[G]-\text{rank}[G_0]}$, such that $H$ and $G$ have the same rank. This allows us to establish a diffeomorphism between the weight space coordinates $x$ of $G$ and $y$ of $H$. We then transform the refined Hilbert series $g_{HS:NON}^{G\left[ {even} \right]}\left( {x,t} \right)$, calculated from \ref{eq:gmodh8}, to an HWG by projection onto a character generating function for $G$:

\begin{equation}
\label{eq:gmodh10}
g_{HWG:NON}^{G[even]} \left( {m,t} \right) = \oint\limits_G {{d \mu}^G\left( x \right)} \text{ } g_{{\chi}}^G\left( {x^*,m} \right)g_{HS:NON}^{G\left[ {even} \right]}\left( {x,t} \right)
\end{equation}
The Haar measure ${{d \mu}^G\left( x \right)}$ for $G$ can be factorised to separate off the Haar measure ${{d \mu}^H\left( x \right)}$ of the $H$ subgroup:
\begin{equation}
\label{eq:gmodh11}
\begin{aligned}
{d \mu}^G\left( x \right) & = \frac{1}{{\left| {W_{G}} \right|}}\left( {\prod\limits_{i = 1}^{\text{rank}[G]} {\frac{{d{x_i}}}{{{x_i}}}} } \right)\left( {\prod\limits_{\alpha  \in \Phi _G^{}} {\left( {1 - {z^\alpha }} \right)} } \right)\\
& = \frac{1}{\left|{W_{G_0}}\right|}\left( {\prod\limits_{i = 1}^{\text{rank}[G]} {\frac{{d{x_i}}}{{{x_i}}}} } \right)\left( {\prod\limits_{\beta  \in \Phi _{G_0}^{}} {\left( {1 - {z^\beta }} \right)} } \right)\frac{{\left| {W_{G_0}} \right|}}{{\left| {W_{G}} \right|}}\left( {\prod\limits_{\alpha  \in \Phi _{G/G_0}^{}} {\left( {1 - {z^\alpha }} \right)} } \right)\\
&  = {d \mu}^H\left( x \right)\frac{{\left| {W_{G_0}}\right|}}{{\left| {W_{G}}\right|}}\left( {\prod\limits_{\alpha  \in \Phi _{G/G_0}^{}} {\left( {1 - {z^\alpha }} \right)} } \right).
\end{aligned}
\end{equation}
Under the fugacity simplification $t \to 1$, \ref{eq:gmodh8} reduces to:
\begin{equation}
\label{eq:gmodh12}
\begin{aligned}
g_{HS:NON}^{G[even]} \left( {x,1} \right) & = \sum\limits_{ w \in W_{G/G_0}}^{} w \cdot \left( {\prod\limits_{\alpha  \in \Phi _{G/G_0}^ + } {\frac{1}{{\left( {1 - {z^\alpha }} \right)\left( {1 - {z^{ - \alpha }}} \right)}}} } \right)\\
&  = \sum\limits_{w \in W_{G/G_0}}^{} w \cdot \left( {\prod\limits_{\alpha  \in \Phi _{G/G_0}^{}} {\frac{1}{{\left( {1 - {z^\alpha }} \right)}}} } \right)\\
\end{aligned}
\end{equation}
Inserting \ref{eq:gmodh11} and \ref{eq:gmodh12} into \ref{eq:gmodh10}, we obtain:
\begin{equation}
\label{eq:gmodh13}
\begin{aligned}
g_{HWG:NON}^{G[even]} \left( {m,1} \right) 
&  =\oint\limits_{} {d{\mu ^H}\left( x \right)}  {~} g_\chi ^G\left( {{x^*},m} \right)  \frac{{\left| {{W_{{G_0}}}} \right|}}{{\left| {{W_G}} \right|}}  \prod\limits_{\alpha  \in \Phi _{G/{G_0}}^{}} {\left( {1 - {z^\alpha }} \right)}\\
& {~}{~}{~}{~}{~}{~}{~}{~}{~}{~}  \times \sum\limits_{w \in {W_{G/{G_0}}}} w  \cdot \left( {\prod\limits_{\alpha  \in \Phi _{G/{G_0}}^{}} {\frac{1}{{\left( {1 - {z^\alpha }} \right)}}} } \right) \\
&= \oint\limits_{} {{d \mu}^H\left( x \right)}{~}  g_{{\chi}}^G\left( {x^*,m} \right)  \frac{\left|{W_{G_0}}\right|}{\left|{W_{G}}\right|}\frac{\left|{W_{G}}\right|}{\left|{W_{G_0}}\right|}\\
 &  = \oint\limits_H {{d \mu}^H\left( y \right)} ~ g_{{\chi}}^G\left( {y,m} \right)\\
&=g^{T^*(G/H)}\left( {m} \right),
\end{aligned}
\end{equation}
where we have replaced one (redundant) Weyl group summation $\sum\limits_{w \in W_{G/G_0}}$ by $\frac{{\left| {{W_G}} \right|}}{{\left| {{W_{{G_0}}}} \right|}}$, and transformed the conjugate $x^*$ coordinates of $G$ to the $y$ coordinates of $H$. Thus, $g^{T^*(G/H)}\left( {m} \right)$ is a specialisation to $t=1$ of $g_{HWG:NON}^{G[even]} \left( {m,t} \right)$.

\subsection{Classical Group Orbits from the $NON$ formula}
\label {sec:ClassNo}

The Classical group moduli spaces obtained from the $NON$ formula \ref{eq:gmodh6} all have palindromic Hilbert series and are similar in this regard to the Coulomb branch constructions from the unitary monopole formula. For a Classical group nilpotent orbit that is normal, as defined in Appendix \ref{apxNO}, the $NON$ formula yields the same moduli space as its Higgs branch construction, or, where available, Coulomb branch construction\footnote{ These are tabulated up to rank 4 in \cite{Hanany:2016gbz}.}. For a non-normal orbit, the $NON$ formula yields a moduli space that is either (i) a normal component, or (ii) a normalisation of the nilpotent orbit.

The cases that require discussion are the \emph{non-normal} nilpotent orbits \cite{Kraft:1982fk}. The number of these increases with rank; their Characteristics, up to rank 5, are listed in Appendix \ref{apxNO} and the moduli spaces obtained from the $NON$ formula are summarised, up to rank 4, in table \ref{tab:ClassNON1}.

\begin{sidewaystable}[htp]
\begin{center}
\small
\begin{tabular}{|c|c|c|c|c|}
\hline
Orbit & Dim. & Unrefined HS & Character HWG & $mHL$ HWG \dummy\\
\hline
$g_{\text{NON}}^{D_2 [20]} $ & $2$ & $\frac{1+t}{(1-t)^2}   $&$\frac{ 1}{(1 - {m_1}^2 t) }$& \scriptsize $ 1 - {h_2}^2 t $ \\
\hline
$g_{\text{NON}}^{D_4 [0020]} $&$12   $
&
$ \frac{ (1 + 14 t + 36 t^2 + 14 t^3 + t^4)}{(1 - t)^{12} (1 + t)^{-2}} $
&
$\frac{1}{{( {1 - {m_2} t} )( {1 - {m_3}^2 t })}} $
&
\scriptsize $
\begin{array}{c}
1-{{h_1}}^2 t^2-{h_4}^2 t^2-2 {h_2} t^3\\
+{h_1} {h_3} {h_4} t^3+{h_1}^2 t^4+{h_3}^2 t^4\\
+{h_4}^2 t^4+2 {h_1} {h_3} {h_4} t^4-{h_2} {h_3}^2 t^4\\
-{h_2} t^5-2 {h_2}^2 t^5-{h_2} {h_3}^2 t^5\\
+2 {h_1} {h_3} {h_4} t^5+{h_3}^4 t^5-{h_1}^2 {h_4}^2 t^5\\
-{h_1}^2 {h_4}^2 t^6+{h_1} {h_2} {h_3} {h_4} t^6\\
-{h_2}^3 t^7-{h_1} {h_3} {h_4} t^7+2 {h_1} {h_2} {h_3} {h_4} t^7\\
-{h_1} {h_3}^3 {h_4} t^7+{h_1} {h_2} {h_3} {h_4} t^8\\
+{h_2}^2 {h_3}^2 t^8-{h_1} {h_3}^3 {h_4} t^8\\
 \end{array}
$ \\
\hline
$g_{\text{NON}}^{D_4 [0220]} $&$20   $
&
$\frac{ (1+6 t+21 t^2+28 t^3+21 t^4+6 t^5+t^6)}{(1-t)^{20} (1+t)^{-2} (1+t^2)^{-1}} $
&
\tiny$
\frac{{\left( \begin{array}{c}
1 + {m_1}{m_3}{m_4}{t^4} \\
+ {m_1}{m_3}{m_4}{t^5} + {m_1}{m_2}{m_3}{m_4}{t^5}\\
 + {m_2}{m_3}^2{t^6} + {m_1}{m_2}{m_3}{m_4}{t^7}\\
 + {m_1}^2{m_2}{m_4}^2{t^8} - {m_1}{m_2}^2{m_3}{m_4}{t^8}\\
 + {m_1}^2{m_3}^2{m_4}^2{t^9} - {m_1}{m_2}{m_3}^3{m_4}{t^9}\\
 - {m_1}^2{m_2}{m_3}^2{m_4}^2{t^{10}} - {m_1}^3{m_2}{m_3}{m_4}^3{t^{11}}\\
 - {m_1}^2{m_2}{m_3}^2{m_4}^2{t^{12}} - {m_1}^2{m_2}^2{m_3}^2{m_4}^2{t^{12}}\\
 - {m_1}^2{m_2}^2{m_3}^2{m_4}^2{t^{13}} - {m_1}^3{m_2}^2{m_3}^3{m_4}^3{t^{17}}
\end{array} \right)}}{{\left( \begin{array}{c}
(1 - {m_2}t)(1 - {m_1}^2{t^2})(1 - {m_3}^2{t^2})\\
  \times(1 - {m_4}^2{t^2})(1 - {m_2}{t^3})(1 - {m_1}{m_3}{m_4}{t^3})\\
  \times (1 - {m_2}^2{t^4})(1 - {m_3}^2{t^4})(1 - {m_1}^2 {m_4}^2{t^6})
\end{array} \right)}}
$
&
\scriptsize$ 1 - {h_2} t^3 + {h_3}^2 t^4 - {h_2} t^5$ \\
\hline
\hline
$g_{\text{NON}}^{B_3 [101]} $&$12 $&
$ \frac{ (1+14 t+36 t^2+14 t^3+t^4)}{(1-t)^{12} (1+t)^{-2}}   $
&
$ \frac{1}{(1-{m_1} t) (1-{m_2} t) (1-{m_3}^2 t^2)}   $
&
\scriptsize $
\begin{array}{c}
1 +{h_1} t -{h_3}^2 t^2 -{h_1} t^3 \\
-2 {h_2} t^3 +{h_3}^2 t^4 +{h_1} {h_3}^2 t^4\\
-{h_2}^2 t^5 +{h_1} {h_3}^2 t^5\\
 \end{array}
$ \\
\hline
$g_{\text{NON}}^{C_4 [0200]} $&$  22 $
&
\scriptsize $\frac{\left(
\begin{array}{c}
 1+13 t+133 t^2+608 t^3+1478 t^4 +2002 t^5\\
 +1478 t^6+608 t^7+133 t^8+13 t^9+t^{10} 
  \end{array}
\right)}{\begin{array}{c}(1-t)^{22}(1+t)^{-1}\\ \end{array}}   $
&
\scriptsize $ \frac{\begin{array}{c} 1+{m_1} {m_2} {m_3} t^4 \\ \end{array}}
{\left(
\begin{array}{c}(1-{m_1}^2 t) (1-{m_2}^2 t) (1-{m_2} t^2)\\
\times (1-{m_3}^2 t^3) (1-{m_4} t^2) (1-{m_1} {m_3} t^3)
\end{array}
\right)} $
&
\scriptsize $
\begin{array}{c}
1 + {h_4} t^2 - {h_1}^2 t^3 - {h_1} {h_3} t^3 - {h_2} t^4\\
+ {h_1}^2 {h_2} t^4 - {h_1} {h_3} t^4 - {h_4} t^4 \\
- {h_1}^4 t^5 + {h_1}^2 {h_2} t^5 + {h_2}^2 t^5 + {h_2} {h_4} t^5\\
- {h_2} t^6 + {h_1} {h_3} t^6 - {h_3}^2 t^6 + {h_4} t^6\\
- {h_1}^2 {h_4} t^6 + {h_2} {h_4} t^6 - {h_2}^2 t^7 - {h_2}^3 t^7 \\
+ 2 {h_1} {h_3} t^7 - {h_1}^2 {h_4} t^7 - {h_2} {h_4} t^7 \\
+ {h_1}^2 {h_2}^2 t^8 + {h_3}^2 t^8 - {h_2} {h_4} t^8 \\
+ {h_1}^3 {h_3} t^9 + {h_2}^2 {h_4} t^9 - {h_1}^2 {h_2} {h_4} t^{10}  
 \end{array}
$ \\
\hline
$g_{\text{NON}}^{B_4 [2101]} $&$ 26  $
&
\scriptsize $\frac{\left(
\begin{array}{c}
1+7 t+39 t^2 +152 t^3+340 t^4 +410 t^5 \\
+340 t^6 +152 t^7 +39 t^8 +7 t^9+  t^{10}
\end{array}
\right)}
{\begin{array}{c}(1-t)^{26} (1+t)^{-3} \\ {~}\end{array} }$
&
$\ldots $
&
\scriptsize $
\begin{array}{c}
1+{h_1} t-{h_3}^2 t^2 -{h_1} t^3-2 {h_2} t^3\\
+{h_3}^2 t^4+{h_1} {h_3}^2 t^4\\
-{h_2}^2 t^5+ {h_1} {h_3}^2 t^5\\
\end{array}$ \\
\hline
\end{tabular}
\end{center}
\text{D series moduli spaces are normal components of spinor pair nilpotent orbits; B/C moduli spaces are normalisations.}\\
\text{$NON$/Coulomb branch HS counted by $t$ correspond to Higgs branch HS counted by $t^2$.}\\
\normalsize
\caption{Classical Non-Normal Orbit Components and Normalisations}
\label{tab:ClassNON1}
\end{sidewaystable}

In the case of the $D_{2r}$ spinor pairs of nilpotent orbits, discussed in \cite{Hanany:2016gbz}, the $NON$ formula gives the individual palindromic spinor moduli spaces, according to the Characteristic chosen. Examples in table \ref{tab:ClassNON1} include $D_2 [20]$, $D_4 [0020]$ and $D_4 [0220]$. The moduli spaces of the conjugate spinors have identical unrefined Hilbert series and their HWGs are related by the exchange of spinor fugacities. These spinor moduli spaces are the normal components of the non-normal nilpotent orbits, constructed on the Higgs branch, which are their unions:

\begin{equation}
\label{eq:ClassNON1}
\begin{aligned}
g_{Higgs}^{{D_{2r}}[ \ldots 20]I/II} = g_{Higgs}^{{D_{2r}}[ \ldots 02]I/II} = g_{NON}^{{D_{2r}}[ \ldots 20]} + g_{NON}^{{D_{2r}}[ \ldots 02]} - g_{NON}^{{D_{2r}}[ \ldots 20] \cap {D_{2r}}[ \ldots 02]\} }.
\end{aligned}
\end{equation}
Also, for spinor pair orbits of Characteristic Height 2, the HS from the $NON$ formula match those obtained from Coulomb branch constructions.

For all the other non-normal nilpotent Classical group orbits, the $NON$ formula yields a normalisation. Examples in table \ref{tab:ClassNON1} include $B_3 [101]$, $B_4 [2101]$ and $C_4 [0200]$.
In each case the (non-normal) Higgs branch construction can be recovered from the normalisation by excluding those elements that fall outside the nilpotent cone $\cal N$. Thus:
\begin{equation}
\label{eq:ClassNON2}
\begin{aligned}
g_{Higgs}^{B_3 \left[ {101} \right]} &= g_{NON}^{B_3 \left[ {101} \right]} - g_{NON}^{B_3 \left[ {200} \right]} \left[ {{x_{1}}t} \right],\\
g_{Higgs}^{B_4 \left[ {2101} \right]} &= g_{NON}^{B_4 \left[ {2101} \right]} - g_{NON}^{B_4 \left[ {2200} \right]}\left[ {{x_{1}}{t^2}} \right],\\
g_{Higgs}^{C_4 \left[ {0200} \right]} &= g_{NON}^{C_4 \left[ {0200} \right]} - g_{NON}^{C_4 \left[ {0002} \right]}\left[ {{x_{4}}{t^2}} \right].
\end{aligned}
\end{equation}
\FloatBarrier
The elements of the moduli spaces in \ref{eq:ClassNON2} that fall outside the nilpotent cone can be described by the charged $NON$ formula \ref{eq:gmodh16}. The nilpotent orbit upon which each of these charged moduli spaces is built is related to its parent orbit by the $A_{2r-1} \cup A_{2r-1}$ Kraft-Procesi degeneration and lies beneath the parent orbit in the Hasse diagram. 

It can be seen from table \ref{tab:ClassNON1}, that $g_{\text{NON}}^{D_4 [0020]}$ and $g_{Higgs}^{B_3 \left[ {101} \right]}$ have the same unrefined Hilbert series. This is an example of a branching relationship between two nilpotent orbits.

Importantly, the equality between the normal Classical group nilpotent orbits constructed on Higgs or Coulomb branches and the moduli spaces obtained from the $NON$ formula, justifies the use of the $NON$ formula to construct Exceptional group nilpotent orbits. Also, the relationship between non-normal Classical group orbits and their normalisations, or normal components, obtained from the $NON$ formula, provides some insight into the relationships between non-normal orbits and their normalisations.

\FloatBarrier

\subsection{Exceptional Group Orbits from the $NON$ Formula}
\label {sec:ExNo}

The construction of Exceptional group nilpotent orbits poses a number of challenges. Firstly, Exceptional groups do not act in a similar manner to $SL(n,\mathbb{R/C})$ rotation matrices on their fundamental vector spaces, so the Higgs branch method is not available \cite{Kraft:1982fk}. This limits the construction methods to those based on the Coulomb branch or $NON$ formulae. These in turn have their own limitations; the unitary monopole formula only works for minimal and near minimal orbits; and the $NON$ formula yields the normalisation of a nilpotent orbit, which only equals the orbit if it is normal. Finally, the high dimensions of the Weyl groups of the E series entail that explicit calculations, using the methods developed during this study, are not always feasible in terms of computing memory and/or time.

In principle, however, those Exceptional group Characteristics for which the $NON$ formula does yield nilpotent orbits can be identified by verifying that the moduli spaces are entirely contained within the nilpotent cone $\cal N$, which is known for every group. Such results can be cross-checked for completeness by comparison with the known non-normal orbits listed in Appendix \ref{apxNO}. Moreover, even without a systematic formula for calculating the non-normal Exceptional group orbits, it is often possible to find candidates on a case by case basis, by restricting their normalisations to exclude charged nilpotent orbits of lower dimension, as will be shown. The findings presented below are derived from a combination of established results, full HS and HWG calculations, and inferences based on unrefined HS expansions, checked to the highest order practicable.

In this study, Exceptional group nilpotent orbits are labelled by their Characteristics for various reasons. Firstly, a Characteristic provides the structure and parameters of the Coulomb branch and $NON$ formulae. Secondly, while a Characteristic provides a clear and unambiguous specification of a nilpotent element $X$, the same is not true of the various alternative labelling methods based on sub-groups, developed, inter alia, by Dynkin \cite{Dynkin:1957um}, Bala-Carter \cite{Bala:1976kl, Bala:1976tg}, Hesselink \cite{HESSELINK:1978fp}. Amongst these, the method that is closest to the use of Characteristics is given by Hesselink, who identifies the semi-simple subgroup $G_0$ under which a nilpotent element $X$ is invariant; this labelling method works for most Richardson orbits, but is not sufficiently general to embrace other types.

As is clear from the discussion on the variants of the $NON$ formula in section \ref{sec:NON}, there is often a choice to be made as to whether an orbit is calculated directly from the roots of $G$, or induced from an orbit of a subgroup $H$, using \ref{eq:gmodh14}. Either choice leads to the same refined HS under the $NON$ formula, but the induction method permits the incorporation, for example, of a non-normal nilpotent orbit of $H$ calculated on the Higgs branch.

The following sub-sections analyse the nilpotent orbits of Exceptional groups, starting from the Characteristics of $SU(2)$ homomorphisms, classifying the type of each orbit, giving its constructions, calculating, where practicable, unrefined HS, character HWGs and $mHL$ HWGs, and identifying whether the moduli spaces are nilpotent orbits or normalisations of non-normal orbits. For $G_2$, $F_4$ and $E_6$, nilpotent orbit Hasse diagrams are drawn, based on moduli space inclusion relations.

\subsubsection{Orbits of $G_2$}
\label {subsec:G2NO}

Table \ref{tab:G2NO1} classifies the 5 nilpotent orbits of $G_2$ and gives their unrefined HS. Table \ref{tab:G2NO2} gives the character and $mHL$ HWGs, calculated from the refined HS. To comment on the various orbits:

\begin{description}

\item $[10]$: 6 dimensional nilpotent orbit. This is the minimal nilpotent orbit and is both rigid and normal. It can be calculated either from a Coulomb branch quiver theory built on the affine Dynkin diagram, as discussed in section \ref{sec:Coulomb}, or from the $NON$ formula. Both its HS and character HWG are palindromic.

\item $[01]$: 8 dimensional nilpotent orbit. This next to minimal orbit is rigid, but not normal. It has Characteristic Height 3 and does not have a Coulomb branch construction. The $NON$ formula yields a normalisation. The non-normal orbit is found by excluding from this normalisation a subspace expressed in terms of the charged $NON$ formula for the minimal nilpotent orbit:

\begin{equation}
\label{eq:exno1}
\begin{aligned}
g_{NO}^{{G_2 \left[ {01} \right]}}{} = g_{NON}^{{G_2}\left[ {01} \right]} - g_{NON}^{{G_2}\left[ {10} \right]}\left[ {x_{2}t} \right].
\end{aligned}
\end{equation}
\item $[20]$: 10 dimensional nilpotent orbit. The sub-regular nilpotent orbit is distinguished and has an invariant subgroup $G_0=A_1$. It can be calculated from the $NON$ formula \ref{eq:gmodh8}. Both its HS and character HWG are palindromic.

\item $[22]$: 12 dimensional nilpotent orbit. The maximal nilpotent orbit is distinguished. It can be calculated from the $NON$ formula. Both its HS and character HWG are palindromic, in the latter case with degree ${m_1}^3 {m_2}^5 t^{18}$.

\end{description}
The normalisation $g_{NON}^{{G_2}\left[ {01} \right]}$ has the same unrefined Hilbert series (up to $t$ counting conventions) as $g_{Higgs}^{{B_3}\left[ {010} \right]}$, as tabulated in \cite{Hanany:2016gbz}, and can be obtained from this using a character map that folds the $B_3$ vector and spinor together \cite{brylinski_kostant_1994}.

\begin{figure}[htbp]
\begin{center}
\includegraphics[scale=.7]{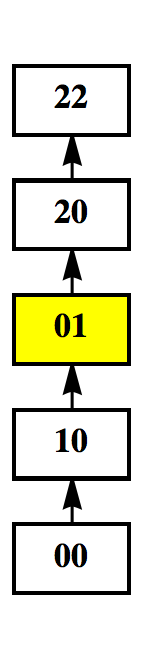}
\caption[$G_2$ Nilpotent Orbit Hasse Diagram]{$G_2$ Nilpotent Orbit Hasse Diagram. The diagram is derived from Hilbert series inclusion relations, with the yellow node indicating a non-normal nilpotent orbit.}
\label{fig:G2Hasse}
\end{center}
\end{figure}

It can easily be checked, both from the unrefined HS and the character HWGs, that these nilpotent orbits satisfy the expected inclusion relations $g_{NO}^{{G_2}\left[ {00} \right]} \subset g_{NO}^{{G_2}\left[ {10} \right]} \subset g_{NO}^{{G_2}\left[ {01} \right]} \subset g_{NO}^{{G_2}\left[ {20} \right]} \subset g_{NO}^{{G_2}\left[ {22} \right]}$, providing that the non-normal 8 dimensional nilpotent orbit is used. These inclusion relations are graphed in the Hasse diagram in figure \ref{fig:G2Hasse}.

\begin{sidewaystable}[htp]
\begin{center}
\begin{tabular}{|c|c|c|c|c|}
\hline
 \text{Characteristic} & \text{Type} & \text{Construction} & \text{Dim.} & \text{Hilbert Series} \dummy \\
 \hline
$ [00] $& \text{Even} & \text{Trivial} &$ 0$ &$ 1$ \\
\hline
$ [10]$ & {Rigid} &
$\begin{array}{*{20}{c}}
{g_{{Coulomb}}^{{G_2 \left[ {10} \right]}}}
\\\text{or}\\{g_{NON}^{{G_2 \left[ {10} \right]}}} \end{array}$
 &$ 6 $&$ \frac{\left(1+t\right) \left(1+7 t+t^2\right)}{\left(1-t\right)^6}$ \\
 \hline
$ [0 1] $& {Rigid, Non-normal} &
$g_{NON}^{{G_2 \left[ {0 1} \right]}} - g_{NON}^{{G_2 \left[ {1 0} \right]}}\left[ {{x_2 }t} \right] $
&$ 8 $&$ \frac{1+6 t+20 t^2+43 t^3-7 t^4-7 t^{5}}{\left(1-t\right)^8}$ \\
\hline
$ [2 0] $& {Distinguished} &
 $g_{NON}^{{G_2 \left[ {2 0} \right]}}$
&$ 10 $&$ \frac{\left(1+t\right) \left(
1 + 3 t + 6 t^2 + 3 t^3 + t^4
\right)}{\left(1-t\right)^{10}} $\\
\hline
$ [2 2] $& {Distinguished} &
 $g_{NON}^{{G_2 \left[ {2 2} \right]}} $
&$ 12 $&$ \frac{\left(1-t^2\right) \left(1-t^{6}\right)}{\left(1-t\right)^{14}}$ \\
\hline
\hline
 $[01] $& {Rigid, Normalisation} &
 $g_{NON}^{{G_2 \left[{0 1} \right]}} $
 &$ 8 $&$ \frac{1+13 t+28 t^2+13 t^3+t^4}{\left(1-t\right)^8} $\\

\hline
\end{tabular}
\end{center}
\text{Both the non-normal 8 dimensional nilpotent orbit and its normalisation are shown.} 
\caption{$G_2$ Orbit Constructions and Hilbert Series}
\label{tab:G2NO1}
\end{sidewaystable}
\begin{sidewaystable}[htp]
\begin{center}
\begin{tabular}{|c|c|c|}
 \hline
 \text{Characteristic} & \text{Character HWG} & \text{$mHL$ HWG} \dummy \\
 \hline
$ [0 0]  $&$ 1  $&$
\begin{array}{c}
1-{h_1} t+{h_2}^3 t^2-{h_1}^2 t^3-{h_2} t^3\\
+{h_1} {h_2} t^3-{h_2}^4 t^3+{h_1} {h_2} t^4-{h_2}^3 t^4\\
+{h_1} {h_2}^3 t^4-{h_1}^3 t^{5}+{h_1} {h_2}^3 t^{5}\\
-{h_2}^5 t^{5}+{h_1}^2 {h_2}^2 t^{6}
 \end{array}
  $\\
  \hline
 $[1 0]  $&$ \frac{1}{1- {m_1} t}  $&$
 \begin{array}{c}
  1-{h_2}^2 t^2-{h_2} t^3+{h_1} {h_2} t^3\\
  +{h_1} {h_2} t^4-{h_2}^3 t^4
 \end{array}
  $\\
   \hline
$
\begin{array}{c}
 [0 1]\\
\text{} 
 \end{array} $&$ \frac{1- {m_2}^6 t^{6}}{\left(1- {m_1} t\right) \left(1- {m_2}^2 t^2\right) \left(1- {m_2}^3 t^3\right)}  $&$
  1-{h_2} t^3-{h_1}^2 t^4+{h_1} {h_2}^2 t^{5}
  $\\
  \hline
$ [2 0]  $&$ \frac{1+ {m_1} {m_2}^3 t^{5}}{\left(1- {m_1} t\right) \left(1- {m_2}^2 t^2\right) \left(1- {m_2}^3 t^3\right) \left(1- {m_1}^2 t^4\right)}  $&$ 1- {h_2} t^3 $\\
  \hline
$ [2 2]  $&\scriptsize $
 \frac{
\left(\begin{array}{c}
  1 -{m_2} t^2 -{m_1} t^4 +{m_2}^2 t^4\\
  +{m_1} {m_2} t^{5} +{m_1} {m_2}^3 t^{5}+{m_1} {m_2} t^{6}+{m_1} {m_2}^2 t^{6}\\
   +{m_1} {m_2} t^{7}-{m_1} {m_2}^2 t^{7}-{m_1} {m_2}^4 t^{7}\\
   +{m_1}^2 t^{8}-{m_1} {m_2}^2 t^{8}-{m_1} {m_2}^3 t^{8} -{m_1} {m_2}^4 t^{8}\\
   -{m_1} {m_2}^2 t^{9}-{m_1}^2 {m_2}^3 t^{9}\\
   -{m_1}^2 {m_2} t^{10}-{m_1}^2 {m_2}^2 t^{10}-{m_1}^2 {m_2}^3 t^{10}+{m_1} {m_2}^5 t^{10}\\
   -{m_1}^2 {m_2} t^{11} -{m_1}^2 {m_2}^3 t^{11}+{m_1}^2 {m_2}^4 t^{11}\\
   +{m_1}^2 {m_2}^3 t^{12}+{m_1}^2 {m_2}^4 t^{12} +{m_1}^2 {m_2}^2 t^{13}+{m_1}^2 {m_2}^4 t^{13}\\
   +{m_1}^3 {m_2}^3 t^{14}-{m_1}^2 {m_2}^5 t^{14} -{m_1}^3 {m_2}^4 t^{16} +{m_1}^3 {m_2}^5 t^{18}\\
\end{array}\right)
 }
 {\left( \begin{array}{c} \left(1- {m_1} t\right) \left(1- {m_2} t^2\right) \left(1- {m_2}^2 t^2\right) \left(1- {m_2} t^3\right) \\
 \times \left(1- {m_2}^3 t^3\right) \left(1- {m_1} t^4\right) \left(1- {m_1}^2 t^4\right) \left(1- {m_1} t^{5}\right) \end{array}\right)}  $&$ 1$ \\
   \hline
  \hline
$
\begin{array}{c}
 [0 1]\\
\text{(Normalisation)}
 \end{array} $&$ \frac{1}{\left(1- {m_1} t\right) \left(1- {m_2} t\right)}  $&$
  \begin{array}{c}
1+{h_2} t-{h_2}^2 t^2-{h_1} t^3\\
-{h_2} t^3+{h_1} {h_2} t^4 
 \end{array}
  $\\
  \hline
\end{tabular}
\end{center}
\text{Both the non-normal 8 dimensional nilpotent orbit and its normalisation are shown.} 
\caption{$G_2$ Orbits and HWGs}
\label{tab:G2NO2}
\end{sidewaystable}

\FloatBarrier
\subsubsection{Orbits of $F_4$}
\label {subsec:F4NO}

The 16 nilpotent orbits of $F_4$ are described in tables \ref{tab:F4NO1} to \ref{tab:F4NO5}, which give their classification, constructions, unrefined HS and, where practicable, character HWGs and modified Hall Littlewood HWGs. Tables \ref{tab:F4NO3} and \ref{tab:F4NO6} contain similar information for the normalisations of the non-normal orbits. Many orbits have distinctive features:

\begin{description}

\item $[1 0 0 0]$ and $[0 0 0 1]$: 16 dimensional minimal and 22 dimensional next to minimal nilpotent orbits. These orbits are rigid and have the invariant subgroups $C_3$ and $B_3$, respectively. Their Hilbert series can be calculated either (i) from the Coulomb branch of a quiver theory built, as described in section \ref{sec:Coulomb}, on the affine or twisted affine Dynkin diagram of $F_4$ (as shown in figure \ref{fig:CBgrid}) or (ii) from the $NON$ formula \ref{eq:gmodh6}. Their HS and character HWGs are palindromic.

\item $[0 1 0 0]$: 28 dimensional nilpotent orbit. This orbit is rigid and has the invariant subgroup $A_1 \otimes A_2$. Its Hilbert series can be calculated either from the $NON$ formula, or as the intersection of the two 30 dimensional orbits. Both the HS and character HWG are palindromic.

\item $[2 0 0 0]$: 30 dimensional nilpotent orbit. This orbit is even, has the invariant subgroup $C_3$, and is normal. Its Hilbert series can be calculated from the $NON$ formula. Both the HS and character HWG are palindromic.

\item $[0 0 0 2]$: 30 dimensional nilpotent orbit. This orbit is even, has the invariant subgroup $B_3$, and is non-normal. The $NON$ formula yields a normalisation. The candidate for the non-normal orbit is found by excluding from this normalisation a subspace expressed in terms of the charged $NON$ formula for the 28 dimensional orbit:
\begin{equation}
\label{exno2}
g_{NO{{}_{}}}^{{F_4}\left[ {0 0 0 2} \right]}{} = g_{NON}^{{F_4}\left[ {0 0 0 2} \right]} - g_{NON}^{{F_4}\left[ {0 1 0 0} \right]}\left[ {{x_4}{t^2}} \right],
\end{equation}
with notation as per \ref{eq:gmodh16}. Both the HS and character HWG are non-palindromic.

\item $[0 0 1 0]$: 34 dimensional nilpotent orbit. This orbit is rigid and has the invariant subgroup $A_2 \otimes A_1$. It can be calculated from the $NON$ formula. The HS and character HWG (omitted) are palindromic.

\item $[2 0 0 1]$: 36 dimensional nilpotent orbit. This orbit is non-rigid, has the invariant subgroup $B_2 \cong C_2$, and is non-normal. Its normalisation can be calculated from the $NON$ formula. The candidate for the non-normal nilpotent orbit is found by excluding from its normalisation a subspace expressed by applying the charged $NON$ formula to the 34 dimensional orbit:
\begin{equation}
\label{exno3}
g_{NO{{}_{}}}^{{F_4}\left[ {2 0 0 1} \right]}{} = g_{NON}^{{F_4}\left[ {2 0 0 1} \right]} - g_{NON}^{{F_4}\left[ {0 0 1 0} \right]}{}\left[ {{x_1}{t^3}} \right].
\end{equation}
The HS is non-palindromic.

\item $[0 1 0 1]$: 36 dimensional nilpotent orbit. This orbit is rigid, has the invariant subgroup $A_1 \otimes A_1$, and is non-normal. Its normalisation can be calculated from the $NON$ formula. The candidate for the non-normal nilpotent orbit is found by excluding from its normalisation a subspace expressed by applying the charged $NON$ formula to the 34 dimensional orbit:
\begin{equation}
\label{exno4}
g_{NO}^{{F_4}\left[ {0 1 0 1} \right]}{} = g_{NON}^{{F_4}\left[ {0 1 0 1} \right]}{} - g_{NON}^{{F_4}\left[ {0 0 1 0} \right]}{}\left[ {{x_4}{t^2}} \right].
\end{equation}
Note the difference in charges between \ref{exno4} and \ref{exno3}. The HS is non-palindromic.

\item $[1 0 1 0]$: 38 dimensional nilpotent orbit. This orbit is non-rigid, has the invariant subgroup $A_1 \otimes A_1$, and is non-normal. Its normalisation is found from the $NON$ formula. The candidate for the non-normal nilpotent orbit is found by excluding from this normalisation a subspace expressed in terms of charged $NON$ formulae for the two 36 dimensional orbits:
\begin{equation}
\label{exno5}
\begin{aligned}
g_{NO}^{{F_4}\left[ {1 0 1 0} \right]}{} = g_{NON}^{{F_4}\left[ {1 0 1 0} \right]}{}
  & - g_{NON}^{{F_4}\left[ {2 0 0 1} \right]}{}\left[ {{x_1}{t^3} + {x_4}{t^2}} \right] \\
 & - g_{NON}^{{F_4}\left[ {0 1 0 1} \right]}{}\left[ {{x_3}{x_4}{t^6}} \right]
\end{aligned}
\end{equation}
Its HS is non-palindromic.

\item $[1 0 1 2]$: 42 dimensional nilpotent orbit. This orbit is Richardson, has the invariant subgroup $A_1$, and is non-normal. Its normalisation can be calculated from the $NON$ formula. Possible candidates for the non-normal nilpotent orbit can be found either (i) by excluding from its normalisation a subspace expressed in terms of the charged $NON$ formula for the 40 dimensional orbit $g_{NO}^{F_4 \left[ {0 2 0 0} \right]}$, or (ii) by induction (using  \ref{eq:gmodh14}) from $g_{Higgs}^{{B_3}\left[ {1 0 1} \right]}$:
\begin{equation}
\label{exno6}
\begin{aligned}
g_{NO}^{{F_4}\left[ {1012} \right]}{} & = g_{NON}^{{F_4}\left[ {1012} \right]}{} - g_{NON}^{{F_4}\left[ {0200} \right]}{}\left[ {{x_4}{t^2} + {x_3}{t^6}} \right]\\
g_{Induced}^{{F_4}\left[ {1012} \right]}{} & = g_{NON}^{{F_4}\left[ {0002} \right]}{}\left[ {g_{Higgs}^{B3\left[ {101} \right]}} \right]
\end{aligned}
\end{equation}
The former is taken as the candidate for the non-normal orbit $g_{NO}^{{F_4}\left[ {1012} \right]}{}$, on the grounds that it is consistent with the restriction method detailed below, and that it includes $[0200]$, as in the standard Hasse diagram. Its HS is non-palindromic.

\item $[0 2 0 0]$, $[0 2 0 2]$, $[2 2 0 2]$ and $[2 2 2 2]$: 40 dimensional , 44 dimensional, 46 dimensional sub-regular, and 48 dimensional maximal nilpotent orbits. These orbits are distinguished and contain the invariant subgroups $A_1 \otimes A_2$, $A_1 \otimes A_1$, $A_1$ and $\emptyset $, respectively. They are found from the $NON$ formula. Their HS are palindromic.

\end{description}

The above list excludes the moduli space defined by the $SU(2)$ homomorphism which has the root map $[2 0 0 2]$. Detailed calculation of Hilbert series shows that $g_{NON}^{{F_4}\left[ {2 0 0 2} \right]}$ is not a nilpotent orbit, but is an extension of the distinguished $g_{NO}^{{F_4}\left[ {0 2 0 0} \right]}$, and can be described using the charged $NON$ formula:
\begin{equation}
\label{exno6a}
g_{NON{{}_{}}}^{{F_4}\left[ {2002} \right]}{} = g_{NO}^{{F_4}\left[ {0200} \right]}{}\left[ {1 + {x_1} {t^3}+{x_4}  {t^2}} \right].
\end{equation}

It is necessary to make some caveats in relation to the non-normal orbits. Firstly, the method of finding the charged $NON$ formula descriptions that restrict their normalisations to $\cal N$ is partly empirical, guided by unrefined HS and character HWGs, where known. The \emph{restricted $NON$} method used  for $F_4$ has been (i) to fix the moduli space inclusion relations below a non-normal orbit using its normalisation and (ii) to exclude from the normalisation a subspace containing one (or sometimes more) \emph{charged} normalisations of orbits lying immediately below in the Hasse diagram, such that the resulting non-normal moduli space lies within the nilpotent cone. This method is consistent with the Higgs branch constructions of non-normal Classical orbits studied in \ref{sec:ClassNo} and has been sufficient to specify candidates for the non-normal orbits of $F_4$.

Secondly, since the charged $NON$ formula does not generally yield an orthogonal basis, there may be alternative charged $NON$ formula descriptions of the non-normal orbits that give the same result.

Finally, it has only proved possible to calculate character HWGs and to use their Taylor series expansions to check the irrep inclusion relations explicitly up to the 34 dimensional nilpotent orbit; for the 36 dimensional and 38 dimensional non-normal orbits, in particular, the analysis has been largely dimensional in nature and therefore should not be taken as definitive.

\begin{figure}[htbp]
\begin{center}
\includegraphics[scale=0.5]{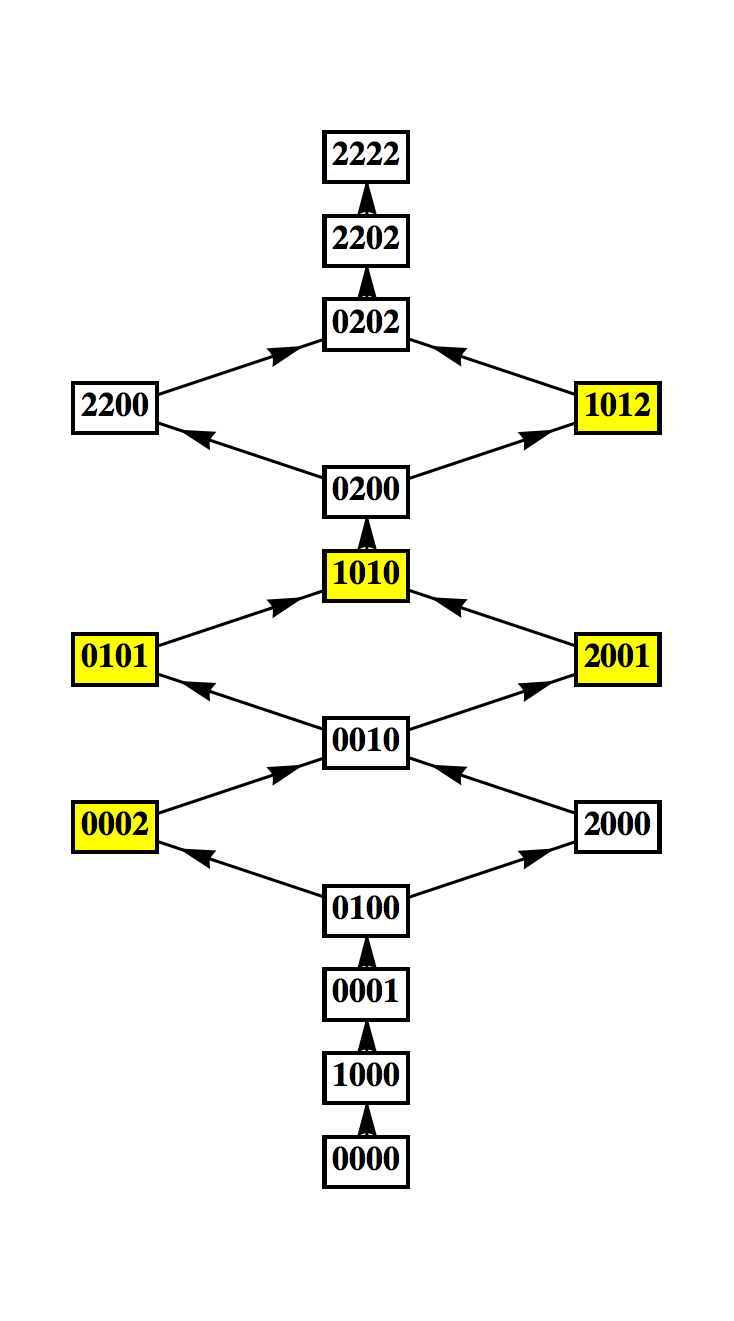}
\includegraphics[scale=0.5]{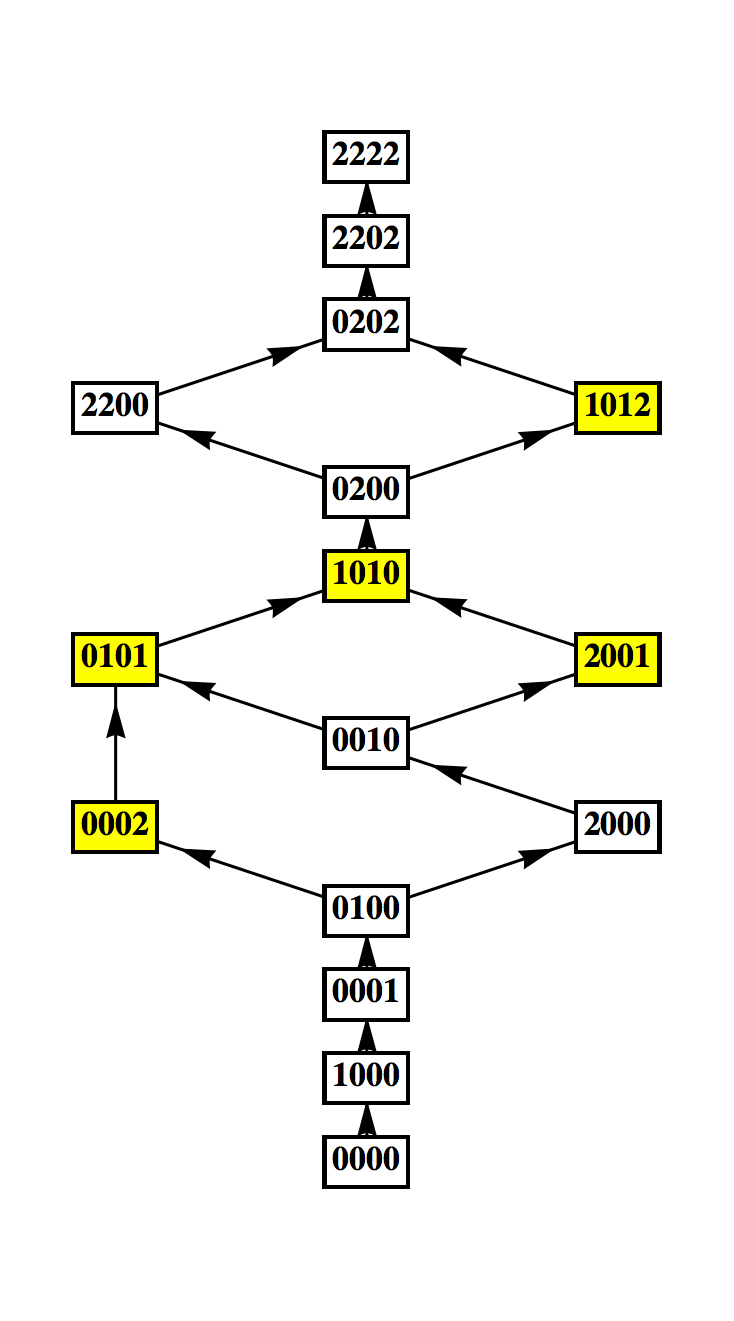}\\
\caption[$F_4$ Nilpotent Orbit Hasse Diagram]{$F_4$ Nilpotent Orbit Hasse Diagram. The left hand diagram is derived from Hilbert series and HWG inclusion relations. The right hand diagram is taken from the mathematical Literature \cite{Adams:jk, Baohua-Fu:2015nr}. Yellow nodes indicate non-normal nilpotent orbits.}
\label{fig:F4Hasse}
\end{center}
\end{figure}

It is interesting to compare the inclusion relations obtained from this analysis of moduli spaces with the standard Hasse diagrams of nilpotent orbits in the mathematical Literature \cite{Collingwood:1993fk, Adams:jk, Baohua-Fu:2015nr}, which are based on earlier work in \cite{spaltenstein_1982}. Figure \ref{fig:F4Hasse} compares the Hasse diagram defined by the inclusion relations amongst the Hilbert series of nilpotent orbits $g_{NO}^{F_4}$ to the standard Hasse diagram.

Unlike the case of Classical group nilpotent orbits, where there is an exact correspondence between the Hasse diagrams (omitted) based upon Hilbert series inclusion relations and the standard diagrams \cite{Collingwood:1993fk, Kraft:1982fk}, there is a discrepancy involving the linking pattern between $F_4[2000]$ and $F_4[0010]$, where the restricted $NON$ method yields an inclusion relationship that is absent in the standard diagram.

One possibility could be that the subtle distinction between normal and non-normal orbits has not been consistently treated in the analyses in the Literature upon which the standard Hasse diagrams are based. In this context, it is worth noting that $g_{NO}^{{F_4}\left[ {0 0 1 0} \right]}$ \emph{does not} include $g_{NON}^{{F_4}\left[ {2 0 0 0} \right]}$,  which is the normalisation of $g_{NO}^{{F_4}\left[ {2 0 0 0} \right]}$.

The nilpotent orbits of $F_4$ include some special orbits, as defined in Appendix \ref{apxNO}. These are summarised in table \ref{tab:F4SDSO}, along with their duals under the Spaltenstein map.

\begin{table}[htp]
\begin{center}
\begin{tabular}{|c|}
\hline
\text {Spaltenstein Dual Orbits}\\
\hline
${[0 0 0 0]}  \Leftrightarrow {[2 2 2 2]}$\\
${[0 0 0 1]} \Leftrightarrow {[2 2 0 2]}$\\
${[0 1 0 0]} \Leftrightarrow {[0 2 0 2]}$\\
${[0 0 0 2]} \Leftrightarrow {[1 0 1 2]}$\\
${[2 0 0 0]} \Leftrightarrow {[2 2 0 0]}$\\
${[0 2 0 0]} \Leftrightarrow {[0 2 0 0]}$\\
\hline
\end{tabular}
\end{center}
\caption{$F_4$ Special Nilpotent Orbits}
\label{tab:F4SDSO}
\end{table}
The special orbit $[0 2 0 0]$ is self dual. The non-normal orbits ${[0 0 0 2]}$ and ${[1 0 1 2]}$ are special and dual to each other. The other non-normal orbits $[2 0 0 1]$, $[0 1 0 1]$ and $[1 0 1 0]$ are not special. It appears that the symmetries of the left hand Hasse diagram based on Hilbert series and HWG inclusion relations are a better fit for the symmetries of the Spaltenstein map.

\begin{sidewaystable}[htp]
\small
\begin{tabular}{|c|c|c|c|c|}
\hline
 {Characteristic} & {Type} & {Construction} & {Dim.} & Unrefined HS \dummy \\
 \hline
$ [0 0 0 0] $& {Even} & 
\scriptsize ${g_{NON}^{{F_4}\left[ {0 0 0 0} \right]}{}}$ &$ 0$ &$ 1$ \\
\hline
$ [1 0 0 0]$ & {Rigid} &
\scriptsize $\begin{array}{*{20}{c}}
{g_{{{Coulomb}}}^{{F_4}\left[ {1 0 0 0} \right]}{}}\\
{{ \text{or}}}\\
{g_{NON}^{{F_4}\left[ {1 0 0 0} \right]}{}}
\end{array}$
 &
 $ 16 $
 &
 \scriptsize $\frac{\left(\begin{array}{c} 1 + 36 t + 341 t^2 + 1208 t^3 + 1820 t^4 \\
 + 1208 t^5 + 341 t^6 + 36 t^7 + t^8 \end{array} \right)}
 {\begin{array}{c}(1-t)^{16}\\ \end{array}}$ \\
\hline
 $[0 0 0 1] $
 &
 {Rigid} 
 &
\scriptsize $\begin{array}{*{20}{c}}
{g_{{{Coulomb}}}^{{F_4}\left[ {0 0 0 1} \right]}{}}\\
{{ \text{or}}}\\
{g_{NON}^{{F_4}\left[ {0 0 0 1} \right]}{}}
\end{array}$
 &
 $ 22 $
 &
 \scriptsize $ \frac{ \left(\begin{array}{c}1 + 29 t + 435 t^2 + 2948 t^3 + 8998 t^4 + 12969 t^5\\
  + 8998 t^6 + 2948 t^7 + 435 t^8 + 29 t^9 + t^{10} \end{array}\right)}
  {\begin{array}{c} (1-t)^{22} (1+t)^{-1} \\  \end{array}}$\\
 \hline
$ [0 1 0 0] $
&
 {Rigid} 
&
\scriptsize ${g_{NON}^{{F_4}\left[ {0 1 0 0} \right]}{}}$
&
$ 28 $
&
\scriptsize $ \frac{\begin{array}{l} \left(1 + 9 t + 19 t^2 + 9 t^3 + t^4\right)\\
\times \left(\begin{array}{c}1 + 13 t + 118 t^2 + 455 t^3 + 716 t^4\\ + 455 t^5 + 118 t^6 + 13 t^7 + t^8 \end{array} \right)\end{array}}
{\begin{array}{c} (1-t)^{28} (1+t)^{-2} \\  \end{array}}$ \\
\hline
$ [2 0 0 0] $
&
 {Even} 
 &
\scriptsize  ${g_{NON}^{{F_4}\left[ {2 0 0 0} \right]}{}}$
&
$ 30 $
&
\scriptsize $\frac{ \left(\begin{array}{c} 1 + 21 t + 231 t^2 + 1498 t^3 + 6219 t^4 + 16834 t^5\\
 + 30420 t^6 +  36972 t^7 + \pal
 + t^{14} \end{array}\right)}
 {\begin{array}{c}  (1-t)^{30} (1+t)^{-1} \\  \end{array}} $\\
\hline

\end{tabular}
\text{}
\normalsize
\caption{$F_4$ Orbit Constructions and Hilbert Series (A)}
\label{tab:F4NO1}
\end{sidewaystable}

\begin{sidewaystable}[htp]
\small
\begin{tabular}{|c|c|c|c|c|}
\hline
 {Characteristic} & {Type} & {Construction} & {Dim.} & Unrefined HS \dummy \\
 \hline
 $ [1 0 1 0] $
&
$\begin{array}{c}
 \text{Non-rigid}\\
 \text{Non-normal}
\end{array} $
&
 \scriptsize $  \begin{array}{c} 
 g_{NON}^{{F_4}\left[ {1 0 1 0} \right]}{}\\
 \\
  - g_{NON}^{{F_4}\left[ {0 1 0 1} \right]}{}\left[ {{x_3}{x_4}{t^6}} \right] \\
  \\
 - g_{NON}^{{F_4}\left[ {2 0 0 1} \right]}{}\left[ {{x_1}{t^3} + {x_4}{t^2}} \right] \\
  \end{array}$
&
$ 38 $
&
\scriptsize $\frac{ \left(   \begin{array}{c}   1 + 13 t + 91 t^2 + 455 t^3 + 1794 t^4 + 5824 t^5 \\
+ 14859 t^6 +  25830 t^7 + 36686 t^8 + 103415 t^9 + 274079 t^{10}\\
 + 418078 t^{11} +  194377 t^{12} - 202566 t^{13} - 231712 t^{14} \\
 - 8813 t^{15} + 52245 t^{16} + 16746 t^{17} - 455 t^{18} \\
 - 92 t^{19} + 170 t^{20} + 40 t^{21} - t^{22}  \end{array}   \right)}
  { \begin{array}{c}  (1-t)^{38}(1+t)^{-1} \\  \end{array}}$ \\
\hline
$ [0 2 0 0] $
&
 {Distinguished} 
&
 \scriptsize ${g_{NON}^{{F_4}\left[ {0 2 0 0} \right]}{}}$
&
$ 40 $
&
\scriptsize $\frac{\left(   \begin{array}{c} 1 + 10 t + 56 t^2 + 230 t^3 + 745 t^4 + 1946 t^5 + 4112 t^6\\
 +  7028 t^7 + 9692 t^8 + 10782 t^9 + \pal
 + t^{18}   \end{array}    \right)}
{\begin{array}{c} (1-t)^{40}(1+t)^{-2} \\  \end{array} }$ \\
\hline
$ [2 2 0 0] $
&
 {Even} 
&
 \scriptsize ${g_{NON}^{{F_4}\left[ {2 2 0 0} \right]}{}}$
&
$ 42 $
&
\scriptsize $\frac{ \left(       \begin{array}{c}       1 + 7 t + 29 t^2 + 91 t^3 + 213 t^4 + 397 t^5\\
 + 591 t^6 + 657 t^7 + \pal 
  + t^{14}      \end{array}     \right)}
  {  \begin{array}{c}  (1-t)^{43}(1+t)^{-2} (1-t^6)^{-1} \\   \end{array} }$ \\
\hline
\end{tabular}
\text{} 
\normalsize
\caption{$F_4$ Orbit Constructions and Hilbert Series (B)}
\label{tab:F4NO2}
\end{sidewaystable}


\begin{sidewaystable}[htp]
\begin{center}
\small
\begin{tabular}{|c|c|c|c|c|}
\hline
 {Characteristic} & {Character HWG} & {$mHL$ HWG}  \dummy \\
 \hline
$ [0 0 0 0] $& $1$ & $\ldots$ \\
\hline
$ [1 0 0 0]$ &$\frac{1}{1-{m_1} t} $&$\ldots$ \\ 
\hline
 $[0 0 0 1] $&$\frac{1}{(1-{m_1} t) (1-{m_4}^2 t^2)} $&$\ldots$ \\ 
 \hline
$ [0 1 0 0] $
&
$\frac{1}{(1-{m_1} t)(1-{m_4}^2 t^2) (1-{m_2} t^3) (1-{m_3}^2 t^4) }$&$\ldots$ \\
\hline
$ [2 0 0 0] $
&
$\frac{1+{m_1} {m_2} t^5}
{(1-{m_1} t)  (1-{m_4}^2 t^2) (1-{m_2} t^3) (1-{m_1}^2 t^4) (1-{m_3}^2 t^4)(1-{m_2}^2 t^6)}$
& \ns  \\
\hline
$ [0 0 0 2] $
&
\footnotesize
$\frac{\left(\begin{array}{c} 1 - {m_4} t^2 - {m_3} t^3 + {m_4}^2 t^4 + 2 {m_3} {m_4} t^5 + {m_3}^2 t^6 \\
+  {m_3} {m_4}^2 t^6 + {m_3}^2 {m_4} t^7 - {m_3} {m_4}^2 t^7 - {m_3}^2 {m_4} t^8 - {m_3}^2 {m_4}^2 t^9 \end{array} \right)}
{\begin{array}{c} (1-{m_1} t)  (1-{m_4} t^2) (1-{m_4}^2 t^2) (1-{m_2} t^3) (1-{m_3} t^3) (1-{m_3}^2 t^4)\\ \end{array}}$
& \ns \\
\hline
$ [0 0 1 0] $
& \ns
&
\footnotesize
 $\begin{array}{c}
 
 1 - {h_4} t^4 - {h_4}^2 t^4 - {h_1} t^5 + {h_3} {h_4} t^6 + {h_2} t^7 + {h_3} t^7 + \\
 {h_1} {h_4} t^7 + {h_3} {h_4} t^7 - {h_4}^3 t^7 + {h_3} t^8 - {h_1} {h_3} t^8 - {h_4} t^8 - \\
 {h_2} {h_4} t^8 - {h_1} {h_4}^2 t^8 + {h_3} t^9 - 2 {h_1} {h_3} t^9 + {h_3}^2 t^9 + \\
 {h_1}^2 {h_4} t^9 - {h_2} {h_4} t^9 - {h_3} {h_4} t^9 - {h_4}^2 t^9 + {h_3} {h_4}^2 t^9 -\\ 
 {h_1} {h_3} t^{10} + {h_3}^2 t^{10} - {h_1} {h_4} t^{10} + {h_1}^2 {h_4} t^{10} + {h_2} {h_4} t^{10} +\\ 
 {h_1} {h_4}^2 t^{10} + {h_3} {h_4}^2 t^{10} - {h_4}^4 t^{10} + {h_3} t^{11} - {h_1} {h_4} t^{11} + \\
 {h_1}^2 {h_4} t^{11} + {h_2} {h_4} t^{11} - {h_1} {h_3} {h_4} t^{11} + {h_2} t^{12} - {h_1} {h_2} t^{12} +\\ 
 {h_2} {h_4} t^{12} - 2 {h_1} {h_3} {h_4} t^{12} - {h_1} {h_2} t^{13} + {h_2} {h_3} t^{13} - \\
 2 {h_1} {h_3} {h_4} t^{13} + {h_2} {h_4}^2 t^{13} + {h_1} {h_4}^3 t^{13} + {h_2} {h_3} t^{14} -\\ 
 {h_3}^2 {h_4} t^{14}
 \end{array} $ \\
\hline
\end{tabular}
\end{center}
\small \text{An $mHL$ HWG of 1 denotes $mHL^{F_4}_{[0000]}(t)$.}\\
\small  \text{The character or $mHL$ HWGs for some quivers are omitted for brevity.}\\
\normalsize
\caption{$F_4$ Orbits and HWGs (A)}
\label{tab:F4NO4}
\end{sidewaystable}


\begin{sidewaystable}[htp]
\begin{center}
\small
\begin{tabular}{|c|c|c|c|c|}
\hline
 {Characteristic} & {Character HWG} & {$mHL$ HWG}  \dummy \\
 \hline
 $ [1 0 1 0] $
&
$\ldots$
&
\footnotesize
$
\begin{array}{c}
1 - {h_4} t^4 - {h_1} t^5 - {h_1}^2 t^6 + {h_3} t^7 + {h_3} t^8 + {h_1} {h_3} t^8 + \\
 {h_3}^2 t^8 - {h_4} t^8 + {h_3} t^9 - {h_4}^2 t^9 - {h_3} {h_4}^2 t^9 + {h_1} {h_3} t^{10} - \\
 {h_2} {h_3} t^{10} - {h_1} {h_4} t^{10} - {h_2} {h_4} t^{10} - {h_1} {h_3} {h_4} t^{10} + {h_3} t^{11} + \\
 {h_1}^2 {h_3} t^{11} + {h_3}^2 t^{11} - {h_1} {h_4} t^{11} - {h_2} {h_4} t^{11} - {h_1} {h_3} {h_4} t^{11} + \\
 {h_3}^2 {h_4} t^{11} + {h_2} {h_4}^2 t^{11} + {h_1} {h_4}^3 t^{11} + {h_2} t^{12} - {h_2}^2 t^{12} + \\
 {h_1}^2 {h_3} t^{12} + {h_2} {h_3} t^{12} - {h_1}^2 {h_4}^2 t^{12} + {h_2} {h_4}^2 t^{12} + \\
 {h_1} {h_4}^3 t^{12} - {h_3} {h_4}^3 t^{12} + {h_1}^2 {h_2} t^{13} + {h_2} {h_3} t^{13} - \\
 {h_1} {h_3}^2 t^{13} - {h_1} {h_2} {h_4} t^{13} + {h_2} {h_3} {h_4} t^{13} - {h_3}^2 {h_4} t^{13} - \\
 {h_1}^2 {h_4}^2 t^{13} - {h_2} {h_4}^2 t^{13} + {h_2}^2 t^{14} - {h_3}^3 t^{14} - \\
 {h_1} {h_2} {h_4} t^{14} - {h_1}^2 {h_3} {h_4} t^{14} + {h_2} {h_3} {h_4} t^{14} - {h_2} {h_4}^2 t^{14} + \\
 {h_1} {h_3} {h_4}^2 t^{14} - {h_2} {h_4}^3 t^{14} + {h_1} {h_3}^2 t^{15} - {h_1}^2 {h_3} {h_4} t^{15} + \\
 {h_1} {h_3} {h_4}^2 t^{15} + {h_3}^2 {h_4}^2 t^{15} + {h_1}^2 {h_4}^3 t^{15} + {h_1} {h_2} {h_3} t^{16} - \\
 {h_2} {h_3} {h_4} t^{16} - {h_1} {h_2} {h_4}^2 t^{17}\\
 \end{array}
$ \\
\hline
$ [0 2 0 0] $
&
$\ldots$
&
$
\begin{array}{c}
1 - {h_4} t^4 - {h_1} t^5 + {h_3} t^7 + {h_3} t^8 - {h_4} t^8 + {h_3} t^9 - {h_4}^2 t^9 - \\
 {h_1} {h_4} t^{10} + {h_3} t^{11} - {h_1} {h_4} t^{11} + {h_2} t^{12}\\
 \end{array}
$ \\
\hline
$ [2 2 0 0] $
&
$\ldots$
&
$\begin{array}{c}
1 - {h_4} t^4 + {h_3} t^8 - {h_4} t^8 - {h_4}^2 t^9 + {h_3} t^{11}\\
 \end{array}$ \\
\hline
\end{tabular}
\end{center}
\small \text{An $mHL$ HWG of 1 denotes $mHL^{F_4}_{[0000]}(t)$.}\\
\normalsize
\caption{$F_4$ Orbits and HWGs (B)}
\label{tab:F4NO5}
\end{sidewaystable}


\begin{sidewaystable}[htp]
\begin{center}
\small
\begin{tabular}{|c|c|c|c|c|}
\hline
 {Characteristic} & {Type} & {Construction} & {Dim.} & Unrefined HS \dummy \\
 \hline
$ [0 0 0 2] $
&
$\begin{array}{c}
 \text{Even}\\
 \text{Normalisation}
\end{array} $
&
 ${g_{NON}^{{F_4}\left[ {0 0 0 2} \right]}{}}$
&
$ 30 $
&
\scriptsize $\frac{\left(\begin{array}{c}1 + 21 t + 257 t^2 + 2018 t^3 + 9573 t^4 + 28261 t^5\\
 + 53781 t^6 + 66651 t^7 + \pal
 + t^{14}\end{array}\right)}
 {\begin{array}{c} (1-t)^{30} (1+t)^{-1}\\ \end{array} }$ \\
\hline
$ [2 0 0 1] $
&
$\begin{array}{c}
 \text{Non-rigid}\\
 \text{Normalisation}
\end{array} $
&
 $g_{NON}^{{F_4}\left[ {2 0 0 1} \right]}{}$
&
$ 36 $
&
\scriptsize$\frac{ \left(  \begin{array}{c} 1 + 14 t + 106 t^2 + 626 t^3 + 2811 t^4 + 9363 t^5\\
 + 21662 t^6 + 35663 t^7 + 41812 t^8 + \pal
  + t^{16} \end{array}   \right)}
 { \begin{array}{c} (1-t)^{36}(1+t)^{-2}\\  \end{array}}$ \\
\hline
$ [0 1 0 1] $
&
$\begin{array}{c}
 \text{Rigid}\\
 \text{Normalisation}
\end{array} $
&
 $g_{NON}^{{F_4}\left[ {0 1 0 1} \right]}{}$
&
$ 36 $
&
\scriptsize$\frac{ \left( \begin{array}{c}   1 + 14 t + 132 t^2 + 912 t^3 + 4528 t^4 + 15655 t^5 \\
+ 37940 t^6 +  64575 t^7 + 77161 t^8 + \pal
 + t^{16}    \end{array}  \right)}
{\begin{array}{c} (1-t)^{36}(1+t)^{-2}\\  \end{array}}$ \\
\hline
$ [1 0 1 0] $
&
$\begin{array}{c}
 \text{Non-rigid}\\
 \text{Normalisation}
\end{array} $
&
 $g_{NON}^{{F_4}\left[ {1 0 1 0} \right]}{}$
&
$ 38 $
&
\scriptsize$\frac{ \left(   \begin{array}{c}   1 + 13 t + 117 t^2 + 819 t^3 + 4121 t^4 + 15171 t^5\\
 + 41431 t^6 +  84642 t^7 + 129597 t^8 + 149240 t^9 + \pal
 + t^{18}    \end{array}    \right)}
 { \begin{array}{c} (1-t)^{38} (1+t)^{-1} \\  \end{array}  }$ \\
\hline
\end{tabular}
\end{center}
\small \text{In addition to the normalisations of non-normal nilpotent orbits, the extension $F_4[2 0 0 2]$ is shown (see text).}\\
 \normalsize
\caption{$F_4$ Nilpotent Orbit Normalisations and Hilbert Series}
\label{tab:F4NO3}
\end{sidewaystable}

\begin{sidewaystable}[htp]
\begin{center}
\small
\begin{tabular}{|c|c|c|c|c|}
\hline
  {Characteristic} & {Character HWG} & {$mHL$ HWG}  \dummy \\
 \hline
$[0 0 0 2] $
&
 \scriptsize $\frac{\begin{array}{c} 1+{m_3} {m_4} t^4 \\  \end{array}}
 {\left(  \begin{array}{c} (1-{m_1} t) (1-{m_4} t^2) (1-{m_4}^2 t^2)\\
 \times (1-{m_2} t^3) (1-{m_3} t^3) (1-{m_3}^2 t^4) \end{array} \right) }$
& \ns \\
\hline
$ [2 0 0 1] $
&
$\ldots$
&
 \scriptsize $\begin{array}{c} 
 
 1 + {h_1} t^3 - {h_4} t^4 - {h_4}^2 t^4 - {h_1} t^5 - {h_3} t^5 + {h_1} {h_3} t^7 + \\
 {h_1} {h_4} t^7 + {h_3} {h_4} t^7 + {h_3} t^8 - {h_4} t^8 + {h_1} {h_4} t^8 - {h_2} {h_4} t^8 + \\
 {h_3} {h_4} t^8 - {h_1} {h_4}^2 t^8 + {h_3} t^9 - {h_1} {h_3} t^9 + {h_1} {h_4} t^9 - \\
 {h_2} {h_4} t^9 - {h_4}^2 t^9 + {h_3} {h_4}^2 t^9 - {h_4}^3 t^9 - {h_2} t^{10} - \\
 {h_1} {h_3} t^{10} + {h_3}^2 t^{10} - {h_1} {h_4} t^{10} - {h_2} {h_4} t^{10} + {h_3} {h_4}^2 t^{10} - \\
 {h_4}^4 t^{10} + {h_1} {h_2} t^{11} + {h_3} t^{11} - {h_1} {h_3} t^{11} + {h_3}^2 t^{11} - \\
 {h_1} {h_4} t^{11} + {h_1}^2 {h_4} t^{11} + {h_2} t^{12} + {h_2} {h_4} t^{12} - {h_1} {h_3} {h_4} t^{12} + \\
 {h_1} {h_4}^2 t^{12} - {h_1} {h_2} t^{13} - {h_1} {h_3} {h_4} t^{13} + {h_1} {h_4}^3 t^{13}\\
 
 \end{array}$ \\
\hline
\end{tabular}
\end{center}
\text{In addition to the normalisations of non-normal nilpotent orbits, the extension $F_4[2 0 0 2]$ is shown (see text).}\\
\normalsize
\caption{$F_4$ Nilpotent Orbit Normalisations and HWGs}
\label{tab:F4NO6}
\end{sidewaystable}

\FloatBarrier

\subsubsection{Orbits of $E_6$}
\label {subsec:E6NO}

The 21 nilpotent orbits of $E_6$ are described in tables \ref{tab:E6NO1} to \ref{tab:E6NO3}, which give their classification, constructions and unrefined HS. Table \ref{tab:E6NO4} contains the same information for the normalisations of the non-normal nilpotent orbits. Table \ref{tab:E6NO5} analyses the three extra root maps that were identified in Appendix \ref{apxNO}.

Unlike $F_4$, it has not proved practicable to resolve many Hilbert series into HWGs, other than for near minimal and maximal orbits, so much of the analysis is based upon unrefined HS. In table \ref{tab:E6NO6}, the character HWGs and $mHL$ HWGs are given for those orbits where it has been possible to complete the calculations.

The normal and non-normal orbits exactly match those listed in \cite{Broer:1998qf} (see Appendix \ref{apxNO}). The tables contain candidates for the constructions of the non-normal orbits. These have been obtained by restricting their normalisations to the nilpotent cone $\cal N$ through the subtraction of sub-spaces, similar to the method used for $g_{NO}^{G_2}$ and $g_{NO}^{F_4}$. Much of the analysis is, however, based on unrefined Hilbert series and should not be taken as definitive. The picture that emerges can be summarised:

\begin{description}

\item $[0 0 0 0 0 1]$ and $[1 0 0 0 1 0]$: 22 dimensional minimal and 32 dimensional next to minimal nilpotent orbits. These orbits have the invariant subgroups $A_5$ and $D_4$ respectively. The orbits can be calculated either (i) from the Coulomb branch of a quiver theory built on the affine Dynkin diagram or Characteristic, or (ii) from the $NON$ formula. The HS and character HWGs are palindromic, and the latter are freely generated.

\item $[0 0 1 0 0 0]$ and $[0 0 0 0 0 2]$: 40 and 42 dimensional nilpotent orbits. These orbits have the invariant subgroups $A_2 \otimes A_2 \otimes A_1$ and $A_5$, respectively. The orbits are calculated from the $NON$ formula. The HS and character HWGs are palindromic, and the latter are freely generated or complete intersections.

\item $[1 0 0 0 1 1]$,  $[2 0 0 0 2 0]$, $[1 0 0 0 1 2]$,  $[0 1 0 1 0 1]$ and $[2 0 0 0 2 2]$: 46, 48, 52, 56 and 60 dimensional nilpotent orbits. These orbits have the invariant subgroups $A_3$, $D_4$, $A_3$, $A_1^3$ and $A_3$ respectively. The orbits are non-normal and candidates for the orbits are found by excluding sub-spaces, as shown in the tables, from their normalisations obtained from the $NON$ formula. The Hilbert series are non-palindromic.

\item The remaining orbits are normal, with palindromic Hilbert series. The decompositions into $mHL$ functions are shown for the 66 dimensional orbit upwards.

\end{description}

\begin{figure}[htbp]
\begin{center}
\includegraphics[scale=0.7]{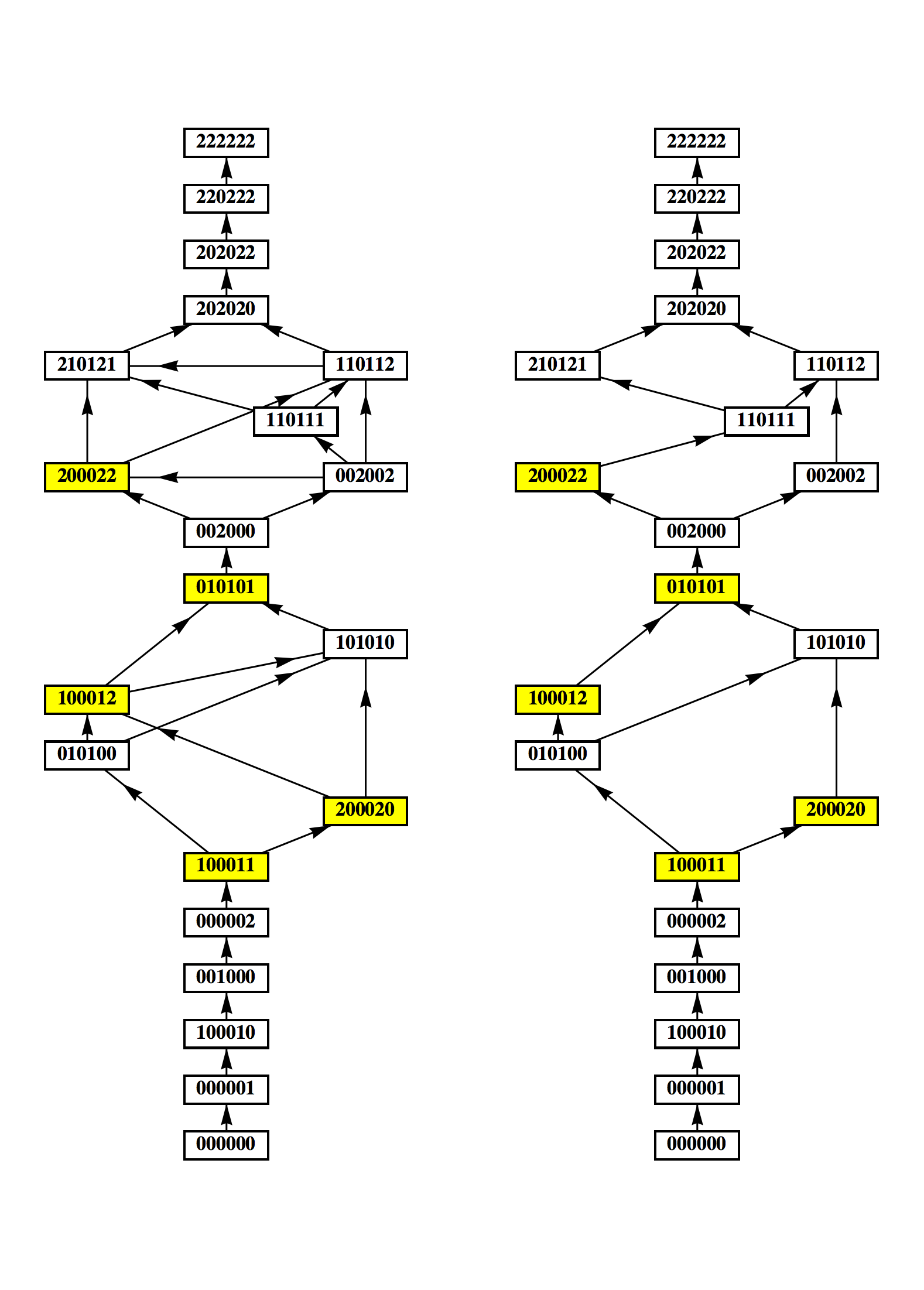}
\caption[$E_6$ Nilpotent Orbit Hasse Diagram]{$E_6$ Nilpotent Orbit Hasse Diagram. The left hand diagram is indicative, being partly derived from \emph{unrefined} Hilbert series, with arrows indicating inclusion relations and yellow nodes indicating non-normal nilpotent orbits. The right hand diagram is taken from the mathematical Literature.}
\label{fig:E6Hasse}
\end{center}
\end{figure}

The Hasse diagram based on the inclusion relationships between unrefined Hilbert series is compared in figure \ref{fig:E6Hasse} with the standard diagram in the mathematical Literature \cite{Adams:jk, Baohua-Fu:2015nr}. 

The two diagrams are broadly consistent. Some of the extra links appearing in the left hand diagram might disappear if the moduli space calculations could be repeated with refined (rather than unrefined) Hilbert series, or with character HWGs. However, the left hand diagram does not have a link (i.e. inclusion relation) between the non-normal $[200022]$ and the normal $[110111]$; considering that unrefined Hilbert series cannot miss an inclusion relation, this may indicate an anomaly in the standard diagram; alternatively, there may be other restrictions of the normalisation of $[200022]$ that should be considered.

Turning to the three extra root maps, whose unrefined HS are set out in table \ref{tab:E6NO5}: two of these maps, $[1 1 1 1 1 0]$ and $[0 2 0 2 0 2]$, have identical \emph{refined} Hilbert series to the nilpotent orbits with Characteristics $[1 1 0 1 1 1]$ and $[2 0 2 0 2 0]$, respectively; these provide examples of dualities, with different $SU(2)$ homomorphisms generating the same nilpotent orbit. The third map, $[1 1 0 1 1 0]$, is non-normal, containing elements outside the nilpotent cone; it can be restricted to the nilpotent cone, by excluding a subspace defined by the charged $NON$ formula, whereupon it appears to be an extension of $[0 0 2 0 0 0]$, the distinguished nilpotent orbit of the same dimension:

\begin{equation}
\label{eq:e6a}
g_{NON}^{{E_6}\left[ {1 1 0 1 1 0} \right]}{} = g_{NON}^{{E_6}\left[ {0 0 2 0 0 0} \right]}{}\left[ {1 + {x_6}{t^3} + {x_3}{t^6}} \right]
\end{equation}

The Weyl group of $E_6$ has 25 irreps and conjugacy classes. In \cite{beynon_spaltenstein_1984}, the 21 nilpotent orbits are identified as these conjugacy classes, modulo some actions of the symmetric groups $S_2$ or $S_3$. Two of the three extra root maps, $[1 1 0 1 1 0]$ and $[0 2 0 2 0 2]$, appear to correspond to other members of these conjugacy classes; however, these are only identified in \cite{beynon_spaltenstein_1984} by Bala Carter labels, so the correspondence with root maps or Characteristics is unclear.

\begin{sidewaystable}[htp]
\begin{center}
\small
\begin{tabular}{|c|c|c|c|c|}
\hline
 {Characteristic} & {Type} & {Construction} & {Dim.} & Unrefined HS \dummy \\
 \hline
$ [0 0 0 0 0 0] $& {Even} & 
\scriptsize ${g_{NON}^{{E_6}\left[ {0 0 0 0 0 0} \right]}{}}$ &$ 0$ &$ 1$ \\
\hline
$ [0 0 0 0 0 1]$ & {Rigid} &
\scriptsize $\begin{array}{*{20}{c}}
{g_{{{Coulomb}}}^{{E_6}\left[ {0 0 0 0 0 1} \right]}{}}\\
{{ \text{or}}}\\
{g_{NON}^{{E_6}\left[ {0 0 0 0 0 1} \right]}{}}
\end{array}$
 &
 $ 22 $
 &
 \scriptsize
 $\frac{ \left( \begin{array}{c} 1 + 55 t + 890 t^2 + 5886 t^3 + 17929 t^4 + 26060 t^5 \\
 + \pal + t^{10} \end{array} \right)}
 {\begin{array}{c} (1-t)^{22}(1+t)^{-1} \\ \end{array}}$ \\
\hline

$ [1 0 0 0 1 0]$ & {Richardson} &
\scriptsize $\begin{array}{*{20}{c}}
{g_{{{Coulomb}}}^{{E_6}\left[ {1 0 0 0 1 0} \right]}{}}\\
{{ \text{or}}}\\
{g_{NON}^{{E_6}\left[ {1 0 0 0 1 0} \right]}{}}
\end{array}$
 &
 $ 32 $
 &
 \scriptsize
 $\frac{ \left(   \begin{array}{c}   1 + 44 t + 991 t^2 + 11649 t^3 + 75681 t^4 + 285009 t^5 \\
 + 632941 t^6 + 827309 t^7 + \pal
  + t^{14}  \end{array} \right)}
 { \begin{array}{c} (1-t)^{32} (1+t)^{-2} \\ \end{array} }$ \\
\hline

$ [0 0 1 0 0 0]$ & {Rigid} &
\scriptsize $
{g_{NON}^{{E_6}\left[ {0 0 1 0 0 0} \right]}{}}
$
 &
 $ 40 $
 &
 \scriptsize
 $\frac{ \left(   \begin{array}{c}   1 + 38 t + 740 t^2 + 9192 t^3 + 76974 t^4 + 446483 t^5 \\
 + 1832104 t^6 + 5408077 t^7 + 11624874 t^8 + 18343974 t^9 + 21346158 t^{10}\\
  + \pal
 + t^{20}  \end{array} \right)}
 { \begin{array}{c} (1-t)^{40} \\ \end{array} }$ \\
\hline

$ [{0  0  0  0  0  2}]$ & {Even} &
\scriptsize $
{g_{NON}^{{E_6}\left[ {0  0  0  0  0  2} \right]}{}}
$
 &
 $ 42 $
 &
 \scriptsize
 $\frac{ \left(  \begin{array}{c}   1 + 35 t + 630 t^2 + 7120 t^3 + 54640 t^4 + 294385 t^5 \\+ 
 1139307 t^6 + 3216888 t^7 + 6702843 t^8 + 10382781 t^9 + 
 12008160 t^{10}\\ + \pal
 + t^{20}  \end{array} \right)}
 { \begin{array}{c}  (1-t)^{42}(1+t)^{-1}  \\ \end{array}}$ \\
\hline

$ [{1  0  0  0  1  1}]$ &$ \begin{array}{c} \text{Non-rigid}\\ \text{ Non-normal}  \end{array}$ &
\scriptsize $
 \begin{array}{c}
g_{NON}^{{E_6}\left[ {1 0 0 0 1 1} \right]}{} \\
 \\
- g_{NON}^{{E_6}\left[ {0 0 0 0 0 2} \right]}{}\left[ {{x_6}{t^2}} \right]
\end{array}
$
 &
 $ 46 $
 &
 \scriptsize
 $\frac{ \left(   \begin{array}{c} 1 + 31 t + 496 t^2 + 5456 t^3 + 45648 t^4 + 303442 t^5\\ + 
 1570349 t^6 + 6114940 t^7 + 17426810 t^8 + 35866167 t^9 + 
 53265869 t^{10}\\ + 58043959 t^{11} + 48411638 t^{12} + 32923009 t^{13} + 
 18740108 t^{14} + 8202867 t^{15} \\+ 2156737 t^{16} + 17233 t^{17} - 
 221603 t^{18} - 85366 t^{19} - 16244 t^{20} \\- 1658 t^{21} - 77 t^{22} \end{array}  \right)}
 { \begin{array}{c}  (1-t)^{46}  (1+t)^{-1}  \\ \end{array}}$ \\
\hline

$ [{2  0  0  0  2  0}]$ &$ \begin{array}{c} \text{Even}\\ \text{Non-normal} \end{array}$
&
\scriptsize $
 \begin{array}{c}
g_{NON}^{{E_6}\left[ {2 0 0 0 2 0} \right]}{} \\
 \\
- g_{NON}^{{E_6}\left[ {0 0 0 0 0 2} \right]}{}\left[ {{x_6}{t^2}} \right]
\end{array}
$
 &
 $ 48 $
 &
 \scriptsize
 $\frac{ \left(   \begin{array}{c}
    1 + 29 t + 435 t^2 + 4495 t^3 + 35882 t^4 + 232149 t^5\\ + 
 1171457 t^6 + 4383879 t^7 + 11766232 t^8 + 22712054 t^9 + 
 33186041 t^{10}\\ + 40726538 t^{11} + 44327421 t^{12} + 38695845 t^{13} + 
 22831110 t^{14} + 8234564 t^{15} \\+ 2786494 t^{16} + 2013171 t^{17} + 
 1068403 t^{18} + 146307 t^{19} - 109837 t^{20} \\- 60051 t^{21} - 13489 t^{22} - 
 1532 t^{23} - 78 t^{24}   \end{array}  \right)}
 { \begin{array}{c}
  (1-t)^{48} (1+t)^{-1}    \end{array}   }$ \\
\hline

$ [{0  1  0  1  0  0}]$ & {Richardson} &
\scriptsize $
{g_{NON}^{{E_6}\left[ {0  1  0  1  0  0} \right]}{}}
$
 &
 $ 50 $
 &
 \scriptsize
 $\frac{\left(   \begin{array}{c}  1 + 27 t + 378 t^2 + 3654 t^3 + 26677 t^4 + 151633 t^5\\ + 679211 t^6 + 
 2411434 t^7 + 6802614 t^8 + 15252535 t^9 + 27167919 t^{10} \\+ 
 38421369 t^{11} + 43128016 t^{12} + \pal
 + t^{24}    \end{array}    \right)}
 { \begin{array}{c} (1-t)^{50} (1+t)^{-1} \\ \end{array}}$ \\
\hline

\end{tabular}
\end{center}
\text{}
\normalsize
\caption{$E_6$ Orbit Constructions and Hilbert Series (A)}
\label{tab:E6NO1}
\end{sidewaystable}


\begin{sidewaystable}[htp]
\begin{center}
\small
\begin{tabular}{|c|c|c|c|c|}
\hline
 {Characteristic} & {Type} & {Construction} & {Dim.} & Unrefined HS \dummy \\
 \hline
 
 $ [{1  0  0  0  1  2}]$ &$ \begin{array}{c} \text{Richardson}\\ \text{ Non-normal}  \end{array} $&
\scriptsize $
 \begin{array}{c}
g_{NON}^{{E_6}\left[ {1 0 0 0 1 2} \right]}{} \\
 \\
 - g_{NON}^{{E_6}\left[ {0 0 0 0 0 2} \right]}{}\left[ {{x_6}{t^3}} \right]
  \end{array} 
$
 &
 $ 52 $
 &
 \scriptsize
 $\frac{\left(    \begin{array}{c}  1 + 25 t + 325 t^2 + 2925 t^3 + 20397 t^4 + 116805 t^5\\ + 562496 t^6 + 
 2201433 t^7 + 6660591 t^8 + 15233736 t^9 + 27936753 t^{10}\\ + 
 44893175 t^{11} + 60568689 t^{12} + 56711757 t^{13} + 35968001 t^{14} + 
 28729791 t^{15}\\ + 27096886 t^{16} + 5811534 t^{17} - 7028305 t^{18} + 
 1316117 t^{19} + 4719834 t^{20} \\- 231609 t^{21} - 1389570 t^{22} - 
 78065 t^{23} + 249547 t^{24} + 51338 t^{25} \\- 15378 t^{26} - 7854 t^{27} - 
 1221 t^{28} - 78 t^{29}   \end{array}   \right)}
 { \begin{array}{c} (1-t)^{52}(1+t)^{-1} \\  \end{array}}$ \\
\hline

$ [{1  0  1  0  1  0}]$ & {Rigid} &
\scriptsize $
{g_{NON}^{{E_6}\left[ {1  0  1  0  1  0} \right]}{}}
$
 &
 $ 54 $
 &
 \scriptsize
 $\frac{\left(  \begin{array}{c}   1 + 21 t + 233 t^2 + 1813 t^3 + 11013 t^4 + 55097 t^5 \\+ 231652 t^6 + 
 814942 t^7 + 2352615 t^8 + 5460885 t^9 + 10045178 t^{10}\\ + 
 14521163 t^{11} + 16426194 t^{12} + \pal
 + t^{24}  \end{array}  \right)}
 {\begin{array}{c} (1-t)^{54} (1+t)^{-3} \\ \end{array} }$ \\
\hline

$ [{0  1  0  1  0  1}]$ &$ \begin{array}{c} \text{Non-rigid}\\ \text{Non-normal} \end{array}$ &
\scriptsize $
 \begin{array}{c}
g_{NON}^{{E_6}\left[ {0 1 0 1 0 1} \right]}{} \\
 \\
 - g_{NON}^{{E_6}\left[ {1 0 1 0 1 0} \right]}{}\left[ {{x_6}{t^3}} \right]
 \end{array}$
 &
 $ 56 $
 &
 \scriptsize
 $\frac{ \left(   \begin{array}{c} 1 + 21 t + 231 t^2 + 1771 t^3 + 10548 t^4 + 51492 t^5\\ + 212289 t^6 + 
 752158 t^7 + 2315893 t^8 + 6248619 t^9 + 14884446 t^{10} \\+ 
 31537643 t^{11} + 58591608 t^{12} + 91163801 t^{13} + 111457472 t^{14} + 
 99817922 t^{15} \\ + 58890046 t^{16} + 16128539 t^{17} - 5514308 t^{18} - 
 7464927 t^{19} - 3286296 t^{20}\\  - 760330 t^{21} - 172433 t^{22} - 
 120068 t^{23} - 76638 t^{24} - 30650 t^{25} \\ - 8184 t^{26} - 1539 t^{27} - 
 210 t^{28} - 20 t^{29} - t^{30}   \end{array} \right)}
 { \begin{array}{c} (1-t)^{56} (1+t)^{-1} \\  \end{array} }$ \\
\hline

$ [0  0  2  0  0  0]$ & {Even} &
\scriptsize $
{g_{NON}^{{E_6}\left[0  0  2  0  0  0 \right]}{}}
$
 &
 $ 58 $
 &
 \scriptsize
 $\frac{ \left(\begin{array}{c} 
 1 + 17 t + 155 t^2 + 1003 t^3 + 5076 t^4 + 21012 t^5\\ 
 + 72753 t^6 + 212554 t^7 + 526005 t^8 + 1104427 t^9 + 1968242 t^{10}\\
  + 2976459 t^{11} + 3816658 t^{12} + 4147046 t^{13} \\ +3816658 t^{14} + \pal
 + t^{26}  \end{array}   \right)}
 {\begin{array}{c} (1-t)^{58} (1+t)^{-3}\\  \end{array}}$ \\
\hline

$ [{0  0  2  0  0  2}]$ & {Even} &
\scriptsize $
{g_{NON}^{{E_6}\left[{0  0  2  0  0  2} \right]}{}}
$
 &
 $ 60 $
 &
 \scriptsize
 $\frac{  \left(\begin{array}{c} 
 1 + 14 t + 107 t^2 + 588 t^3 + 2515 t^4 + 8716 t^5 \\+ 24956 t^6 + 
 58860 t^7 + 114672 t^8 + 184443 t^9 + 244674 t^{10} \\+ 269178 t^{11} + \pal
 + t^{22}   \end{array} \right)}
 {\begin{array}{c}  (1-t)^{64}(1 - t^2)^{-3} (1 - t^6)^{-1} \\  \end{array} }$ \\
\hline

$ [{2  0  0  0  2  2}]$ & $ \begin{array}{c} \text{Even}\\\text{Non-normal} \end{array}$ &
\scriptsize $
\begin{array}{c} 
g_{NON}^{{E_6}\left[ {2 0 0 0 2 2} \right]}{} \\\\- g_{NON}^{{E_6}\left[ {0 0 2 0 0 0} \right]}{}\left[ {{x_6}{t^3}} \right]  \end{array} $
 &
 $ 60 $
 &
 \scriptsize
 $\frac{ \left( \begin{array}{c} 1 + 16 t + 137 t^2 + 832 t^3 + 4013 t^4 + 16257 t^5\\ + 57091 t^6 + 
 180151 t^7 + 521432 t^8 + 1375550 t^9 + 3229355 t^{10}\\ + 
 6552743 t^{11} + 11221499 t^{12} + 15931890 t^{13} + 18517801 t^{14} + 
 17407937 t^{15} \\+ 13045728 t^{16} + 7612781 t^{17} + 3291500 t^{18} + 
 911351 t^{19} + 42335 t^{20}\\ - 105546 t^{21} - 61493 t^{22} - 19309 t^{23} - 
 3591 t^{24} - 259 t^{25}\\ + 59 t^{26} + 16 t^{27} + t^{28}  \end{array}   \right)}
 { \begin{array}{c}  (1-t)^{60} (1+t)^{-2} \\  \end{array}}$ \\
\hline

\end{tabular}
\end{center}
\text{}
\normalsize
\caption{$E_6$ Orbit Constructions and Hilbert Series (B)}
\label{tab:E6NO2}
\end{sidewaystable}

\begin{sidewaystable}[htp]
\begin{center}
\small
\begin{tabular}{|c|c|c|c|c|}
\hline
 {Characteristic} & {Type} & {Construction} & {Dim.} & Unrefined HS \dummy \\
 \hline
 $ [{1  1  0  1  1  1}]$ & {Richardson} &
\scriptsize $
{g_{NON}^{{E_6}\left[{1  1  0  1  1  1} \right]}{}}
$
 &
 $ 62 $
 &
 \scriptsize
 $\frac{ \left(\begin{array}{c}   1 + 12 t + 80 t^2 + 389 t^3 + 1536 t^4 + 5133 t^5 \\+ 14863 t^6 + 
 37773 t^7 + 84597 t^8 + 166302 t^9 + 284667 t^{10} \\ + 421063 t^{11} + 
 534371 t^{12} + 579012 t^{13} + \pal
 + t^{26}   \end{array}  \right)}
 {\begin{array}{c}  (1-t)^{66} (1 - t^2)^{-3} (1 - t^3)^{-1} \\  \end{array}}$ \\
\hline

$ [{2  1  0  1  2  1}]$ & {Non-rigid} &
\scriptsize $
{g_{NON}^{{E_6}\left[{2  1  0  1  2  1} \right]}{}}
$
 &
 $ 64 $
 &
 \scriptsize
 $\frac{ \left(\begin{array}{c} 1 + 10 t + 56 t^2 + 232 t^3 + 791 t^4 + 2343 t^5\\ + 6228 t^6 + 
 15100 t^7 + 33650 t^8 + 69224 t^9 + 129347 t^{10} \\+ 213929 t^{11} + 
 298121 t^{12} + 335808 t^{13} + \pal
 + t^{26}   \end{array}  \right)}
 {\begin{array}{c} (1-t)^{68}(1 - t^2)^{-2} (1 - t^3)^{-2} \\ \end{array}}$ \\
\hline

$ [{1  1  0  1  1  2}]$ & {Richardson} &
\scriptsize $
{g_{NON}^{{E_6}\left[{1  1  0  1  1  2} \right]}{}}
$
 &
 $ 64 $
 &
 \scriptsize
 $\frac{\left( \begin{array}{c} 1 + 11 t + 67 t^2 + 298 t^3 + 1079 t^4 + 3366 t^5\\ + 9362 t^6 + 
 23671 t^7 + 54328 t^8 + 112202 t^9 + 205531 t^{10} \\+ 330265 t^{11} + 
 463957 t^{12} + 568853 t^{13} + 608454 t^{14} +
  568853 t^{15}\\ + \pal
 + t^{28}    \end{array}  \right)}
 { \begin{array}{c} (1-t)^{67} (1 - t^2)^{-2} (1 - t^3)^{-1}\\  \end{array} }$ \\
\hline

$ [{2  0  2  0  2  0}]$ & {Distinguished} &
\scriptsize $
{g_{NON}^{{E_6}\left[{2  0  2  0  2  0} \right]}{}}
$
 &
 $ 66 $
 &
 \scriptsize
 $\frac{ \left(   \begin{array}{c}  1 + 7 t + 30 t^2 + 100 t^3 + 283 t^4 + 710 t^5\\ + 1623 t^6 + 
 3364 t^7 + 6314 t^8 + 10710 t^9 + 16269 t^{10} \\+ 22197 t^{11} + 
 26940 t^{12} + 28824 t^{13} +\pal
  + t^{26}  \end{array}  \right)}
 {  \begin{array}{c} (1-t)^{71} (1 - t^2)^{-3} (1 - t^3)^{-2}\\ \end{array}}$ \\
\hline

$ [{2  0  2  0  2  2}]$ & {Even} &
\scriptsize $
{g_{NON}^{{E_6}\left[{2  0  2  0  2  2} \right]}{}}
$
 &
 $ 68 $
 &
 \scriptsize
 $\frac{ \left(     \begin{array}{c}   1 + 6 t + 22 t^2 + 62 t^3 + 149 t^4 + 319 t^5 \\+ 626 t^6 + 1146 t^7 + 
 1905 t^8 + 2883 t^9 + 3941 t^{10}\\ + 4824 t^{11} + 5087 t^{12} + \pal 
 + t^{24}    \end{array}  \right)}
 {  \begin{array}{c} (1-t)^{72} (1 - t^2)^{-2} (1 - t^4)^{-1} (1 - t^6)^{-1}  \\ \end{array}}$ \\
\hline

$ [{2  2  0  2  2  2}]$ & {Distinguished} &
\scriptsize $
{g_{NON}^{{E_6}\left[{2  2  0  2  2  2}\right]}{}}
$
 &
 $ 70 $
 &
 \scriptsize
 $\frac{ \left(  \begin{array}{c} 
 1 + 3 t + 6 t^2 + 11 t^3 + 19 t^4 + 30 t^5 + 45 t^6 + 65 t^7 \\+ 
 90 t^8 + 120 t^9 + 156 t^{10} + \pal
 + t^{20}   \end{array} \right)}
 { \begin{array}{c} (1-t)^{75} (1 - t^2) ^{-1}(1 - t^3)^{-1} (1 - t^4)^{-1} (1 - t^5)^{-1} (1 - t^6)^{-1} \\  \end{array}}$ \\
\hline

$ [{2  2  2  2  2  2}]$ & {Distinguished} &
\scriptsize $
{g_{NON}^{{E_6}\left[{2  2  2  2  2  2}\right]}{}}
$
 &
 $ 72 $
 &
 \scriptsize
 $\frac{ \begin{array}{c} (1-t^2) (1-t^5) (1-t^6) (1-t^8) (1-t^9) (1-t^{12}) \\  \end{array}}
 {\begin{array}{c} (1-t)^{78} \\  \end{array}}$ \\
\hline

\end{tabular}
\end{center}
\normalsize
\caption{$E_6$ Orbit Constructions and Hilbert Series (C)}
\label{tab:E6NO3}
\end{sidewaystable}

\begin{sidewaystable}[htp]
\begin{center}
\small
\begin{tabular}{|c|c|c|c|c|}
\hline
 {Characteristic} & {Type} & {Construction} & {Dim.} & Unrefined HS \dummy \\
 \hline
 $ [{1  0  0  0  1  1}]$ &$ \begin{array}{c} \text{Non-rigid}\\ \text{Normalisation}  \end{array} $
&
\scriptsize $
{g_{NON}^{{E_6}\left[ {1  0  0  0  1  1} \right]}{}}
$
 &
 $ 46 $
 &
 \scriptsize
 $\frac{ \left(   \begin{array}{c}   
 1 + 31 t + 574 t^2 + 7145 t^3 + 62466 t^4 + 395953 t^5\\ 
 +  1854418 t^6 + 6493660 t^7 + 17124491 t^8 + 34156960 t^9 \\
 +  51650252 t^{10} + 59277910 t^{11} + \pal 
 + t^{22}   \end{array}  \right)}
 { \begin{array}{c} (1-t)^{46} (1+t)^{-1}  \\ \end{array}}$ \\
\hline

$ [{2  0  0  0  2  0}]$ &$ \begin{array}{c} \text{Even}\\ \text{Normalisation}  \end{array} $&
\scriptsize $
{g_{NON}^{{E_6}\left[ {2  0  0  0  2  0} \right]}{}}
$
 &
 $ 48 $
 &
 \scriptsize
 $\frac{ \left(  \begin{array}{c}    
 1 + 27 t + 457 t^2 + 5059 t^3 + 38341 t^4 + 205456 t^5\\ 
 + 794669 t^6 + 2248381 t^7 + 4698986 t^8 + 7296802 t^9 \\ + 8446562 t^{10} +  \pal
 + t^{20}    \end{array}  \right)}
 { \begin{array}{c}  (1-t)^{51} (1 - t^2)^{-2} (1 - t^3)^{-1}   \\ \end{array}}$ \\
\hline

$ [{1  0  0  0  1  2}]$ &$ \begin{array}{c} \text{Richardson}\\ \text{Normalisation}  \end{array} $&
\scriptsize $
{g_{NON}^{{E_6}\left[ {1  0  0  0  1  2} \right]}{}}
$
 &
 $ 52 $
 &
 \scriptsize
 $\frac{ \left(  \begin{array}{c} 
 1 + 24 t + 301 t^2 + 2702 t^3 + 18916 t^4 + 105743 t^5+ 472131 t^6 \\
 + 1677965 t^7 + 4733104 t^8 + 10579022 t^9 + 18750304 t^{10} \\
 + 26396098 t^{11} + 29577416 t^{12} + \pal
  + t^{24}   \end{array}  \right)}
 { \begin{array}{c}  (1-t)^{52} (1+t)^{-2}  \\  \end{array}}$ \\
\hline

$ [{0  1  0  1  0  1}]$ &$ \begin{array}{c} \text{Non-rigid}\\ \text{Normalisation}  \end{array} $&
\scriptsize $
{g_{NON}^{{E_6}\left[{0  1  0  1  0  1} \right]}{}}
$
 &
 $ 56 $
 &
 \scriptsize
 $\frac{\left(\begin{array}{c} 
 1 + 20 t + 211 t^2 + 1638 t^3 + 10469 t^4 + 56733 t^5 + 260036 t^6\\
  + 993325 t^7 + 3125563 t^8 + 8036658 t^9 + 16802409 t^{10} \\
 + 28491536 t^{11} + 39129101 t^{12} + 43499048 t^{13}\\ + \pal
 + t^{26}  \end{array}   \right)}
 {\begin{array}{c} (1-t)^{56} (1+t)^{-2} \\ \end{array} }$ \\
\hline

$ [{2  0  0  0  2  2}]$ &$ \begin{array}{c} \text{Even}\\ \text{Normalisation}  \end{array} $&
\scriptsize $
{g_{NON}^{{E_6}\left[{2  0  0  0  2  2} \right]}{}}
$
 &
 $ 60 $
 &
 \scriptsize
 $\frac{\left(  \begin{array}{c}   
 1 + 16 t + 136 t^2 + 894 t^3 + 5046 t^4 + 24136 t^5 \\
 + 96384 t^6 + 318938 t^7 + 873668 t^8 + 1984329 t^9 + 3747603 t^{10}\\ + 
 5898185 t^{11} + 7743629 t^{12} + 8479209 t^{13} + 7743629 t^{14} \\+ \pal 
 + t^{26}  \end{array}   \right)}
 {\begin{array}{c}   (1-t)^{60} (1+t)^{-2} (1+t^2)^{-1} \\ \end{array} }$ \\
\hline
 
\end{tabular}
\end{center}
\text{}\\
\normalsize
\caption{$E_6$ Nilpotent Orbit Normalisations and Hilbert Series}
\label{tab:E6NO4}
\end{sidewaystable}

\begin{sidewaystable}[htp]
\begin{center}
\small
\begin{tabular}{|c|c|c|c|c|}
\hline
 {Characteristic} & {Type} & {Construction} & {Dim.} & Unrefined HS \dummy \\
 \hline

$ [{1  1  0  1  1  0}]$ &$ \begin{array}{c} \text{}\\ \text{Cover}  \end{array} $&
\scriptsize $
{g_{NON}^{{E_6}\left[{1  1  0  1  1  0} \right]}{}}
$
 &
 $ 58 $
 &
 \scriptsize
 $\frac{ \left( \begin{array}{c}
 1 + 17 t + 155 t^2 + 1159 t^3 + 7570 t^4 + 41208 t^5 \\
 + 186890 t^6 + 699308 t^7 + 2146085 t^8 + 5383228 t^9 + 11015451 t^{10}\\ + 
 18374964 t^{11} + 24976338 t^{12} + 27669872 t^{13} + 24976338 t^{14}\\ + \pal
 + t^{26}   \end{array}   \right)}
 {\begin{array}{c} (1-t)^{58} (1+t)^{-3} \\ \end{array} }$ \\
\hline

$ [{1  1  1  1  1  0}]$ &$ \begin{array}{c} \text{Richardson}\\ \text{Dual}  \end{array} $  &
\scriptsize $
{g_{NON}^{{E_6}\left[{1  1  1  1  1  0} \right]}{}}
$
 &
 $ 62 $
 &
 \scriptsize
 $\frac{ \left(\begin{array}{c}   
 1 + 12 t + 80 t^2 + 389 t^3 + 1536 t^4 + 5133 t^5 \\
 + 14863 t^6 + 37773 t^7 + 84597 t^8 + 166302 t^9 + 284667 t^{10} \\ 
 + 421063 t^{11} + 534371 t^{12} + 579012 t^{13} + \pal
 + t^{26}   \end{array}  \right)}
 {\begin{array}{c}  (1-t)^{66} (1 - t^2)^{-3} (1 - t^3)^{-1} \\  \end{array}}$ \\
\hline

$ [{0  2  0  2  0  2}]$ &$ \begin{array}{c} \text{Distinguished}\\ \text{Dual}  \end{array} $ &
\scriptsize $
{g_{NON}^{{E_6}\left[{0  2  0  2  0  2} \right]}{}}
$
 &
 $ 66 $
 &
 \scriptsize
 $\frac{ \left(   \begin{array}{c}  1 + 7 t + 30 t^2 + 100 t^3 + 283 t^4 + 710 t^5\\ + 1623 t^6 + 
 3364 t^7 + 6314 t^8 + 10710 t^9 + 16269 t^{10} \\+ 22197 t^{11} + 
 26940 t^{12} + 28824 t^{13} + \pal
 + t^{26}  \end{array}  \right)}
 {  \begin{array}{c} (1-t)^{71} (1 - t^2)^{-3} (1 - t^3)^{-2} \\ \end{array}}$ \\
\hline
 
\end{tabular}
\end{center}
\small \text{These moduli spaces are associated with SU(2) homomorphisms but do not represent additional nilpotent orbits (see text).}\\
\normalsize
\caption{$E_6$ Extra Moduli Spaces and Hilbert Series}
\label{tab:E6NO5}
\end{sidewaystable}

\begin{sidewaystable}[htp]
\begin{center}
\begin{tabular}{|c|c|c|}
 \hline
 \text{Characteristic} & \text{Character HWG} & \text{$mHL$ HWG} \dummy \\
 \hline
$ [0  0  0  0  0  0]  $&$ 1  $&$\ldots  $\\
  \hline
 $[0  0  0  0  0  1]  $&$ \frac{1}{1- {m_6} t}  $&$\ldots  $\\
  \hline
 $[1  0  0  0  1  0]$&$ \frac{1}{(1-{m_6} t) (1-{m_1} {m_5} t^2)}  $&$\ldots  $\\
  \hline
$ [0  0  1  0  0  0]  $&$\frac{1}{(1-{m_6} t)  (1-{m_1} {m_5} t^2)  (1-{m_3} t^3) (1-{m_2} {m_4} t^4)} $&$\ldots  $\\
  \hline
$ [0  0  0  0  0  2]  $& $\frac{1 + {m_3} {m_6} t^5}{(1-{m_6} t)  (1-{m_1} {m_5} t^2) (1-{m_3} t^3)(1-{m_6}^2 t^4) (1-{m_2} {m_4} t^4)  (1-{m_3}^2 t^6) } $&$\ldots  $ \\
  \hline
  $\ldots  $& $\ldots  $&$\ldots  $ \\
  \hline
       $ [2 0 2 0 2 0]  $& $\ldots  $&$\begin{array}{c} 1 - h_6 t^7 - h_6 t^8 - h_6 t^{11} +\\
        h_1 h_5 t^{11} + h_1 h_5 t^{12} + h_1 h_5 t^{13} - h_3 t^{16} \end{array}$ \\
  \hline
       $ [2 0 2 0 2 2]  $& $\ldots  $&$1 - h_6 t^8 - h_6 t^{11} + h_1 h_5 t^{13}$ \\
  \hline
     $ [2 2 0 2 2 2]  $& $\ldots  $&$1 - h_6 t^{11}$ \\
  \hline
   $ [2 2 2 2 2 2]  $& $\ldots  $&$1$ \\
  \hline
\end{tabular}
\end{center}
\small \text{An $mHL$ HWG of 1 denotes $mHL^{E_6}_{[000000]}(t)$.}\\
\small \text{Orbits in the centre of the Hasse diagram remain to be calculated.} 
\caption{$E_6$ Orbits and HWGs}
\label{tab:E6NO6}
\end{sidewaystable}

\FloatBarrier

\subsubsection{Orbits of $E_7$ and $E_8$}
\label {subsec:E78NO}

A comparable analysis for the 45 nilpotent orbits of $E_7$ and the 70 orbits of $E_8$ poses computational challenges and it is only possible to present a partial picture. Tables \ref{tab:E7NO1} to \ref{tab:E8NOhwg} set out those Hilbert series and HWGs that have been calculated, along with details of the constructions. Unrefined HS for normal nilpotent orbits of $E_7$ and $E_8$ are shown in tables \ref{tab:E7NO1} to \ref{tab:E7NO6} and \ref{tab:E8NO1a} to \ref{tab:E8NO1b}; the normalisations of the 10 non-normal nilpotent orbits of $E_7$ are shown in tables \ref{tab:E7NO8} and \ref{tab:E7NO9}; the 8 extra root maps of $E_7$ are analysed in tables \ref{tab:E7NO10} and \ref{tab:E7NO11}; and some HWGs for near minimal and near maximal orbits of $E_7$ and $E_8$ are shown in tables \ref{tab:E7NOhwg} and \ref{tab:E8NOhwg}, respectively.

The pattern is similar to that for $E_6$. The near-minimal orbits are normal with palindromic Hilbert series and have character HWGs that are freely generated or complete intersections. All these orbits can be constructed using the $NON$ formula. The minimal and next to minimal orbits (and the next to next to minimal $E_7$ orbit) also have Coulomb branch constructions. In all the cases calculated, the normal orbits are consistent with the established classification, as described in Appendix \ref{apxNO}.

The $mHL$ HWGs for the sub-regular orbits of $E_7$ and $E_8$ have been inferred from a result in \cite{2017arXiv170604820J}, which invites the conjecture that the sub-regular orbit of any group has a $mHL$ HWG given by $1 - {h_\phi }{t^{ht\left( \phi  \right)}}$, where $\phi$ is the irrep whose highest weight is the shortest dominant root of $G$, and $ht\left( \phi  \right)$ counts the number of simple roots within $\phi.$\footnote{ For ADE groups this is the Coxeter number.} This is consistent with the $mHL$ HWGs for the sub-regular orbits of other Exceptional groups and Classical groups \cite{Hanany:2016gbz}.

The 8 extra root maps of $E_7$ include further examples of dualities, with at least three giving copies of nilpotent orbits: $E_7[2020000]$ and $E_7[0110100]$ are normal, with their unrefined HS matching $E_7[0200200]$ and $E_7[0020000]$, respectively; $E_7[2000002]$ appears to be non-normal, with its unrefined HS matching $E_7[0100011]$. Amongst the others, four generate extensions that do not match either the orbits or their normalisations, and one remains to be calculated.

The Weyl group of $E_7$ has 60 irreps and conjugacy classes. In \cite{beynon_spaltenstein_1984}, the 45 nilpotent orbits are identified as these conjugacy classes, modulo some actions of the symmetric groups $S_2$ or $S_3$. Six of the eight extra root maps, appear to correspond to other members of these conjugacy classes; however, these are only identified in \cite{beynon_spaltenstein_1984} by Bala Carter labels, so their root maps or Characteristics are unclear.

The Weyl group of $E_8$ has 112 irreps and conjugacy classes. In \cite{beynon_spaltenstein_1984}, the 70 nilpotent orbits are identified as these conjugacy classes, modulo some actions of the symmetric groups $S_2$, $S_3$ or $S_5$. It can be anticipated that most of the 39 extra root maps of $E_8$ correspond to other members of these conjugacy classes.


\begin{sidewaystable}[htp]
\begin{center}
\small
\begin{tabular}{|c|c|c|c|c|}
\hline
 {Characteristic} & {Type} & {Construction} & {Dim.} & Unrefined HS \dummy \\
 \hline
$ [0 0 0 0 0 0 0] $& {Even} & 
\scriptsize ${g_{NON}^{{E_7}\left[ {0 0 0 0 0 0 0} \right]}{}}$ &$ 0$ &$ 1$ \\

\hline
$ [1 0 0 0 0 0 0]$ & {Rigid} &
\scriptsize $\begin{array}{*{20}{c}}
{g_{{{Coulomb}}}^{{E_7}\left[ {1 0 0 0 0 0 0} \right]}{}}\\
{{ \text{or}}}\\
{g_{NON}^{{E_7}\left[ {1 0 0 0 0 0 0} \right]}{}}
\end{array}$
 &
 $ 34 $
 &
 \scriptsize
 $\frac{ \left(   \begin{array}{c}  1 + 98 t + 3312 t^2 + 53305 t^3 + 468612 t^4 \\
 + 2421286 t^5 + 7664780 t^6 + 15203076 t^7 + 19086400 t^8 \\+ \pal
 + t^{16} \end{array} \right)}
 { \begin{array}{c} (1-t)^{34} (1+t)^{-1} \\ \end{array} }$ \\
\hline

$ [0 0 0 0 1 0 0]$ & {Rigid} &
\scriptsize $\begin{array}{*{20}{c}}
{g_{{{Coulomb}}}^{{E_7}\left[ {0 0 0 0 1 0 0} \right]}{}}\\
{{ \text{or}}}\\
{g_{NON}^{{E_7}\left[ {0 0 0 0 1 0 0} \right]}{}}
\end{array}$
 &
 $ 52 $
 &
 \scriptsize
 $\frac{ \left(   \begin{array}{c}  1 + 79 t + 3161 t^2 + 75291 t^3 + 1158376 t^4 \\+ 12099785 t^5 + 
 88650725 t^6 + 465895118 t^7 + 1783653576 t^8\\ + 5026645901 t^9 + 
 10497603729 t^{10} + 16309233956 t^{11} + 18885794304 t^{12} \\
 + \pal
 + t^{24}  \end{array} \right)}
 { \begin{array}{c} (1 - t)^{52} (1 + t)^{-2} \\ \end{array} }$ \\
\hline

$ [0 0 0 0 0 2 0]$ & {Even} &
\scriptsize $\begin{array}{*{20}{c}}
{g_{{{Coulomb}}}^{{E_7}\left[ {0 0 0 0 0 2 0} \right]}{}}\\
{{ \text{or}}}\\
{g_{NON}^{{E_7}\left[ {0 0 0 0 0 2 0} \right]}{}}
\end{array}$
 &
 $ 54 $
 &
 \scriptsize
 $\frac{ \left(   \begin{array}{c}  1 + 76 t + 2928 t^2 + 67583 t^3 + 1012266 t^4 \\ + 10332067 t^5 + 
 74214232 t^6 + 383547072 t^7 + 1448282149 t^8 \\ + 4037523484 t^9 + 
 8366120760 t^{10} + 12936087566 t^{11} + 14955872444 t^{12} \\+\pal 
 + t^{24}  \end{array} \right)}
 { \begin{array}{c} (1 - t)^{54} (1 + t)^{-3} \\ \end{array} }$ \\
\hline

\end{tabular}
\end{center}
\text{} 
\normalsize
\caption{$E_7$ Orbit Constructions and Hilbert Series (A)}
\label{tab:E7NO1}
\end{sidewaystable}


\begin{sidewaystable}[htp]
\begin{center}
\small
\begin{tabular}{|c|c|c|c|c|}
\hline
 {Characteristic} & {Type} & {Construction} & {Dim.} & Unrefined HS \dummy \\
 \hline

$ [1  0  0  0  1  0  0]$ & {Non-rigid} &
\scriptsize ${g_{NON}^{{E_7}\left[ {1  0  0  0  1  0  0} \right]}{}}$
 &
 $ 76 $
 &
 \scriptsize
 $\frac{ \left(   \begin{array}{c} 
 1 + 55 t + 1541 t^2 + 29315 t^3 + 424272 t^4 + 4952595 t^5 \\+ 
 47877722 t^6 + 386190616 t^7 + 2596530729 t^8 + 14510866523 t^9 \\+ 
 67296385685 t^{10} + 259002809519 t^{11} + 828279369291 t^{12} \\+ 
 2205076546265 t^{13} + 4896965388648 t^{14} + 9088788793591 t^{15}\\ + 
 14120229879183 t^{16} + 18383918865257 t^{17} + 20072546522168 t^{18} \\
 +\pal  
 + t^{36} \end{array} \right)}
 { \begin{array}{c} (1 - t)^{76} (1 + t)^{-2} \\ \end{array} }$ \\
\hline

$ [0  0  1  0  0  0  0]$ & {Rigid} &
\scriptsize ${g_{NON}^{{E_7}\left[ {0  0  1  0  0  0  0} \right]}{}}$
 &
 $ 82 $
 &
 \scriptsize
 $\frac{ \left(   \begin{array}{c} 
 1 + 48 t + 1178 t^2 + 19696 t^3 + 250716 t^4 + 2565971 t^5 \\+ 
 21762497 t^6 + 155518280 t^7 + 944204003 t^8 + 4887315987 t^9\\ + 
 21592275087 t^{10} + 81450603000 t^{11} + 262392783373 t^{12}\\ + 
 722133713475 t^{13} + 1698619550612 t^{14} + 3416812201696 t^{15}\\ + 
 5880606714062 t^{16} + 8663598664204 t^{17} + 10929523033749 t^{18} \\+ 
 11809319976778 t^19 + \pal
 + t^38 \end{array} \right)}
 { \begin{array}{c} (1 - t)^{82} (1 + t)^{-3} \\ \end{array} }$ \\
\hline

\end{tabular}
\end{center}
\text{} 
\normalsize
\caption{$E_7$ Orbit Constructions and Hilbert Series (B)}
\label{tab:E7NO2}
\end{sidewaystable}


\begin{sidewaystable}[htp]
\begin{center}
\small
\begin{tabular}{|c|c|c|c|c|}
\hline
 {Characteristic} & {Type} & {Construction} & {Dim.} & Unrefined HS \dummy \\
 \hline

$[{0 1 0 0 1 0 0}]$ & {Rigid} &
\scriptsize ${g_{NON}^{{E_7}\left[{0 1 0 0 1 0 0}  \right]}{}}$
 &
 $ 90 $
 &
 \scriptsize
 $\frac{ \left(   \begin{array}{c} 
1 + 40 t + 822 t^2 + 11560 t^3 + 125053 t^4 + 1107492 t^5 + \\
 8324729 t^6 + 54205995 t^7 + 308921360 t^8 + 1547294387 t^9 + \\
 6817553073 t^{10} + 26416129223 t^{11} + 89960249299 t^{12} + \\
 269160204736 t^{13} + 707461384957 t^{14} + 1633679518004 t^{15} + \\
 3315057850714 t^{16} + 5912586107207 t^{17} + 9270875065762 t^{18} + \\
 12781960001328 t^{19} + 15497517399838 t^{20} + 16525204589536 t^{21}\\ + \pal 
 + t^{42}     \end{array} \right)}
 { \begin{array}{c} (1 - t)^{90} (1 + t)^{-3} \\ \end{array} }$ \\
\hline

$[{1 0 1 0 0 0 0}]$ &$ \begin{array}{c} \text{Rigid}\\ \text{Non-normal}  \end{array} $ & ? & $ 92 $ & \scriptsize $\ldots$ \\
\hline

$[{0 2 0 0 0 0 0}]$ & {Even} &
\scriptsize ${g_{NON}^{{E_7}\left[{0 2 0 0 0 0 0}  \right]}{}}$
 &
 $ 94 $
 &
 \scriptsize
 $\frac{ \left(   \begin{array}{c} 
1 + 36 t + 668 t^2 + 8508 t^3 + 83586 t^4 + 673392 t^5 + \\
 4607032 t^6 + 27312586 t^7 + 141941392 t^8 + 650985986 t^9 + \\
 2645699417 t^{10} + 9554923437 t^{11} + 30727827270 t^{12} + \\
 88136822566 t^{13} + 225764838482 t^{14} + 516956432313 t^{15} + \\
 1058929210661 t^{16} + 1941429840126 t^{17} + 3186932272113 t^{18} + \\
 4685112655122 t^{19} + 6169119469044 t^{20} + 7276373148733 t^{21} + \\
 7687953729238 t^{22} + \pal
 + t^{44}
      \end{array} \right)}
 { \begin{array}{c} (1 - t)^{94} (1 + t)^{-3} \\ \end{array} }$ \\
\hline

$[ {1 0 0 1 0 1 0}]$ &
$ \begin{array}{c} \text{Non-rigid}\\ \text{Non-normal}  \end{array} $ & ? & $ 94 $ & \scriptsize $\ldots$ \\
\hline

\end{tabular}
\end{center}
\text{} 
\normalsize
\caption{$E_7$ Orbit Constructions and Hilbert Series (C)}
\label{tab:E7NO3}
\end{sidewaystable}

\begin{sidewaystable}[htp]
\begin{center}
\small
\begin{tabular}{|c|c|c|c|c|}
\hline
 {Characteristic} & {Type} & {Construction} & {Dim.} & Unrefined HS \dummy \\
 \hline

$[ {0 0 0 2 0 0 0}]$ &
$ \begin{array}{c} \text{Even}\\ \text{}  \end{array} $ &
\scriptsize ${g_{NON}^{{E_7}\left[ {0 0 0 2 0 0 0} \right]}{}}$
 &
 $ 100 $
 &
 \scriptsize
 $\frac{ \left(   \begin{array}{c} 
1 + 29 t + 438 t^2 + 4582 t^3 + 37271 t^4 + 250862 t^5 + \\
 1450141 t^6 + 7367761 t^7 + 33375509 t^8 + 135968656 t^9 + \\
 500664993 t^{10} + 1670928197 t^{11} + 5061667873 t^{12} + \\
 13926555453 t^{13} + 34810610044 t^{14} + 79050125417 t^{15} + \\
 163069587001 t^{16} + 305534193394 t^{17} + 519874019101 t^{18} + \\
 803213777947 t^{19} + 1126718360132 t^{20} + 1434898159505 t^{21} + \\
 1658939075660 t^{22} + 1741143029810 t^{23} + \pal
 + t^{46}      \end{array} \right)}
 { \begin{array}{c} (1 - t)^{100} (1 + t)^{-4} \\ \end{array} }$ \\
\hline

$[ {2 0 0 0 2 0 0}]$ &
$ \begin{array}{c} \text{Even}\\ \text{Non-normal}  \end{array} $ &  ? & $ 100 $ &\scriptsize $\ldots$ \\
\hline

$[ {2 0 0 0 2 2 0}]$ &
$ \begin{array}{c} \text{Even}\\ \text{Non-normal}  \end{array} $ & ? & $ 102 $ &\scriptsize $\ldots$ \\
\hline

$[ {2 1 0 0 0 1 1}]$ &
$ \begin{array}{c} \text{Non-rigid}\\ \text{}  \end{array} $ &
\scriptsize ${g_{NON}^{{E_7}\left[ {2 1 0 0 0 1 1} \right]}{}}$
 &
 $ 102 $
 &
 \scriptsize
 $\frac{ \left(   \begin{array}{c} 
1 + 28 t + 408 t^2 + 4116 t^3 + 32280 t^4 + 209580 t^5 + \\
 1171719 t^6 + 5794499 t^7 + 25834071 t^8 + 105236571 t^9 + \\
 395177977 t^{10} + 1374921878 t^{11} + 4440514461 t^{12} + \\
 13302464973 t^{13} + 36864118261 t^{14} + 94136235017 t^{15} + \\
 220539780742 t^{16} + 471978064553 t^{17} + 919131363037 t^{18} + \\
 1623431344293 t^{19} + 2593892844620 t^{20} + 3741627500468 t^{21} + \\
 4865497692158 t^{22} + 5698077622856 t^{23} + 6006501802258 t^{24} \\+ \pal
 + t^{48}
  \end{array} \right)}
 { \begin{array}{c} (1 - t)^{102} (1 + t)^{-3}  \\ \end{array} }$ \\
\hline

\end{tabular}
\end{center}
\normalsize
\caption{$E_7$ Orbit Constructions and Hilbert Series (D)}
\label{tab:E7NO4}
\end{sidewaystable}


\begin{sidewaystable}[htp]
\begin{center}
\small
\begin{tabular}{|c|c|c|c|c|}
\hline
 {Characteristic} & {Type} & {Construction} & {Dim.} & Unrefined HS \dummy \\
 \hline
 
$[{2 0 1 0 1 0 0}]$ &
$ \begin{array}{c} \text{Richardson}  \end{array} $ &
\scriptsize ${g_{NON}^{{E_7}\left[{2 0 1 0 1 0 0} \right]}{}}$
 &
 $ 106 $
 &
 \scriptsize
 $\frac{ \left(   \begin{array}{c} 
1 + 23 t + 278 t^2 + 2347 t^3 + 15528 t^4 + 85677 t^5 + 409818 t^6\\ + 
 1744604 t^7 + 6733298 t^8 + 23872838 t^9 + 78471083 t^{10}\\ + 
 240598830 t^{11} + 690346420 t^{12} + 1854303338 t^{13} \\+ 
 4651709274 t^{14} + 10851525702 t^{15} + 23419398905 t^{16}\\ + 
 46527253037 t^{17} + 84742390056 t^{18} + 141081479581 t^{19} \\+ 
 214295496214 t^{20} + 296699661744 t^{21} + 374312393511 t^{22} \\+ 
 430286790879 t^{23} + 450739862708 t^{24} + \pal
 + t^{48}
   \end{array} \right)}
 { \begin{array}{c} (1 - t)^{110} (1 - t^2)^{-3} (1 - t^3)^{-1} \\ \end{array} }$ \\
\hline

$[{2 0 0 2 0 0 0}]$ &
$ \begin{array}{c} \text{Even}  \end{array} $ &
\scriptsize ${g_{NON}^{{E_7}\left[{2 0 0 2 0 0 0} \right]}{}}$
 &
 $ 108 $
 &
 \scriptsize
 $\frac{ \left(   \begin{array}{c} 
1 + 20 t + 213 t^2 + 1601 t^3 + 9511 t^4 + 47457 t^5 + 206535 t^6\\ + 
 804110 t^7 + 2849798 t^8 + 9301048 t^9 + 28157657 t^{10}\\ + 
 79371845 t^{11} + 208613748 t^{12} + 511155245 t^{13} + 1166404057 t^{14}\\ + 
 2475221676 t^{15} + 4877328017 t^{16} + 8910423434 t^{17}\\ + 
 15071159308 t^{18} + 23570050287 t^{19} + 34043293243 t^{20}\\ + 
 45364765295 t^{21} + 55726054790 t^{22} + 63063271014 t^{23}\\ + 
 65720671708 t^{24} + \pal
 + t^{48}
   \end{array} \right)}
 { \begin{array}{c} (1 - t)^{113} (1 - t^2)^{-4} (1 - t^3)^{-1} \\ \end{array} }$ \\
\hline

\end{tabular}
\end{center}
\small \text{Some orbits in the upper Hasse diagram remain to be calculated.} 
\normalsize
\caption{$E_7$ Orbit Constructions and Hilbert Series (E)}
\label{tab:E7NO5}
\end{sidewaystable}

\begin{sidewaystable}[htp]
\begin{center}
\small
\begin{tabular}{|c|c|c|c|c|}
\hline
 {Characteristic} & {Type} & {Construction} & {Dim.} & Unrefined HS \dummy \\
 \hline

$[{0 0 2 0 0 2 0}]$ &
$ \begin{array}{c} \text{Distinguished} \end{array} $ &
\scriptsize ${g_{NON}^{{E_7}\left[{0 0 2 0 0 2 0}\right]}{}}$
 &
 $ 112 $
 &
 \scriptsize
 $\frac{ \left(   \begin{array}{c} 
1 + 16 t + 138 t^2 + 849 t^3 + 4168 t^4 + 17338 t^5 + 63414 t^6\\ + 
 208931 t^7 + 630376 t^8 + 1761854 t^9 + 4597735 t^{10} \\+ 
 11259794 t^{11} + 25951083 t^{12} + 56346651 t^{13} + 115226073 t^{14}\\ + 
 221674893 t^{15} + 400557834 t^{16} + 678589198 t^{17} + 1075849008 t^{18}\\ + 
 1593563475 t^{19} + 2202070691 t^{20} + 2835436943 t^{21} \\+ 
 3398852391 t^{22} + 3790337376 t^{23} + 3930789260 t^{24} \\+ \pal
 + t^{48}   \end{array} \right)}
 { \begin{array}{c} (1 - t)^{117} (1 - t^2)^{-3}  (1 - t^3)^{-1} (1 - t^4)^{-1} \\ \end{array} }$ \\
\hline

$[{2 2 0 0 2 0 0}]$ &
$ \begin{array}{c} \text{Even} \end{array} $ &
\scriptsize ${g_{NON}^{{E_7}\left[{2 0 0 0 2 0 2}\right]}{}}$
 &
 $ 112 $
 &
 \scriptsize
  $\frac{ \left(   \begin{array}{c} 
1 + 17 t + 154 t^2 + 986 t^3 + 5000 t^4 + 21352 t^5 + 79766 t^6 \\+ 
 267478 t^7 + 820082 t^8 + 2329206 t^9 + 6180894 t^{10} \\+ 
 15395201 t^{11} + 36035767 t^{12} + 79165241 t^{13} + 162774319 t^{14} \\+ 
 312189809 t^{15} + 556698600 t^{16} + 920602987 t^{17} + 1409431708 t^{18} \\+ 
 1996091906 t^{19} + 2614816619 t^{20} + 3169351708 t^{21} \\+ 
 3556104836 t^{22} + 3695053378 t^{23} + \pal
 + t^{46}
 \end{array} \right)}
 { \begin{array}{c} (1 - t)^{116} (1 - t^2)^{-2} (1 - t^4)^{-1}  (1 - t^6)^{-1} \\ \end{array} }$ \\
\hline

\end{tabular}
\end{center}
\small \text{Some orbits in the upper Hasse diagram remain to be calculated.} 
\normalsize
\caption{$E_7$ Orbit Constructions and Hilbert Series (F)}
\label{tab:E7NO6}
\end{sidewaystable}


\begin{sidewaystable}[htp]
\begin{center}
\small
\begin{tabular}{|c|c|c|c|c|}
\hline
 {Characteristic} & {Type} & {Construction} & {Dim.} & Unrefined HS \dummy \\
 \hline
 
 $[ {2 0 0 0 1 0 0}]$ &
$ \begin{array}{c} \text{Non-rigid}\\ \text{Normalisation}  \end{array} $ &
\scriptsize ${g_{NON}^{{E_7}\left[ {2 0 0 0 1 0 0} \right]}{}}$
 &
 $ 84 $
 &
 \scriptsize
 $\frac{ \left(   \begin{array}{c} 
1 + 47 t + 1129 t^2 + 18604 t^3 + 236007 t^4 + 2439134 t^5 + \\
 21162857 t^6 + 156442728 t^7 + 991136142 t^8 + 5390236639 t^9 + \\
 25171607172 t^{10} + 100975434413 t^{11} + 348257944507 t^{12} + \\
 1033980718497 t^{13} + 2646472834763 t^{14} + 5847508202694 t^{15} + \\
 11167569105434 t^{16} + 18453105745417 t^{17} + 26401938893655 t^{18} + \\
 32725550735483 t^{19} + 35152494567578 t^{20} + \pal
 + t^{40}
     \end{array} \right)}
 { \begin{array}{c} (1 - t)^{84} (1 + t)^{-2} \\ \end{array} }$ \\
\hline
 
 $[ {2 0 0 0 0 2 0}]$ &$ \begin{array}{c} \text{Even}\\ \text{Normalisation}  \end{array} $ &
\scriptsize ${g_{NON}^{{E_7}\left[ {2 0 0 0 0 2 0}\right]}{}}$
 &
 $ 86 $
 &
 \scriptsize
 $\frac{ \left(   \begin{array}{c} 
1 + 44 t + 992 t^2 + 15401 t^3 + 186066 t^4 + 1858596 t^5 + \\
 15801050 t^6 + 115491687 t^7 + 726319079 t^8 + 3923623495 t^9 + \\
 18189961165 t^{10} + 72405189290 t^{11} + 247811206611 t^{12} + \\
 730559598720 t^{13} + 1858314990721 t^{14} + 4084659605334 t^{15} + \\
 7767711560354 t^{16} + 12792450962244 t^{17} + 18258619963592 t^{18} + \\
 22598363867429 t^{19} + 24262180770322 t^{20} + \pal
 + t^{40}
     \end{array} \right)}
 { \begin{array}{c} (1 - t)^{86} (1 + t)^{-3} \\ \end{array} }$ \\
\hline
  
\end{tabular}
\end{center}
\text{}\\
\normalsize
\caption{$E_7$ Nilpotent Orbit Normalisations and Hilbert Series (A)}
\label{tab:E7NO8}
\end{sidewaystable}


\begin{sidewaystable}[htp]
\begin{center}
\small
\begin{tabular}{|c|c|c|c|c|}
\hline
 {Characteristic} & {Type} & {Construction} & {Dim.} & Unrefined HS \dummy \\
 \hline
 
 $[{0 0 1 0 1 0 0}]$ &
$ \begin{array}{c}\text{Non-rigid}\\ \text{Normalisation}  \end{array} $ &
\scriptsize ${g_{NON}^{{E_7}\left[ {0 0 1 0 1 0 0} \right]}{}}$
 &
 $ 98 $
 &
 \scriptsize
 $\frac{ \left(   \begin{array}{c} 
1 + 32 t + 530 t^2 + 6181 t^3 + 57541 t^4 + 453663 t^5 + \\
 3113473 t^6 + 18837445 t^7 + 101120527 t^8 + 483453140 t^9 + \\
 2063999692 t^{10} + 7883746897 t^{11} + 26978747805 t^{12} + \\
 82795280050 t^{13} + 228030235944 t^{14} + 563908554084 t^{15} + \\
 1252637000445 t^{16} + 2500204122427 t^{17} + 4485045975803 t^{18} + \\
 7232498422887 t^{19} + 10486086399152 t^{20} + 13671033536263 t^{21} + \\
 16028614605883 t^{22} + 16901469671998 t^{23} +\\
 \pal + t^{46}
  \end{array} \right)}
 { \begin{array}{c} (1 - t)^{98} (1 + t)^{-3} \\ \end{array} }$ \\
\hline
 
 $[ {2 0 0 0 2 0 0}]$ &
$ \begin{array}{c} \text{Even}\\ \text{Normalisation}  \end{array} $ &
\scriptsize ${g_{NON}^{{E_7}\left[ {2 0 0 0 2 0 0} \right]}{}}$
 &
 $ 100 $
 &
 \scriptsize
 $\frac{ \left(   \begin{array}{c} 
1 + 31 t + 496 t^2 + 5589 t^3 + 50366 t^4 + 384882 t^5 + \\
 2559623 t^6 + 14987542 t^7 + 77732943 t^8 + 358507573 t^9 + \\
 1474780676 t^{10} + 5424817521 t^{11} + 17880645370 t^{12} + \\
 52902604370 t^{13} + 140698668543 t^{14} + 336770031480 t^{15} + \\
 726143059501 t^{16} + 1411518293411 t^{17} + 2475048767521 t^{18} + \\
 3916605082526 t^{19} + 5595119586673 t^{20} + 7217473926998 t^{21} + \\
 8408241643667 t^{22} + 8847237672754 t^{23} + \pal
  + t^{46}
 \end{array} \right)}
 { \begin{array}{c} (1 - t)^{100} (1 + t)^{-2} (1 + t^2)^{-1} \\ \end{array} }$ \\
\hline

\end{tabular}
\end{center}
\small \text{Some orbits in the upper Hasse diagram remain to be calculated.} 
\normalsize
\caption{$E_7$ Nilpotent Orbit Normalisations and Hilbert Series (B)}
\label{tab:E7NO9}
\end{sidewaystable}


\begin{sidewaystable}[htp]
\begin{center}
\small
\begin{tabular}{|c|c|c|c|c|}
\hline
 {Characteristic} & {Type} & {Construction} & {Dim.} & Unrefined HS \dummy \\
 \hline
 
  $[{2 0 0 0 0 0 2}]$ &
$ \begin{array}{c}\text{Even} \\ \text{Normalisation} \\ \text{Dual} \end{array} $ &
\scriptsize ${g_{NON}^{{E_7}\left[ {2 0 0 0 0 0 2} \right]}{}}$
 &
 $ 96 $
 &
 \scriptsize
 $\frac{ \left(   \begin{array}{c} 
1 + 33 t + 564 t^2 + 6777 t^3 + 64849 t^4 + 524725 t^5 + \\
 3696270 t^6 + 22952836 t^7 + 126186057 t^8 + 614879502 t^9 + \\
 2656800340 t^{10} + 10184095054 t^{11} + 34652624372 t^{12} + \\
 104728696487 t^{13} + 281292351713 t^{14} + 671789330508 t^{15} + \\
 1427206725858 t^{16} + 2698313629951 t^{17} + 4541504048285 t^{18} + \\
 6806645160112 t^{19} + 9086346089774 t^{20} + 10805286175350 t^{21} + \\
 11447593810728 t^{22} + \pal
 + t^{44}
 \end{array} \right)}
 { \begin{array}{c} (1 - t)^{96} (1 + t)^{-4} \\ \end{array} }$ \\
\hline
 
  $[{0 0 0 0 2 0 2}]$ &
$ \begin{array}{c} \text{Even}\\  \text{Cover}  \end{array} $ &
\scriptsize ${g_{NON}^{{E_7}\left[ {0 0 0 0 2 0 2} \right]}{}}$
 &
 $ 104 $
 &
 \scriptsize
 $\frac{ \left(   \begin{array}{c} 
1 + 27 t + 379 t^2 + 3681 t^3 + 27917 t^4 + 176916 t^5 + 977308 t^6 \\+ 
 4835965 t^7 + 21801267 t^8 + 90393556 t^9 + 346150921 t^{10}\\ + 
 1225347941 t^{11} + 4006584460 t^{12} + 12082557933 t^{13}\\ + 
 33549402062 t^{14} + 85635361526 t^{15} + 200649655775 t^{16} \\+ 
 431024819812 t^{17} + 847995374640 t^{18} + 1526658506276 t^{19}\\ + 
 2513332265924 t^{20} + 3781664655782 t^{21} + 5198296443145 t^{22} \\+ 
 6526055274327 t^{23} + 7481012613977 t^{24} + 7829538747780 t^{25}\\ + \pal
 + t^{50}  \end{array} \right)}
 { \begin{array}{c} (1 - t)^{104} (1 + t)^{-2}  \\ \end{array} }$ \\
\hline
 
\end{tabular}
\end{center}
\text{These moduli spaces are associated with SU(2) homomorphisms but do not represent additional nilpotent orbits (see text).}\\
\normalsize
\caption{$E_7$ Extra Moduli Spaces and Hilbert Series (A)}
\label{tab:E7NO10}
\end{sidewaystable}

\begin{sidewaystable}[htp]
\begin{center}
\small
\begin{tabular}{|c|c|c|c|c|}
\hline
 {Characteristic} & {Type} & {Construction} & {Dim.} & Unrefined HS \dummy \\
 \hline
  
 $[{0 1 1 0 1 0 0}]$ &
$ \begin{array}{c} \text{Richardson} \\ \text{Dual}  \end{array} $ &
\scriptsize ${g_{NON}^{{E_7}\left[{0 1 1 0 1 0 0} \right]}{}}$
 &
 $ 106 $
 &
 \scriptsize
 
$\frac{ \left(   \begin{array}{c} 
1 + 22 t + 254 t^2 + 2048 t^3 + 12949 t^4 + 68223 t^5 + 310470 t^6 \\+ 
 1248798 t^7 + 4509005 t^8 + 14770398 t^9 + 44209400 t^{10} \\+ 
 121447827 t^{11} + 306952315 t^{12} + 714370802 t^{13} + 1530391654 t^{14} \\+ 
 3014498298 t^{15} + 5450883816 t^{16} + 9031969079 t^{17} \\+ 
 13689807076 t^{18} + 18950786410 t^{19} + 23928392089 t^{20} \\+ 
 27532705839 t^{21} + 28852797782 t^{22} + \pal
 + t^{44} 
 \end{array} \right)}
 { \begin{array}{c} (1 - t)^{111} (1 - t^2)^{-2} (1 - t^3)^{-2} (1 - t^4)^{-1}  \\ \end{array} }$\\

\hline
 
 $[{2 0 2 0 0 0 0}]$ &
$ \begin{array}{c} \text{Even}\\ \text{Dual}  \end{array} $ &
\scriptsize ${g_{NON}^{{E_7}\left[{2 0 2 0 0 0 0}\right]}{}}$
 &
 $ 110 $
 &
 \scriptsize
 $\frac{ \left(   \begin{array}{c} 
1 + 19 t + 191 t^2 + 1350 t^3 + 7526 t^4 + 35208 t^5 + 143642 t^6\\ + 
 524421 t^7 + 1744507 t^8 + 5355360 t^9 + 15303933 t^{10} \\+ 
 40934228 t^{11} + 102767564 t^{12} + 242325486 t^{13} + 536176791 t^{14}\\ + 
 1110998287 t^{15} + 2150326632 t^{16} + 3877166750 t^{17} \\+ 
 6495975323 t^{18} + 10090787477 t^{19} + 14505728258 t^{20} \\+ 
 19267371072 t^{21} + 23618674467 t^{22} + 26697216460 t^{23} \\+ 
 27811736966 t^{24} +\pal
 + t^{48}   \end{array} \right)}
 { \begin{array}{c} (1 - t)^{114} (1 - t^2)^{-2} (1 - t^3)^{-1} (1 - t^4)^{-1} \\ \end{array} }$ \\
\hline

\end{tabular}
\end{center}
\small \text{These moduli spaces are associated with SU(2) homomorphisms but do not represent additional nilpotent orbits (see text).}\\
\normalsize
\caption{$E_7$ Extra Moduli Spaces and Hilbert Series (B)}
\label{tab:E7NO11}
\end{sidewaystable}

\begin{sidewaystable}[htp]
\begin{center}
\small
\begin{tabular}{|c|c|c|c|c|}
\hline
 {Characteristic} & {Type} & {Construction} & {Dim.} & Unrefined HS \dummy \\
 \hline
$ [0 0 0 0 0 0 0 0] $& {Trivial} & 
\scriptsize ${g_{NON}^{{E_8}\left[ {0 0 0 0 0 0 0 0} \right]}{}}$ &$ 0$ &$ 1$ \\
\hline

$ [0 0 0 0 0 0 1 0]$ & {Rigid} &
\scriptsize $\begin{array}{*{20}{c}}
{g_{{{Coulomb}}}^{{E_8}\left[ {0 0 0 0 0 0 1 0} \right]}{}}\\
{{ \text{or}}}\\
{g_{NON}^{{E_8}\left[ {0 0 0 0 0 0 1 0} \right]}{}}
\end{array}$
 &
 $ 58 $
 &
 \scriptsize
 $\frac{ \left(   \begin{array}{c}  
 1 + 189 t + 14080 t^2 + 562133 t^3 + 13722599 t^4+ 220731150 t^5\\
  + 2454952400 t^6 + 19517762786 t^7 + 113608689871 t^8 \\
 + 492718282457 t^9 + 1612836871168 t^{10} \\
 + 4022154098447 t^{11} +7692605013883 t^{12} + 11332578013712 t^{13} \\
 + 12891341012848 t^{14} + \pal
 + t^{28}       \end{array} \right)}
 { \begin{array}{c} (1-t)^{58}   (1 + t) ^{-1}        \\ \end{array} }$ \\
\hline

$ [1 0 0 0 0 0 0 0]$ & {Rigid} &
\scriptsize $\begin{array}{*{20}{c}}
{g_{{{Coulomb}}}^{{E_8}\left[ {1 0 0 0 0 0 0 0} \right]}{}}\\
{{ \text{or}}}\\
{g_{NON}^{{E_8}\left[ {1 0 0 0 0 0 0 0} \right]}{}}
\end{array}$
 &
 $ 92 $
 &
 \scriptsize
 $\frac{ \left(   \begin{array}{c}     
 1 + 154 t + 11936 t^2 + 590394 t^3 + 20506501 t^4 + 527204320 t^5\\ 
 + 10378075500 t^6 + 160202160870 t^7 + 1974075401833 t^8\\
  + 19687582048248 t^9 + 160663069166772 t^{10} + 1082397634755580 t^{11} \\
 + 6063957293140705 t^{12} + 28419656685639506 t^{13} \\
 + 111973818366411496 t^{14} + 372403620091866888 t^{15}\\
 + 1048956129150029406 t^{16} + 2509111945366416404 t^{17}\\ 
 + 5107852437413443490 t^{18} + 8864217905887836806 t^{19} \\
 + 13130135397834893988 t^{20} + 16614983532652987182 t^{21} \\
 + 17970189038072829240 t^{22} +\pal
 + t^{44}        \end{array} \right)}
 { \begin{array}{c} (1 - t)^{92}     (1 + t) ^{-2}        \\ \end{array} }$ \\
\hline

\end{tabular}
\end{center}
\small \text{} 
\normalsize
\caption{$E_8$ Orbit Constructions and Hilbert Series (A)}
\label{tab:E8NO1a}
\end{sidewaystable}

\begin{sidewaystable}[htp]
\begin{center}
\small
\begin{tabular}{|c|c|c|c|c|}
\hline
 {Characteristic} & {Type} & {Construction} & {Dim.} & Unrefined HS \dummy \\
 \hline
$ [0  0  0  0  0  0  2  0] $& {Even} & 
\scriptsize ${g_{NON}^{{E_8}\left[ {0 0 0 0 0 0 2 0} \right]}{}}$ &$ 114$ &\scriptsize
 $\frac{ \left(   \begin{array}{c}  
1 + 133 t + 8911 t^2 + 400995 t^3 + 13486580 t^4 + 356584057 t^5\\
 +  7623391966 t^6 + 134206634140 t^7 + 1971437827690 t^8 \\+ 
 24415159489727 t^9 + 257072913186568 t^{10} + 2317538562914724 t^{11}\\ + 
 17995649940635004 t^{12} + 120979229232553840 t^{13} \\+ 
 707282480739436178 t^{14} + 3609956701308961796 t^{15}\\ + 
 16140434282332124866 t^{16} + 63405524248622811378 t^{17}\\ + 
 219417855826005179940 t^{18} + 670412057684945457526 t^{19} \\+ 
 1812177346079591528612 t^{20} + 4341065907801003784365 t^{21}\\ + 
 9229322963452174348555 t^{22} + 17436647191913535443645 t^{23}\\ + 
 29303858730884657436479 t^{24} + 43844634221326426603948 t^{25} \\+ 
 58440640016940455205864 t^{26} + 69425059021468796578222 t^{27} \\+ 
 73525348635628936776580 t^{28} + \pal
 + t^{56}
     \end{array} \right)}
 { \begin{array}{c} (1 - t)^{114} (1 + t)^{-1}        \\ \end{array} }$ \\
\hline

$\ldots$ & $\ldots$ &$\ldots$&$\ldots$&$\ldots$ \\
\hline

$ [{2 2 2 2 2 2 2 2}]$ & {Distinguished} &
\scriptsize $
{g_{NON}^{{E_8}\left[ {2 2 2 2 2 2 2 2} \right]}{}}
$
 &
 $ 240 $
 &
 \scriptsize
 $\frac{\begin{array}{c} (1-t^2) (1-t^8) (1-t^{12}) (1-t^{14}) (1-t^{18}) (1-t^{20}) (1-t^{24}) (1-t^{30})   \\ \end{array}}
 {\begin{array}{c}  (1-t)^{240}    \\ \end{array}}$ \\
\hline

\end{tabular}
\end{center}
\small \text{Orbits in the centre of the Hasse diagram remain to be calculated.} 
\normalsize
\caption{$E_8$ Orbit Constructions and Hilbert Series (B)}
\label{tab:E8NO1b}
\end{sidewaystable}


\begin{sidewaystable}[htp]
\begin{center}
\begin{tabular}{|c|c|c|}
 \hline
 \text{Characteristic} & \text{Character HWG} & \text{$mHL$ HWG} \dummy \\
 \hline
$ [0  0  0  0  0  0  0]  $&$ 1  $&$\ldots  $\\
  \hline
 $[1  0  0  0  0  0  0]  $&$ \frac{1}{1- {m_1} t}  $&$\ldots  $\\
  \hline
 $[0  0  0  0  1  0  0]$&$ \frac{1}{(1-{m_1} t) (1- {m_5} t^2)}  $&$\ldots  $\\
  \hline
 $[0  0  0  0  0  2  0]$&$ \frac{1}{(1-{m_1} t) (1- {m_5} t^2) (1 - {m_6}^2 t^3)}  $&$\ldots  $\\
  \hline 
  $[0  1  0  0  0  0  0]$&$ \frac{1}{(1-{m_1} t) (1-{m_2} t^3) (1-{m_3} t^4) (1-{m_5} t^2)} $&$\ldots  $\\
  \hline  
    $[2 0  0  0  0  0  0]$&$\frac{1-{m_1}^2 {m_2}^2 t^{10}}{(1-{m_1} t) (1-{m_1}^2 t^4) (1-{m_2} t^3) (1-{m_2}^2 t^6) (1-{m_3} t^4) (1-{m_5} t^2) (1-{m_1} {m_2} t^5)}$&$\ldots  $\\
  \hline 
  $\ldots  $& $\ldots  $&$\ldots  $ \\
  \hline
  $[2 2  0  2  2  2  2]$& $\ldots$ & $1 - h_1 t^{17} $\\
  \hline 
   $ [2 2 2 2 2 2 2]  $& $\ldots  $&$1$ \\   
\hline 

\end{tabular}
\end{center}
\small \text{An $mHL$ HWG of 1 denotes $mHL^{E_7}_{[0000000]}(t)$.}\\
\small \text{Orbits in the centre of the Hasse diagram are omitted.} 
\caption{$E_7$ Orbits and HWGs}
\label{tab:E7NOhwg}
\end{sidewaystable}


\begin{sidewaystable}[htp]
\begin{center}
\begin{tabular}{|c|c|c|}
 \hline
 \text{Characteristic} & \text{Character HWG} & \text{$mHL$ HWG} \dummy \\
 
\hline
$ [0  0  0  0  0  0  0 0]  $&$ 1  $&$\ldots  $\\
  \hline
 $[0  0  0  0  0  0  1  0]  $&$ \frac{1}{1- {m_7} t}  $&$\ldots  $\\
  \hline
 $[1  0  0  0  0  0  0  0]$&$ \frac{1}{(1-{m_7} t) (1- {m_1} t^2)}  $&$\ldots  $\\
  \hline
   $[0  0  0  0  0  1  0  0]$&$ \frac{1}{(1-{m_7} t)  (1-{m_1} t^2) (1-{m_6} t^3) (1-{m_5} t^4)  }  $&$\ldots  $\\
  \hline
     $[0  0  0  0  0  0  2  0]$&$ \frac{1-{m_6}^2 {m_7}^2 t^{10}}{(1-{m_7} t) (1-{m_1} t^2) (1-{m_6} t^3)  (1-{m_5} t^4) (1-{m_7}^2 t^4) (1-{m_6} {m_7} t^5)  (1-{m_6}^2 t^6)  }  $&$\ldots  $\\
  \hline
  $\ldots  $& $\ldots  $&$\ldots  $ \\
  \hline
  $[2 2  0  2  2  2  2 2]$& $\ldots$ & $1 - h_7 t^{29}  $\\
  \hline 
   $ [2 2 2 2 2 2 2 2]  $& $\ldots  $&$1$ \\
  \hline
  
\end{tabular}
\end{center}
\small \text{An $mHL$ HWG of 1 denotes $mHL^{E_8}_{[00000000]}(t)$.}\\
\text{Orbits in the centre of the Hasse diagram are omitted.} 
\caption{$E_8$ Orbits and HWGs}
\label{tab:E8NOhwg}
\end{sidewaystable}


\FloatBarrier

\section{Discussion and Conclusions}
\label{sec:Conclusions}

\paragraph{Coulomb Branch}

Taken together with the Classical group quivers given in \cite{Hanany:2016gbz}, the Exceptional group quivers in section \ref{sec:Coulomb} provide Coulomb branch constructions, using the unitary monopole formula, for all nilpotent orbits with Characteristic Height 2. These Coulomb branch constructions either yield the closures of normal nilpotent orbits, or, in the case of non-normal orbits, their normal components (the unions of which yield the orbits). Other than in the case of the A series, where quivers can be found by 3d mirror symmetry \cite{Intriligator:1996ex}, Coulomb branch constructions based on the monopole formula for nilpotent orbits of Characteristic Height greater than 2 are not (yet) known.

All the nilpotent orbits with Characteristic Height 2 have character HWGs of a \emph{freely generated} type. In the case of nilpotent orbits higher up a Hasse diagram, multiple roots have Characteristic Height $\ge 2$, so the moduli spaces can be complicated by relations between Lie algebra operators, with the result that the HWGs are usually not freely generated.

This leaves open the question as to whether faithful Coulomb branch constructions for a broader class of nilpotent orbits can be found. While preliminary steps have been taken towards developing constructions based on non-unitary versions of the monopole formula in \cite{Cremonesi:2014kwa}, for example, this remains an area for further research.

\paragraph{Nilpotent Orbit Normalisation Formula}

In the absence of quiver theory constructions for Exceptional group nilpotent orbits of Characteristic Height greater than 2, it is a significant finding that a direct plethystic calculation of the closure of any normal nilpotent orbit is possible using the Nilpotent Orbit Normalisation formula developed in section \ref{sec:GmodH}. The $NON$ formula can be viewed as a generalisation of the Weyl character formula. One of its attractions is that it explicates, in a direct manner, the relationship between an $SU(2)$ homomorphism, as described by its Characteristic, its nilpositive element $X$ and the resulting nilpotent orbit (or normalisation). 

Like the Coulomb branch formula, the $NON$ formula yields a moduli space with a palindromic Hilbert series, so the situation surrounding non-normal nilpotent orbits, which have non-palindromic Hilbert series, needs consideration; however, for normal orbits, the Higgs or Coulomb branch (where available) and $NON$ methods all construct the same moduli spaces.

Turning to the established list of non-normal orbits; in all the cases calculated, the $NON$ formula leads to moduli spaces, with palindromic Hilbert series, containing elements outside the nilpotent cone $\cal N$.

For Classical non-normal orbits, the $NON$ formula either yields the normal components of those orbits that are unions, as in the case of $D_{2r}$ spinor pairs, or it yields their normalisations. These normalisations can be restricted to equal the non-normal orbits by excluding sub-spaces described by \emph{charged} orbits of lower dimension.

In the case of Exceptional non-normal orbits, there are no spinor pairs, and the $NON$ formula yields normalisations. By conjecturing relationships, similar to those between Classical non-normal orbits and their normalisations, it has been possible to find restrictions of the normalisations in $G_2$, $F_4$ and $E_6$, that yield Hilbert series lying within $\cal N$, and which, subject to a more definitive analysis, can be viewed as candidates for the non-normal orbits.

This study has not made significant use of Bala-Carter labels \cite{Bala:1976kl, Bala:1976tg, Collingwood:1993fk}. The perspective herein is that a nilpotent element $X$ is more naturally characterised by an extension of the quotient group structure $G/G_0$ that applies to Richardson orbits. The $NON$ formula generalises this structure to non-Richardson orbits, by defining $\tilde \Phi _{G/G_0}$ to exclude the roots in $\Phi _G^{[1]}$ from the positive roots in $\Phi _{G/G_0}$; this appears to be permissible due to the Weyl group invariance of $\Phi _G^{[1]}$ under the subset $W_{G_0}$ of reflections of $\Phi _{G}$.

This analysis of the closures of nilpotent orbits as moduli spaces does, however, leave a few residual puzzles in relation to the narrative in the mathematical Literature regarding the nilpotent orbits of Exceptional groups. Specifically:

\begin{enumerate}

\item A small number of \emph{extra root maps}, which are not listed amongst the Characteristics in standard tables, follow from the $SU(2)$ homomorphisms of $EF$ groups. Some of these extra root maps, such as $E_6[111110]$, $E_6[020202]$, $E_7[2020000]$, $E_7[0110100]$, $E_7[2000002]$ and  $E_7[2020000]$, generate refined Hilbert series that are identical to those from the Characteristics of nilpotent orbits; others give rise to moduli spaces with palindromic HS, that are extensions of nilpotent orbits outside ${\cal N}$. Although several cases for $E_7$ and $E_8$ remain to be calculated, no new nilpotent orbits have been identified. This appears consistent with the perspective that these extra root maps may be related to Weyl group conjugacy classes that are equivalent to nilpotent orbits, modulo certain symmetric group actions \cite{beynon_spaltenstein_1984}.

Nonetheless, these extra roots maps provide examples of dualities, such that certain $SU(2)$ embeddings in $G$ with different root maps or Characteristics lead to identical closures of nilpotent orbits of $G$; such dualities appear to conflict with the standard narrative surrounding the Jacoboson-Morozov theorem in the Literature \cite{Collingwood:1993fk}, which claims a bijection, not just between $SU(2)$ embeddings and nilpotent elements $X$, but also between $SU(2)$ embeddings and nilpotent orbits ${\cal O}_X$.

\item When defining the partial ordering (or topology) of nilpotent orbits within the nilpotent cone $\cal N$, it is important to deal with the orbits, rather than their normalisations. The Hasse diagrams of inclusion relations depend on whether non-normal nilpotent orbits, or their normalisations, are used. This may account for the few discrepancies in linking patterns (to or from non-normal orbits) between the $F_4$ and $E_6$ Hasse diagrams obtained from the moduli space analysis in this study and the standard diagrams in the Literature. Whereas the standard diagrams date from \cite{spaltenstein_1982}, the listing of non-normal orbits of Exceptional groups appears some years later in \cite{Broer:1998qf}. It would be interesting to be able to give a precise account of the source of the differences between the topologies of orbits calculated from the $NON$ formula and the standard diagrams.

\end{enumerate}

The moduli space calculations for Exceptional groups, in particular, have been limited by practical computing constraints and so several tables herein are incomplete, more so in terms of HWG descriptions than unrefined HS. Given continuing developments in computing power, in terms of memory, speed and algorithms for polynomial algebra, it should eventually be possible to fill in the gaps in this analysis of the moduli spaces of quiver theories. This may resolve the open questions about the nilpotent orbits of Exceptional groups.

There remains the problem of how to formulate an unambiguous analytic method for restricting the normalisation of a non-normal Exceptional group nilpotent orbit to the nilpotent cone $\cal N$, as required by \ref{eq:gmodh5}. The analysis for Classical orbits, drawing on Higgs branch results, describes the difference between a non-normal orbit and its normalisation in terms of the charged $NON$ formula for an orbit lower down the Hasse diagram; but what determines the particular charges and coefficients that appear? The solution may be related to the type of degeneration between adjacent orbits, where it is known from \cite{Baohua-Fu:2015nr} that for Exceptional group orbits this is considerably more complicated than the Kraft-Procesi transitions \cite{Kraft:1982fk} between Classical group orbits. 

Although such technical issues remain to be resolved, this paper and its companion \cite{Hanany:2016gbz} go some considerable way towards systematising the intricate relationships between quiver theories, Hilbert series (and their generating functions) and the closures of nilpotent orbits, as well as developing a number of relevant analytical methods and tools. These in turn lay the foundations for the use of such quiver theories for Hilbert series with background charges as canonical building blocks that can be deployed to construct, analyse and/or decompose a much wider range of theories. Looking beyond $4d$ ${\cal N}=2$ Higgs branch and $3d$ ${\cal N}=4$  Coulomb branch theories, these methods should be applicable to other moduli spaces where nilpotent orbits play a central structural role, such as $5d{~}{\cal N} = 1$ theories, $6d{~}(1,0)$ CFTs, ${\cal F}$ theory and class ${\cal S}$ theories, amongst others.



\paragraph{Acknowldgements}
Rudolph Kalveks is grateful to Marcus Sperling, Fakult\"at f\"ur Physik, Universit\"at Wien, to Santiago Cabrera Marquez, Imperial College, London and to Giulia Ferlito, Imperial College, London for valuable discussions.\\%

\appendix

\section{Hilbert Series Transformations}
\label{apx:HST}

A refined Hilbert series $g_{HS}^G\left( {x,t} \right)$ in class functions of $G$ can be transformed by Weyl integration to a HWG based on the characters (or modified Hall Littlewood polynomials) of $G$, with the aid of a generating function for the characters (or mHL) of $G$:

\begin{equation}
\label{eq:HSTx1}
\begin{aligned}
g_{HWG}^G\left( {m,t} \right) & = \oint\limits_G {d{\mu ^G}}{~} g_\chi ^G\left( {x^*,m} \right){~} g_{HS}^G\left( {x,t} \right),\\
g_{HWG}^G\left( {h,t} \right) & = \oint\limits_G {d\mu _{mHL}^G}{~} \overline {g_{mHL}^G\left( {x^*,h,t} \right)}{~} g_{HS}^G\left( {x,t} \right),\\
\end{aligned}
\end{equation}
where $g_\chi ^G\left( {x^*,m} \right)$ is a generating function for conjugate characters and $\overline {g_{(m)HL}^G\left( {x^*,h,t} \right)}$ is a generating function for \emph{orthonormal} modified Hall Littlewood polynomials.

The reverse transformations to recover refined Hilbert series from the HWGS can be implemented by summation over the Weyl group of G:

\begin{equation}
\label{eq:HSTx1}
\begin{aligned}
g_{HS}^G\left( {x,t} \right) & = \sum\limits_{w \in {W_G}} {w \cdot \left( {\left. {g_{HWG}^G{{\left( {m,t} \right)}}} \right|_{m \to x} {~}\prod\limits_{\alpha  \in {\Phi ^ + }} {\frac{1}{{\left( {1 - {z^{ - \alpha }}} \right)}}} } \right)},\\
g_{HS}^G\left( {x,t} \right) & = \sum\limits_{w \in {W_G}} {w \cdot \left( {\left. {g_{HWG}^G{{\left( {h,t} \right)}}} \right|_{h \to x}  {~}\prod\limits_{\alpha  \in {\Phi ^ + }} {\frac{1}{{\left( {1 - {z^{ - \alpha }}} \right)\left( {1 - {z^\alpha }t} \right)}}} } \right)},\\
\end{aligned}
\end{equation}
where the elements $w$ of the Weyl group act on the fugacities $x$ and $z$. For further detail see \cite{Hanany:2015hxa, Hanany:2016gbz}.

\section{Symmetry Factors}
\label{apxPUN}

The determination of symmetry factors from $U(N)$ Casimirs in \ref{eq:mon2} follows \cite{Cremonesi:2013lqa}. These are conditional functions that depend on the partition $\lambda \left( {{q_i}} \right)$ that enumerates the number of monopole charges $q_{i,j}$ that are equal within each monopole flux $q_i$.  Construct a partition of $N_i$ for each node, which counts how many of the charges $q_{i,j}$ are equal, such that $\lambda(q_i)=(\lambda_{i,1},\ldots,\lambda_{i,N_i})$, where $\sum\limits_{j = 1}^{{N_i}} {{\lambda _{i,j}}}  = {N_i}$. The non-zero terms $\lambda_{i,j}$ in the partition give the ranks of the residual $U(N_i)$ symmetries associated with each node.\footnote{ A $U(N)$ group has Casimirs of degrees 1 through $N$. So, for example, if $q_{i,j}= q_{i,k}$ for all $j ,k$, then $\lambda=(N_i)$ and $\{d_{i,1},\ldots d_{i,N_i}\}=\{1,\ldots, N_i$\} and if $q_{i,j}\neq q_{i,k}$ for all $j ,k$, then $\lambda=(1^{N_i})$ and $\{d_{i,1},\ldots d_{i,N_i}\}=\{1,\ldots, 1$\}.} As examples for $U(2)$, $U(3)$ and $U(4)$, we have:

\begin{equation}
\label{eq:apxPUN1}
\small
\begin{aligned}
P_{{q_i}}^{U\left( 2 \right)} & = \frac{1}{{\left( {1 - t} \right)\left( {1 - {t^2}} \right)}} \times \left\{ \begin{array}{l}
1:\lambda=  \left( 2 \right)\\
\left( {1 + t} \right):\lambda=  \left( {1,1} \right)
\end{array} \right.,\\
\end{aligned}
\end{equation}

\begin{equation}
\label{eq:apxPUN2}
\small
\begin{aligned}
P_{{q_i}}^{U\left( 3 \right)} & = \frac{1}{{\left( {1 - t} \right)\left( {1 - {t^2}} \right)\left( {1 - {t^3}} \right)}} \times \left\{ \begin{array}{l}
1:\lambda= \left( 3 \right)\\
\left( {1 + t+t^2} \right):\lambda=  \left( {2,1} \right)\\
\left( {1 + t} \right)\left( {1 + t + {t^2}} \right):\lambda=  \left( {1,1,1} \right)
\end{array} \right.,\\
\end{aligned}
\end{equation}

\begin{equation}
\label{eq:apxPUN3}
\small
\begin{aligned}
P_{{q_i}}^{U\left( 4 \right)} & = \frac{1}{{\left( {1 - t} \right)\left( {1 - {t^2}} \right)\left( {1 - {t^3}} \right)\left( {1 - {t^4}} \right)}} \times \left\{ \begin{array}{l}
1: \lambda=  \left( 4 \right)\\
\left(  {1 + t + {t^2} + {t^3}} \right):\lambda=  \left( {3,1} \right)\\
\left( {1 + {t^2}} \right)\left( {1 + t + {t^2}} \right):\lambda=  \left( {2,2} \right)\\
\left( {1 + t + {t^2}} \right)\left( {1 + t + {t^2} + {t^3}} \right):\lambda=  \left( {2,1,1} \right)\\
\left( {1 + t} \right)\left( {1 + t + {t^2}} \right)\left( {1 + t + {t^2} + {t^3}} \right): \lambda= \left( {1,1,1,1} \right)
\end{array} \right..\\
\end{aligned}
\end{equation}

\section{Background on Nilpotent Orbits}
\label{apxNO}
\subsection{Nilpotent Elements}
\label{sec:nilpotents}

The closure of a nilpotent orbit of $G$ can be considered as a moduli space described by class functions on the representation lattice of $G$. So, as a necessary preliminary to motivating the use of SUSY quiver theories and their moduli spaces in this context, it is useful to review the relationships between a group $G$, the nilpotent elements (or operators) $X$ of its Lie algebra $\mathfrak{g}$, and the nilpotent orbits ${\cal O}_X$ to which they give rise.

A nilpotent matrix $M$ over some field (taken as $\mathbb C$) is one that vanishes at some power $M^k=0$ for $k \ge d$, where $d$ is defined as the nilpotent degree of the matrix. By similarity transformation, all the eigenvalues of $M$ are zero and all its invariants vanish: $\text{det}[M]=0,\ldots, \text{tr}[M]=0$. Examples of nilpotent matrices include strictly upper (or lower) triangular matrices. Thus, a \emph{nilpositive} raising operator $X$ of a Lie algebra $\{H_{i},E_{\alpha}^+, E_{\alpha}^- \}$, defined as ${X} \equiv \sum\limits_\alpha {{u_\alpha }} E_\alpha ^ + $, for some coefficients $u_{\alpha}$, acts as a nilpotent matrix on the vector space of representations. Importantly, elements of $\mathfrak{g}$ obtained by applying a similarity transformation from $G$ to $X$ retain zero eigenvalues and remain nilpotent. This leads naturally to the concept of a nilpotent orbit defined as an equivalence class of nilpotent elements \cite{Collingwood:1993fk}:
\begin{equation}
\label{eq:nilpotents1}
{\cal O}_X=\{M: M=A X A^{-1} \text{ for } A \in G\}.
\end{equation}
The simple restriction that an element $X$ should be nilpotent can be combined with further restrictions, with respect to nilpotent degree, matrix rank, etc., to define a poset (partially ordered set) of equivalence classes of nilpotent matrices. This poset can be graphed to give a distinct \emph{Hasse diagram} for each Lie group. The boundary of all the nilpotent orbits associated with these equivalence classes is known as the \emph{closure of the maximal nilpotent orbit} or \emph{nilpotent cone} $\cal N$. Similarly, each equivalence class ${{ \cal{ O}}_X}$ gives rise to the \emph{closure} of a nilpotent orbit ${{ \cal{\bar O}}_X}$. By a common abuse of terminology, closures of nilpotent orbits ${{ \cal{\bar O}}_X}$ are often referred to simply as nilpotent orbits ${{ \cal{ O}}_X}$, and this is the convention generally adopted herein.

Now consider the Casimir invariants of a Lie group $G$, which are equal in number to its rank, with their degrees $\{d\}$ being amongst the defining properties of $G$ \cite{Fuchs:1997bb}.\footnote{The Casimirs of $G$ are symmetric invariant tensors of the adjoint representation, with their degrees being $A_n:\{2,\ldots,n,n+1\}$, $B_n/C_n:\{2, 4,\ldots,2n \}$, $D_n:\{2,4,\ldots,2n-2,n\}$, $G_2: \{2,6\}$, $F_4: \{2,6,8,12\}$, $E_6:\{2,5,6,8,9,12\}$, $E_7:\{2,6,8,10,12,14,18\}$ and $E_8:\{2,8,12,14,18,20,24,30\}$.} The simple condition that a Lie algebra element $X$ should be nilpotent entails, from the vanishing eigenvalues of $X$, that the Casimir operators formed from the traces of symmetrised products of $X$ should vanish:

\begin{equation}
\label{eq:nilpotents2}
\forall d: d \in \left\{\text{Degrees of Symmetric Casimirs of G} \right\} \to \text{tr}\left[ {{X^d}} \right] = 0.
\end{equation}
Viewed as a moduli space, the nilpotent cone $\cal N$ is therefore the quotient of the moduli space of Lie algebra generators of $G$ (the PE of the adjoint representation) divided by (its subspace of) the moduli space of Casimir invariants. The resulting Hilbert series takes the form:
\begin{equation} 
\label{eq:nilpotents3}
\begin{aligned}
g_{HS}^{{\cal N}} & =\frac{{PE\left[ {\chi _{{[adj.]}}^Gt} \right]}}{{\prod\limits_{d \in {\text{Casimirs}}\left[ G \right]} {PE\left[ {{t^d}} \right]} }} \\
& =\prod\limits_{d \in  {\text{Casimirs}}\left[ {{G}} \right]} {\left( {1 - {t^d}} \right)} PE\left[ {\chi _{{[adj.]}}^Gt} \right]\\
& = mHL^{G}_{[0,\ldots,0]}(t)
 \end{aligned}
\end{equation}
This exactly matches the definition of the modified Hall Littlewood function $mHL^G_{[0,\ldots,0]}$ \cite{Hanany:2016gbz}. So, the Hilbert series of the (closure of the) maximal nilpotent orbit is equal to $mHL^G_{[singlet]}$ and has the dimension:

\begin{equation}
\label{eq:nilpotents4}
|{\cal N}|=|\mathfrak g|-\text{rank}[\mathfrak g].
\end{equation}
The dimension of (the closure of) a general nilpotent orbit $|{\cal O}_X|$ is given by \cite{Collingwood:1993fk}:

\begin{equation}
\label{eq:nilpotents5}
|{\cal O}_X| = |\mathfrak g| - |{\mathfrak g}^X|,
\end{equation} 
where ${\mathfrak g}^X$ is the centraliser of $X$ in ${\mathfrak g}$, defined as ${{\mathfrak g}^X} \equiv \left\{ {{\mathfrak c}:{\mathfrak c} \in {\mathfrak g} {~}\&{~} [X,{\mathfrak c}] = 0} \right\}$, and consistency with $|{\cal O}_X| \le |{\cal N}| $ entails that $|{\mathfrak g}^X| \ge \text{rank}[\mathfrak g]$.

This definition of a nilpotent orbit in Lie algebra terms generalises from the matrix operators of Classical groups to the operators of Exceptional groups.


\subsection{SU(2) Homomorphisms}
\label{sec:homo}
Key methods of identifying and classifying the nilpotent orbits of $G$ follow from their relationships with $SU(2)$ homomorphisms. As described in \cite{Collingwood:1993fk}, the Jacobson-Morozov theorem shows that each nilpotent element $X$ of  ${\mathfrak g}$ falls within some \emph{standard triple} $\{H,X,Y\}$ of some $SU(2)$ subalgebra of $ {\mathfrak g}$. Also, a theorem of Kostant shows that the map from standard triples to nilpotent elements is injective, up to conjugation of the nilpotent elements. Taken together, these theorems establish a bijection between standard triples and conjugacy classes of nilpotent elements. By arguing a bijection between conjugacy classes of nilpotent elements and nilpotent orbits, (Theorem 3.2.10) \cite{Collingwood:1993fk} further claims a bijection between standard triples and (closures of) nilpotent orbits ${\cal O}_X$. Each standard triple $\{H,X,Y\}$ is in turn defined by a homomorphism (or embedding) $\rho$ from $G$ to $SU(2)$ and this implies a bijection between $SU(2)$ homomorphisms $\rho$ and distinct nilpotent orbits ${\cal O}_X$.

The possible embeddings of $SU(2)$ into $G$ were first systematically enumerated, for both Classical and Exceptional groups, by Dynkin \cite{Dynkin:1957um}. From the perspective of character analysis, each such homomorphism $\rho$ corresponds to a fugacity map between the CSA coordinates $\{x_1,\ldots,x_r\}$ of $G$ and $\{x\}$ of $SU(2)$, under which the character  ${\chi^G}$ of each representation of $G$ decomposes into a sum of characters of $SU(2)$ irreps:

\begin{equation} 
\label{eq:homo1}
\begin{aligned}
\rho :\left\{ {{x_1}, \ldots,{x_r}} \right\} & \to \left\{ {{x^{{\omega _1}}}, \ldots,{x^{{\omega _r}}}} \right\},\\
\rho : { \chi^G}\left(x_1, \ldots, x_r \right) & \to \sum \nolimits_n^ \oplus {{a_n}\left[ n \right]} \left( x \right),
\end{aligned}
\end{equation}
where the coefficients $a_n$ are non-negative integers. The exponents $[\omega_1,\ldots,\omega_r]$ in \ref{eq:homo1}, are referred to herein as the \emph{weight map} of $\rho$. The enumeration of nilpotent orbits via $SU(2)$ homomorphisms is therefore equivalent to the problem of identifying all such valid weight maps.

The number of possible homomorphisms is limited by a theorem \cite{Dynkin:1957um}, which entails that $\rho$, when expressed in terms of simple root fugacities $\{ z_1,\dots, z_r \}$ of $G$ and $\{z\}$ of $SU(2)$, must be conjugate under the action of the Weyl group of $G$ to a map of the form:
\begin{equation} 
\label{eq:homo2}
\begin{aligned}
\rho :\{ {{z_1}, \ldots {z_r}} \} \to \{ {{z^{\frac{q_1}{2}}}, \ldots,{z^ {\frac{q_r}{2}}}} \},
 \end{aligned}
\end{equation}
where ${q_i} \in \left\{ {0,1,2} \right\}$. The labels $[q_1,\ldots,q_r]$ are termed the \emph{Characteristic} of a nilpotent orbit \cite{Dynkin:1957um}, also refered to herein as a \emph{root map} of $\rho$.\footnote{The Literature also refers to a Characteristic $G[\rho]$ as the Dynkin labels (of a nilpotent orbit), not to be confused with the weight space Dynkin labels (of irreps) $[n]_{G}$. Since the labels in a Characteristic can only be 0, 1 or 2, it can be convenient to omit the separators ``,".} Thus, there are at most $3^{\text{rank}[G]}$ root maps that need to be tested, which is a straightforward computational procedure for low rank groups.\footnote{Note that root and weight fugacities and maps are related by the Cartan matrix of $G$ as $z=x^A$ and $q=A \omega$, respectively.}

These homomorphisms can also be labelled by the $SU(2)$ decomposition of $\rho(R)$, where $R$ is some representation of $G$. $R$ is usually chosen to be the fundamental representation for $A$ series groups, or the vector representation for $BCD$ series groups. Such decompositions of $\rho(R)$ are conventionally expressed using condensed partition notation, under which each $SU(2)$ irrep $[n]$ with non-zero multiplicity $a_n$ is assigned an element in the partition equal to its dimension, with an exponent equal to its multiplicity:

\begin{equation} 
\label{eq:homo3}
\begin{aligned}
\rho(R) & = \sum\limits_{n = 0}^{n_{\max} } {{a_n}\left[ n \right]}
 & \Leftrightarrow \left( {{{\left| {\left[ {{n_{\max }}} \right]} \right|}^{{a_{{n_{\max }}}}}}, \ldots ,{{\left| {\left[ n \right]} \right|}^{{a_n}}}, \ldots ,{1^{{a_0}}}} \right).
 \end{aligned}
\end{equation}

Additional selection rules are required to ensure that the representations $\rho(R)$ assigned to each irrep $R$ of $G$ are consistent with its bilinear invariants. Recall that an irrep can be classified as (i) real, (ii) pseudo real or (iii) complex, depending, respectively, on whether it has (i) a symmetric bilinear invariant with itself, (ii) an antisymmetric bilinear invariant with itself, or (iii) a bilinear invariant with its contragredient representation (complex conjugate in the case of unitary representations). As shown in \cite{Collingwood:1993fk}, when $R$ has bilinear symmetric or antisymmetric invariants, this requires \emph{irrep selection rules}, to exclude any homomorphisms $\rho$ under which such bilinears change type:

\begin{enumerate}
\item Real $R$. If a partition element (i.e. $SU(2)$ irrep) of even dimension appears, it must appear an even number of times. This ensures that any pseudo real $SU(2)$ irreps come in pairs. These are often referred to as $B$ partitions or $D$ partitions.
\item Pseudo real $R$. If a partition element (i.e. $SU(2)$ irrep) of odd dimension appears, it must appear an even number of times. This ensures that any real $SU(2)$ irreps come in pairs. These are often referred to as $C$ partitions.
\item Complex $R$. Complex irreps have bilinear invariants with their complex conjugates, rather than with themselves. Conjugate pairs of representations have identical $SU(2)$ partitions, so no selection rules apply.
\end{enumerate}
It is important to appreciate that these irrep selection rules depend on the type of representation $R$ of the parent group, upon which $\rho$ acts, and not on the parent group series (as implied in some of the Literature). The Real and Pseudo real rules apply across all representations of both Classical and Exceptional groups.

Appendix \ref{apxHom} tabulates these homomorphisms for Exceptional groups.\footnote{Homomorphisms for Classical groups up to rank 5 were tabulated in \cite{Hanany:2016gbz}.} The homomorphisms are described by their dimensions, their Characteristics (or root maps) and weight maps, and the resulting partitions of the key irreps of $G$. While partial tables are often presented in the Literature \cite{Collingwood:1993fk, Chacaltana:2012zy}, this fuller presentation, including vectors/fundamentals and the adjoint representation, is helpful for the analysis of nilpotent orbits.

In particular, the $SU(2)$ homomorphism of $[adj]_G$ encodes information about the dimension $|{\mathfrak g}^X| $ of the centraliser, and this allows calculation of the dimension of an orbit: since the highest weight of each $SU(2)$ irrep is annihilated by the $SU(2)$ raising operator $X$, $|{\mathfrak g}^X| $ is equal to the number of $SU(2)$ irreps in the partition $\rho([adj]_G)$. Thus \ref{eq:nilpotents5} can be restated in terms of the length of the adjoint partition as:

\begin{equation}
\label{eq:nilpotents5a}
|{\cal O}_X | = |\mathfrak g| - |\rho \left( {{{\left[ {adj} \right]}_G}} \right)|
\end{equation}

As an example, $G_2$ has five nilpotent orbits and these can be referred to uniquely, either by the partition data assigned (under $\rho$) to one of its representations, or by the Characteristic (root map), or by the weight map. Taking the CSA fugacities of $G_2$ as $\{x_1,x_2\}$ and the simple root fugacities as $\{ z_1={x_1^2}/{x_2^3}, z_2={x_2^2}/{x_1}\}$, and those of $SU(2)$ and $\{x\}$ and $\{z=x^2\}$, respectively, the homomorphism $\rho$ with Characteristic $[2 0]$ (and weight map $[4 2]$) can also be identified in any one of the following equivalent ways:
\begin{equation} 
\label{eq:homo4}
\begin{aligned}
\rho :\left( {{z_1},{z_2}} \right) &\to \left( {{z},{1}} \right),\\
\rho :\left( {{x_1},{x_2}} \right) &\to \left( {{x^4},{x^2}} \right),\\
\rho :\left[ {0,1} \right] & \to [2] \oplus [2] \oplus [0],\\
\rho :\left[ {0,1} \right] & \to (3^2,1).\\
\rho :\left[ {1,0} \right] & \to (5,3^3).\\
|{\cal O}_{\rho}|=10.\\
 \end{aligned}
\end{equation}
The dimension of this orbit is 10, equal to the dimension 14 of $G_2$ less the length of the adjoint partition $(5,3^3)$, which contains 4 $SU(2)$ irreps.

Intriguingly, while these $SU(2)$ homomorphisms identify all the Characteristics of Exceptional group nilpotent orbits that appear in standard tables \cite{Dynkin:1957um, Collingwood:1993fk}, this method also leads to a few extra root maps for some Exceptional groups. One extra root map arises in $F_4$; there are 3 in $E_6$, 8 in $E_7$ and 39 in $E_8$. These are highlighted in Appendix \ref{apxHom} and their moduli spaces are examined and discussed in section \ref{sec:GmodH}.

\subsection{Standard Triples}
\label{subsec:triples}

It is useful to summarise the relationship between $SU(2)$ homomorphisms and standard triples $\{H,X,Y\}$, as elaborated in \cite{Dynkin:1957um}. Standard triples are defined by the commutation relations $[H,X]= 2X, [H,Y]= -2Y, [X,Y] = H$. These operators are embedded in the Lie algebra $\mathfrak g$ of $G$, which is given by the operators $\{H_i, E_{\alpha +}, E_{\alpha-} \}$.

Now, consider a Characteristic $[q] \equiv [q_1, \ldots ,q_r]$, with corresponding weight map $[w] \equiv [w_1, \ldots ,w_r]$, related by $[q]=A \cdot [w]$. Each root, $\alpha= \small \sum\nolimits_i {{a_i}{\alpha _i}}$, where $\{\alpha_1,\ldots, \alpha_r\}$ are simple roots, is assigned a \emph{Characteristic root height}:
\begin{equation} 
\label{eq:homo5a}
[\alpha] \equiv \sum\limits_{i = 1}^r {{a_i}} {q_i}.
\end{equation}
The elements of the standard triple $\{H,X,Y\}$ are then chosen as:

\begin{equation} 
\label{eq:homo5}
\begin{aligned}
H & = \sum\limits_{i = 1}^r {{w_i}{H_i}},\\
X & = \sum\limits_{\alpha \in \Phi_G: [\alpha] = 2} {{u_\alpha }E_{\alpha +} },\\
Y & = \sum\limits_{\alpha \in \Phi_G: [\alpha] = 2} {{v_\alpha }E_{\alpha -} },
\end{aligned}
 \end{equation}
for some coefficients $u_\alpha$ and $v_\alpha$. $X$ contains only those roots with $[\alpha]=2$, and each of these satisfies the commutation relations $[H,E_{\alpha +}]= 2E_{\alpha +}$, so $[H,X]= 2X$. Similarly, $Y$ satisfies $[H,Y]= -2Y$. The commutation relation $[X,Y] = H$ constrains $u_\alpha$ and $v_\alpha$.
This analysis generalises to any $SU(2)$ homomorphism of $G$. The nilpotent operators $E_\alpha$ within in the standard triple follow directly from the Characteristic. The coefficients $u_\alpha$ and $v_\alpha$ can then be determined, up to scaling freedoms, from the Lie algebra $\mathfrak g$. 

Notwithstanding the received bijective relationship between standard triples and nilpotent orbits, there is no simple prescription in the Literature for finding the closure of a nilpotent orbit from its standard triple, although its dimension can be obtained from \ref{eq:nilpotents5a}.
\FloatBarrier
\subsection{Terminology}

It is helpful to collect some of the terminology surrounding the classification of nilpotent orbits.

\paragraph{Canonical Orbits}
The dimensions of nilpotent orbits have a partial ordering, which is often expressed using Hasse diagrams. Formally, this partial ordering is defined by inclusion relations amongst the closures $ \cal \bar O$ of nilpotent orbits $\cal O$.\footnote{The closures $ \cal \bar O$ correspond to the quiver theory moduli spaces that are calculated in this study.} There are a number of canonical orbits within this partial ordering:
\begin{enumerate}
\item The \emph{trivial orbit}. This is associated with the partitions $\rho^G(R)=(1^{\left | R \right|})$ and always has zero dimension.
\item The \emph{minimal orbit}. This is the first orbit with non-zero dimension and is always unique. Its complex dimension is equal to twice the sum of the dual Coxeter labels of $G$. This equals the dimension of the reduced single instanton moduli space of $G$.
\item The \emph{sub-regular orbit}. This is the orbit with next to highest dimension. It is always unique, having a complex dimension equal to the number of the roots of $G$, less $2$.
\item The \emph{maximal orbit}. This is the orbit with highest dimension and is always unique. Its complex dimension is equal to the number of roots of $G$. This equals the dimension of the modified Hall Littlewood function $mHL_{[0,\ldots,0]}^G$.

\end{enumerate}

The above orbits are not distinct for low rank groups. For example, in $A_1$, the minimal and maximal orbits coincide, as do the trivial and sub-regular. 

\paragraph{Distinguished Orbits}
A \emph{distinguished} nilpotent element is associated with an $SU(2)$ homomorphism in which $\rho^G(adj.)$ contains no $SU(2)$ singlets \cite{Collingwood:1993fk}. This rule leads to the following list of distinguished nilpotent orbits:\footnote{The list of distinguished Exceptional group Characteristics appears in table 23 of \cite{Dynkin:1957um}.}

\begin{description}
\item $A_r$: Maximal nilpotent orbit only, 
\item $B_r$: Partitions of $2r+1$ into distinct odd parts,
\item $C_r$: Partitions of $2r$ into distinct even parts,
\item $D_r$: Partitions of $2r$ into distinct odd parts,
\item $G_2$: [20] and [22],
\item $F_4$: [0200], [0202], [2202] and [2222],
\item $E_6$: [202020] [220222] and [222222],
\item $E_7$: [0020020], [2020020], [2020220], [2202022], [2202222] and [2222222],
\item $E_8$: [00020000], [00200020], [00200200], [00200220], [20200200], [20200220], [20202020], [20202220], [22020222], [22022222] and [22222222].
\end{description}

\paragraph{Even Orbits}
\label{subsec:EvO}
An \emph{even} nilpotent orbit is one that has a Characteristic containing the labels 0 or 2 only. All distinguished orbits are even \cite{Collingwood:1993fk}.

\paragraph{Richardson Orbits}
\label{subsec:RichO}
A \emph{Richardson} nilpotent orbit is one that can be induced from the trivial nilpotent orbit of a subgroup \cite{Collingwood:1993fk}. Every nilpotent orbit that has a Characteristic containing only the labels 0 or 2 has a quotient group $G/H$ structure and can be induced, as explained in section \ref{sec:NON}, from the trivial nilpotent orbit of the subgroup $H$, whose Dynkin diagram is defined by the 0 labels of the Characteristic. All even orbits are thus Richardson orbits. In addition, some groups have \emph{non-even Richardson} orbits, with the rules for identifying such orbits being given in \cite{Fu:2004uq}. Richardson orbits have polarizations \cite{HESSELINK:1978fp} and symplectic resolutions \cite{Fu:2004uq}. The complete set of Richardson orbits is:

\begin{description}
\item $A_r$: All nilpotent orbits,
\item $B_r$: Partitions of $2r+1$, whose first $q$ parts are odd, where $q$ is \emph{odd}, with the remaining parts even,
\item $C_r$: Partitions of $2r$, whose first $q$ parts are odd, where $q$ is \emph{even}, with the remaining parts even,
\item $D_r$: Partitions of $2r$, whose first $q$ parts are odd, with the remaining parts even, and either (i) $q$ is even but $q \neq 2$, or (ii) $q=2$ and the two odd parts are located at positions $2k-1$ and $2k$ for some integer $k$,
\item $EFG$: All even orbits, plus
\item $F_4$: [1012],
\item $E_6$: [100010], [010100], [100012], [110111] and [110112],
\item $E_7$: [0100011], [1010100], [2010100], [2101101] and [2101021],
\item $E_8$: [01001002], [101010000], [21010220], [01000120], [10101010], [10101020] and [20101020].

\end{description}

\paragraph{Rigid vs Non-Rigid Orbits}
\label{subsec:RigO}
A \emph{non-rigid} nilpotent orbit is one that can be induced from some nilpotent orbit of a subgroup. All Richardson orbits are thus non-rigid, being induced from a trivial nilpotent orbit. Importantly, any orbit whose Characteristic contains 2 can be induced from the orbit of the subgroup defined by the Dynkin diagram and Characteristic that remains after removing one or more nodes with Characteristic 2 from the parent diagram.

Conversely, a \emph{rigid} nilpotent orbit is one that \emph{cannot} be induced from a nilpotent orbit of a subgroup. A rigid nilpotent orbit has a Characteristic containing 0 and 1 only, as a necessary, but not sufficient, condition. Notably, the minimal nilpotent orbits of simple groups, other than those isomorphic to the A series, are rigid \cite{Collingwood:1993fk}. Also, for example, ${D_4}[1011]$ is rigid amongst orbits of low rank groups. Rigid orbits of Exceptional groups are identified in \cite{Adams:jk}.

The inclusion relations between the above types of orbit provide a classification scheme:
\begin{equation}
\label{eq:nilpotentsXXXXXXXXX}
\begin{array}{c}
\left\{ {Nilpotent~Orbits} \right\} = \left\{ {Rigid} \right\} \cup \left\{ {Non - Rigid} \right\}\\
\\
\left\{ {Non - Rigid} \right\} \supset \left\{ {Richardson} \right\} \supset \left\{ {Even} \right\} \supset \left\{ {Distinguished} \right\}
\end{array}
\end{equation}

\paragraph{Special Orbits}
\label{subsec:SO}
A \emph{special} nilpotent orbit is one that is invariant under two applications of the Spaltenstein map. For Classical groups the Spaltenstein map is defined by fundamental/vector partition transposition, followed, if the transpose partition is not valid under the Real/Pseudo real selection rules, by $BCD$-collapse to a lower partition. The Spaltenstein map $d$ is thus many to one, often described as $d^3 = d$, and can lead to the conflation of distinct nilpotent orbits, as discussed in \cite{Hanany:2016gbz}. All $A$ series nilpotent orbits are special. A special $BC$ series nilpotent orbit is one whose Spaltenstein map does not require $BC$ collapse.

A Spaltenstein map can also be defined for Exceptional groups. All Richardson orbits are special, as is any orbit of a higher rank group induced from a special orbit \cite{Collingwood:1993fk}. Some rigid orbits are also special.

\paragraph{Normal vs Non-Normal Orbits}
\label{subsec:NNO}
From the perspective of this study, a more important distinction is that between \emph{normal} and \emph{non-normal} nilpotent orbits. A normal symplectic variety only contains singularities that are rational Gorenstein \cite{Baohua-Fu:2015nr}, and by virtue of a theorem in \cite{stanley_1978}, this entails that it is Calabi-Yau with a palindromic Hilbert series \cite{Gray:2008yu}. Consistent with this, the \emph{normal} nilpotent orbits of Classical groups were found in \cite{Hanany:2016gbz} to have palindromic Hilbert series; however, \emph{non-normal} nilpotent orbits were found to have non-palindromic Hilbert series.

The \emph{normalisation} of a nilpotent orbit can be defined as a palindromic moduli space of the same dimension that forms a covering space. A normal nilpotent orbit is its own normalisation. Normalisations of non-normal nilpotent orbits contain elements outside the nilpotent cone $\cal N$.

For Classical groups, it was shown in \cite{Kraft:1982fk}, based on a geometric analysis, that the vector partition of a non-normal orbit is always related to that of the orbit immediately below it, by a particular degeneration of its Young's diagram. In this degeneration, a pair of even rows in some \emph{sub-diagram}, described by the partition $(2r,2r)$, degenerates to $(2r-1,2r-1,1,1)$; all the rows above and all the columns to the left of the sub-diagram remain unchanged. Such degenerations result from a $D_{2r}$ subalgebra of a $BCD$ series parent (all $A$ series nilpotent orbits are normal) and are termed $A_{2r-1} \cup A_{2r-1}$ degenerations. A $D_{2r}$ group may have several degenerations associated with its spinor pairs, including $A_{2r-1} \cup A_{2r-1}$ and $A_{1} \cup A_{1}$ degenerations.

A similar situation arises in Exceptional groups, where non-normal nilpotent orbits are also associated with particular degenerations of their partitions \cite{Baohua-Fu:2015nr}. The non-normal nilpotent orbits of Exceptional groups are listed in \cite{ Broer:1998qf}, being:

\begin{description}
\item $G_2$: [01],
\item $F_4$: [0002], [2001], [0101], [1010], [1012], \text{(5 cases)},
\item $E_6$: [100011], [200020], [100012], [010101], [200022], \text{(5 cases)},
\item $E_7$:  [2000100], [2000020], [1010000], [1001010], [0100011], [0010100], [2000200], [2000220], [0101021], [2101021], \text{(10 cases)},
\item $E_8$: [10000020], [00001010], [00000220], [0100010], [10001000], [20000020], [00000121], [10001020], [20001010], [00100020], [00000022], [20000200], [20000220], [10100010], [01001010], [01000101], [10010100], [00101000], [10010120], [20002000], [01000121], [00101020], [20002020], [21000121], [20002220], [20101020], [20020020], [01010221], [21010221], \text{(29 cases)}.
\end{description}

The non-normal orbits of Exceptional groups occur amongst all types other than distinguished and their relationships with their normalisations are complicated \cite{Baohua-Fu:2015nr}. It is conjectured in \cite{Broer:1998qf} that all distinguished nilpotent orbits are \emph{normal}.

\clearpage

\section{Exceptional Group Nilpotent Orbits and $SU(2)$ Homomorphisms}
 \label{apxHom}

\subsection{$G_2$}
\begin{center}
~\\
\includegraphics[scale=.7]{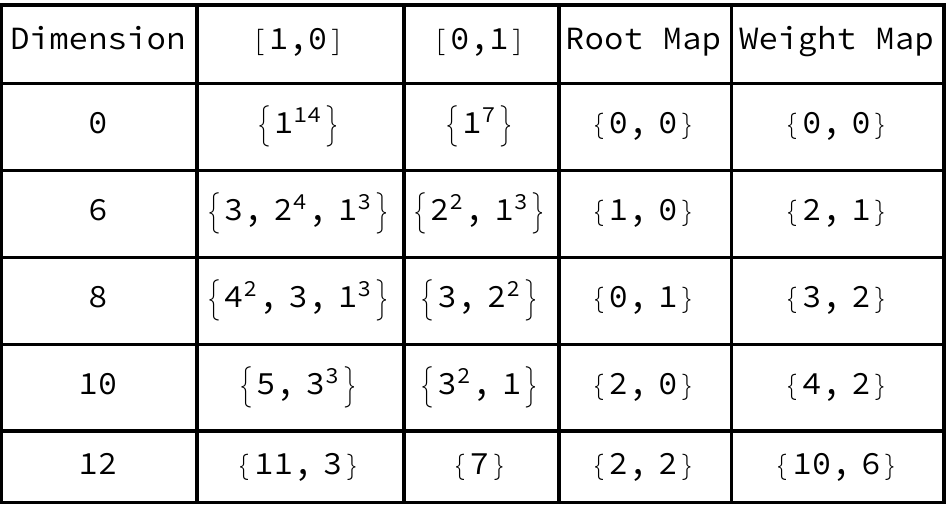}\\
\end{center}

\subsection{$F_4$}
\begin{center}
~\\
\includegraphics[scale=.7]{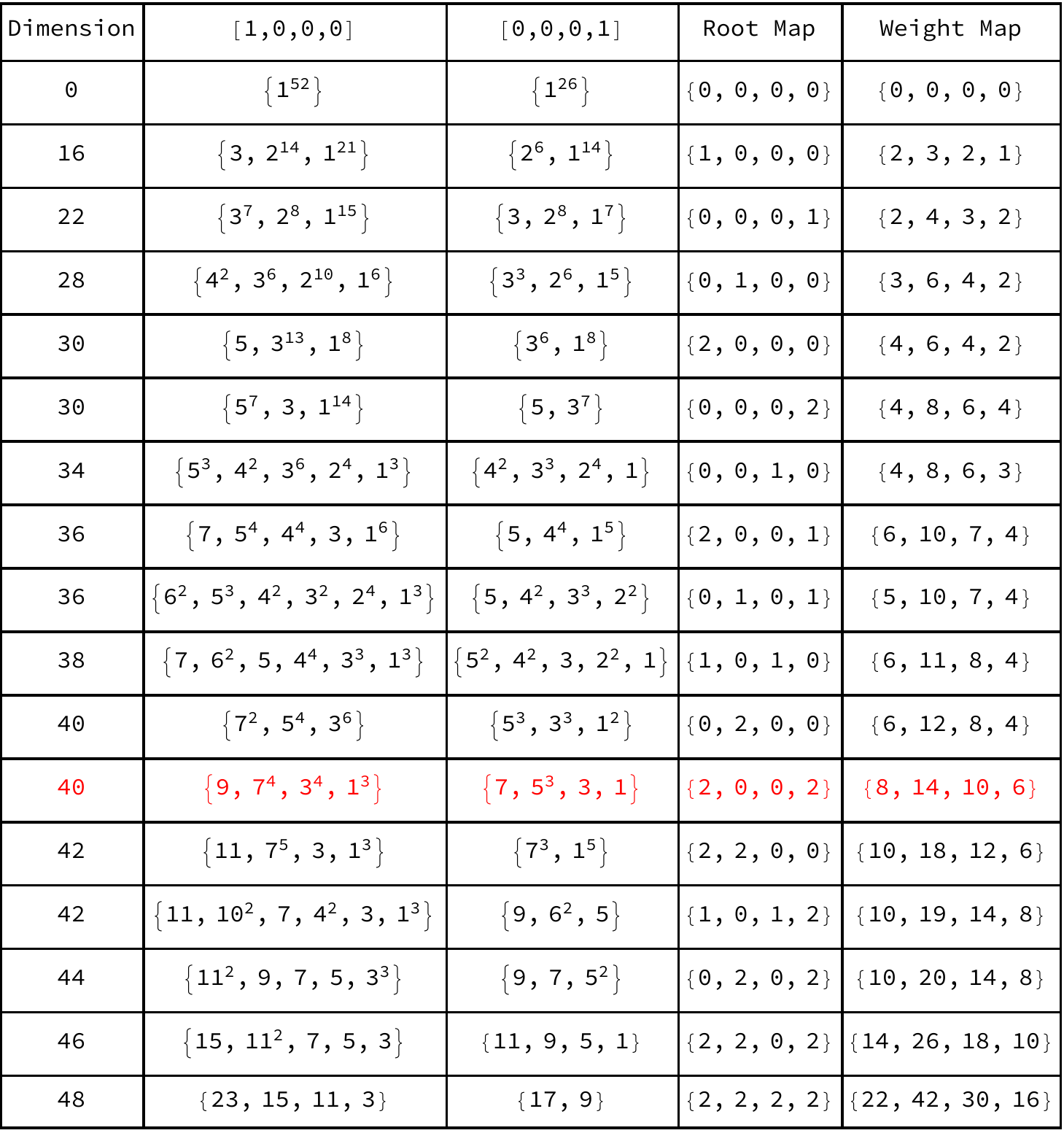}\\
\end{center}
Partitions are shown for the $F_4$ adjoint and vector representations only. Homomorphisms identified include one root map which is not a nilpotent orbit: 40: [2,0,0,2]. This is highlighted in red.
\clearpage

\subsection{$E_6, E_7, E_8$}
\begin{center}
~\\
\includegraphics[scale=.7]{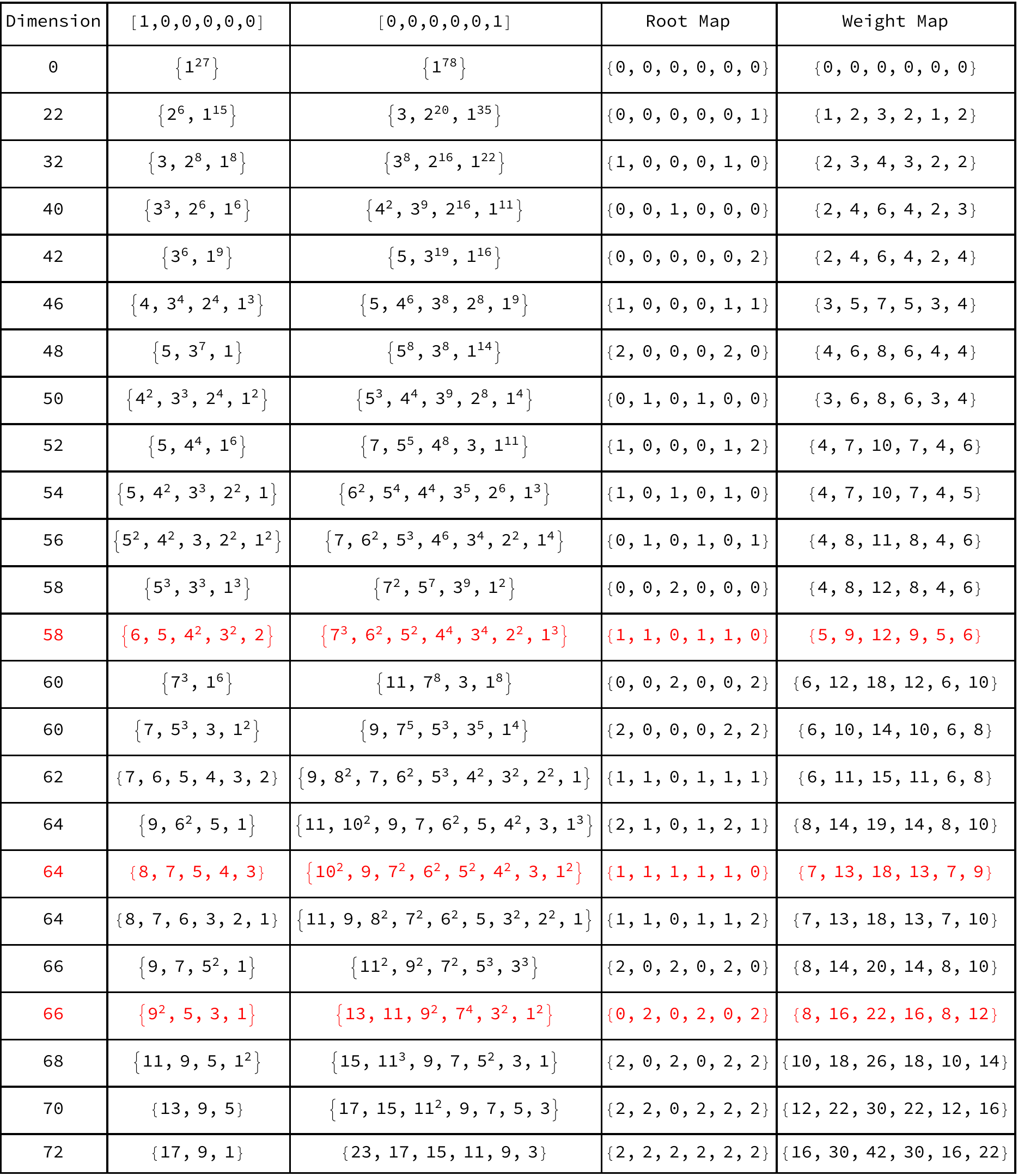}\\
\end{center}
Partitions are shown for the $E_6$ adjoint and fundamental representations only. Homomorphisms identified include three root maps which are not recognised Characteristics of nilpotent orbits: these are highlighted in red.
\clearpage

\begin{center}
~\\
\includegraphics[scale=.6]{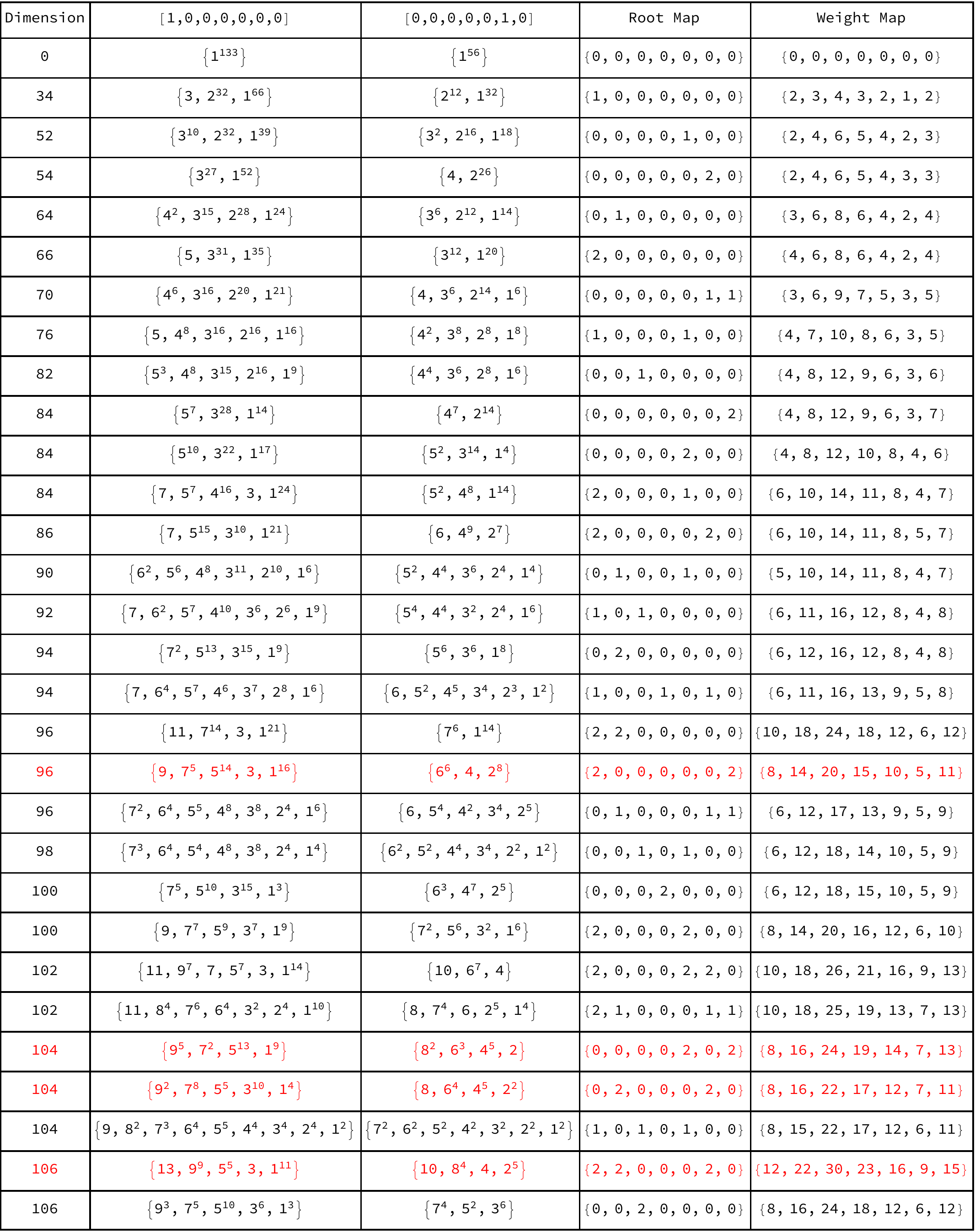}\\
~\\
\includegraphics[scale=.6]{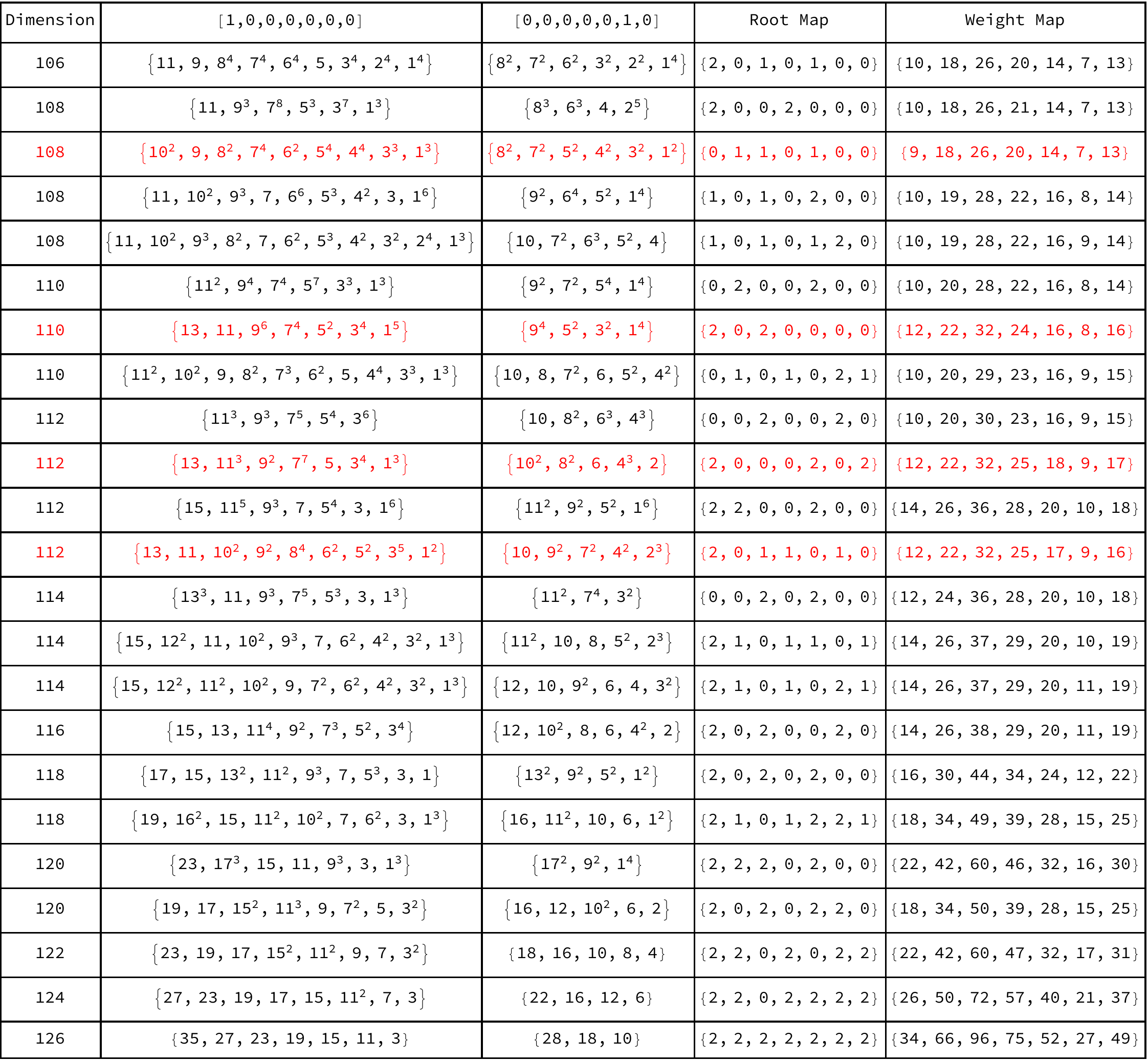}\\
\end{center}
Partitions are shown for the $E_7$ adjoint and vector representations only. Homomorphisms identified include eight root maps which are not recognised Characteristics of nilpotent orbits: these are highlighted in red.
\clearpage

\begin{center}
~\\
\includegraphics[scale=.58]{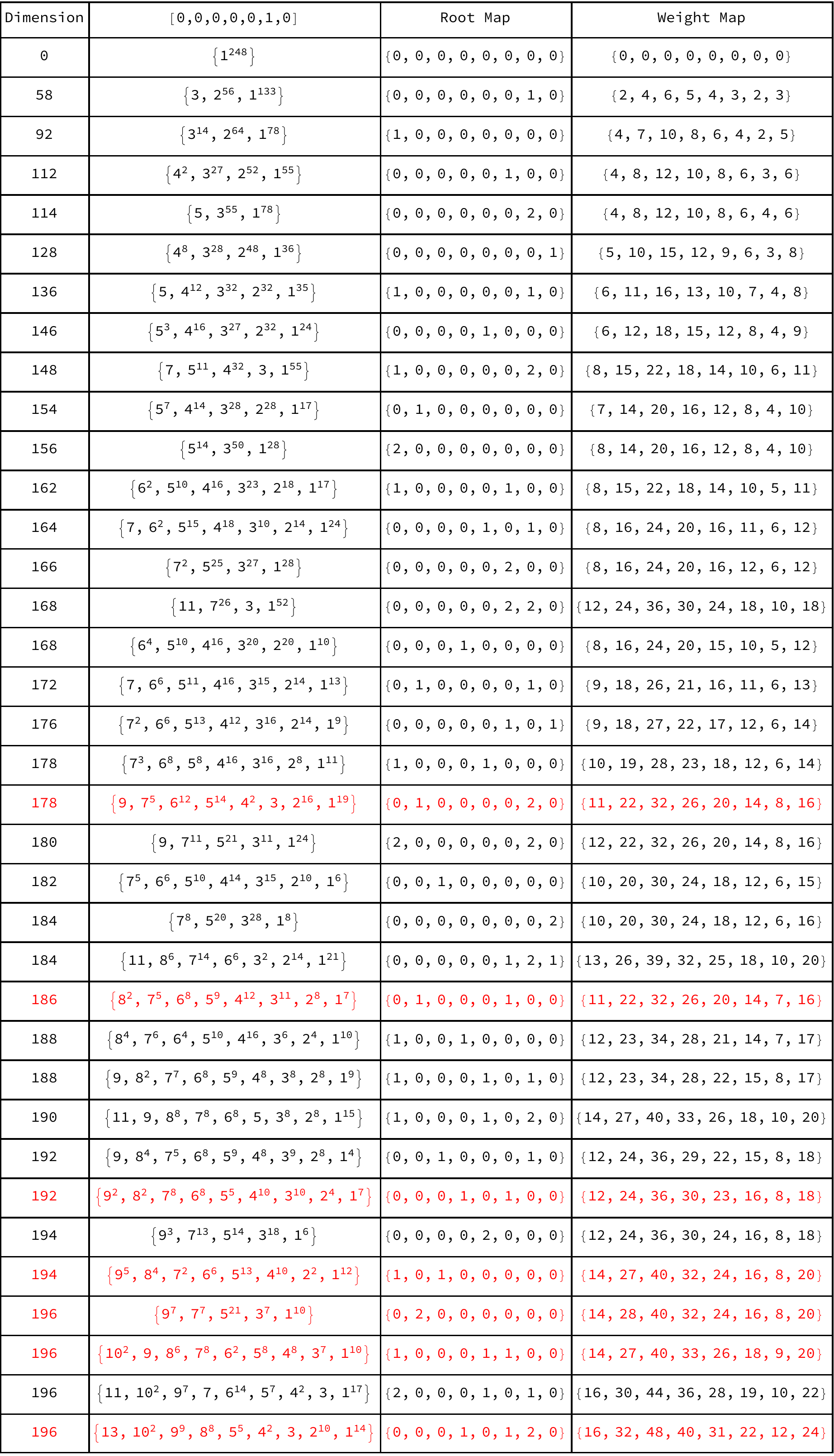}\\
~\\
\includegraphics[scale=.6]{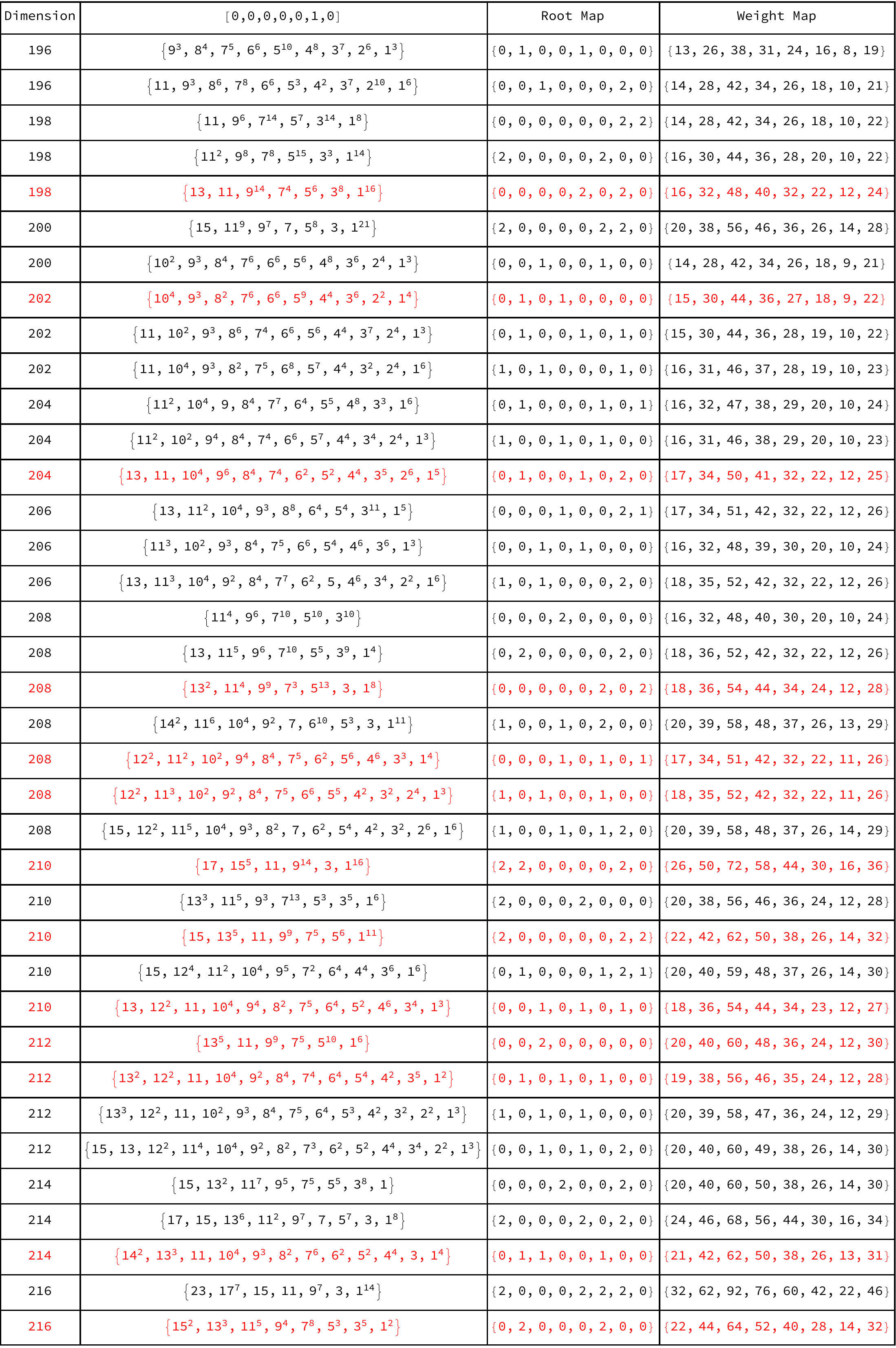}\\
~\\
\includegraphics[scale=.58]{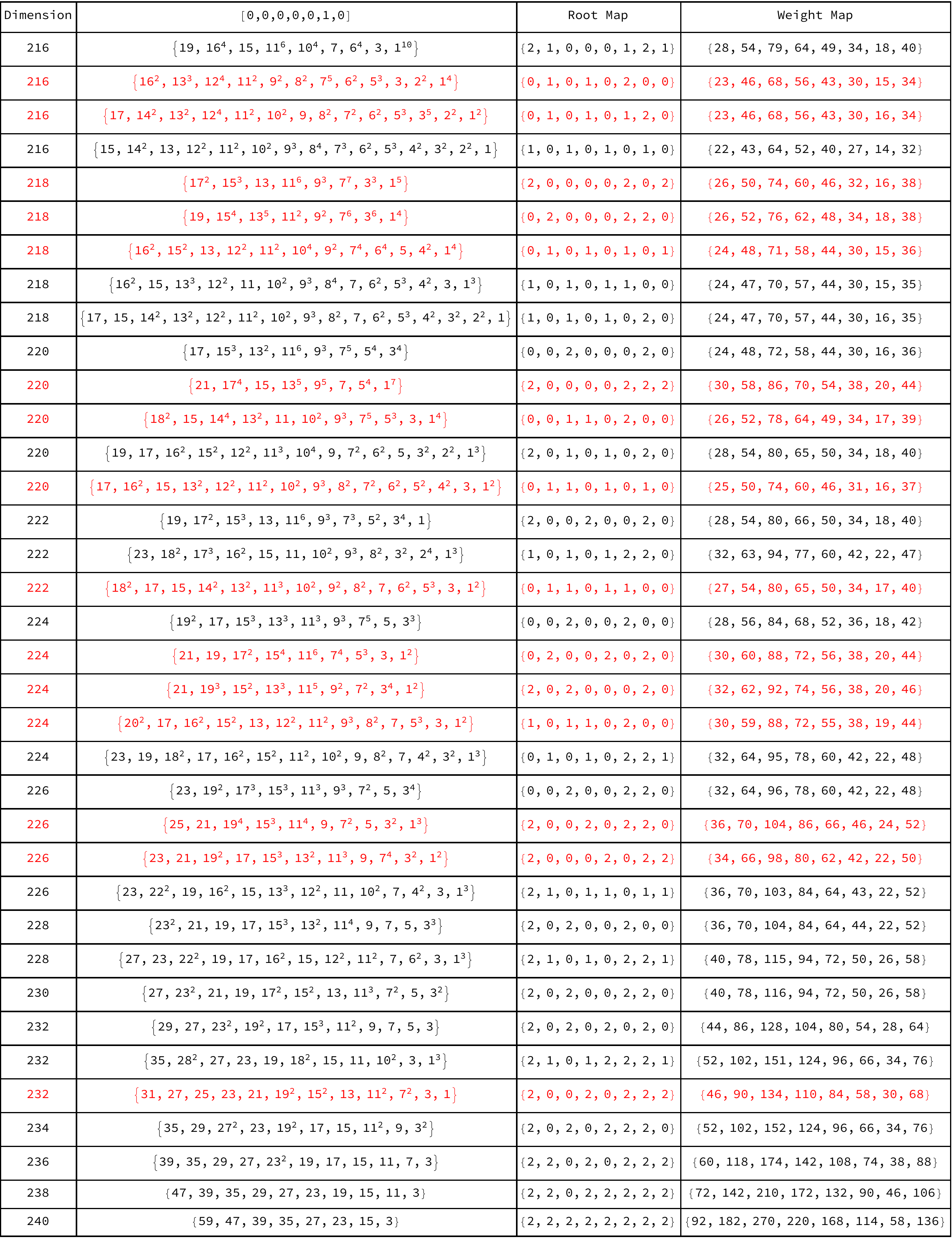}\\
\end{center}
Partitions are shown for the adjoint representation only. Homomorphisms identified include 39 root maps which are not recognised Characteristics of nilpotent orbits: these are highlighted in red.

\clearpage


\bibliographystyle{JHEP}
\bibliography{RJKBibLib}


\end{document}